\documentclass[
 prl,twocolumn,amsmath,amssymb,
 aps,superscriptaddress
]{revtex4-2}

\usepackage[colorlinks=true,allcolors=blue]{hyperref}
\usepackage{hyperxmp}
\usepackage{bm}
\usepackage{float}
\usepackage{booktabs,tabularx,array}
\usepackage{graphicx,color}
\usepackage[dvipsnames]{xcolor}

\usepackage{enumerate}
\usepackage{harpoon}
\usepackage{amsmath,amssymb}
\usepackage{gensymb}
\usepackage{accents}
\usepackage[e]{esvect}
\usepackage[type={CC},modifier={by},version={4.0}]{doclicense}
\usepackage{footmisc}

\definecolor{ctcolor}{rgb}{0.8, 0.0, 0.2}

\begin{document}

\preprint{APS/123-QED}
\vspace*{15px}

\title{Odd elasticity in disordered chiral active materials}
\thanks{\doclicenseThis}
\author{Cheng-Tai Lee}
\email[]{chengtailee@tauex.tau.ac.il}
\affiliation{School of Mechanical Engineering, Tel Aviv University, Tel Aviv 69978, Israel}
\author{Tom C. Lubensky}
\affiliation{Department of Physics and Astronomy, University of Pennsylvania, Philadelphia, PA 19104, USA}
\author{Tomer Markovich}
\affiliation{School of Mechanical Engineering, Tel Aviv University, Tel Aviv 69978, Israel}
\affiliation{Center for Physics and Chemistry of Living Systems, Tel Aviv University, Tel Aviv 69978, Israel}

\begin{abstract}
Chiral active materials are abundant in nature, including the cytoskeleton with attached motor proteins, rotary clusters of bacterial flagella, and self-spinning starfish embryos. These materials break both time reversal and mirror-image (parity) symmetries due to injection of  torques at the microscale. {It was recently discovered} that chiral active materials show a new type of elastic response termed `odd' elasticity. Currently, odd elasticity is understood microscopically only in ordered structures, e.g., lattice designs of metamaterials. It remains to explore how odd elasticity emerges in natural or biological systems, which are usually disordered. To address this, we propose a minimal generic model for disordered `odd solids', using micropolar (Cosserat) elasticity in the presence of local active torques. We find that odd elasticity naturally emerges as a nonlinear effect of \emph{internal} particle rotations. Exploring the viscoelasticity of such a solid, when immersed in an odd fluid, we discover new dynamically unstable regions driven by the odd solid-fluid coupling, and, in the underdamped regime, also by inertia. Remarkably, in the overdamped limit, this odd solid-fluid coupling allows for bulk wave propagation near these unstable regions.
\end{abstract}

\maketitle

Chiral active materials~\cite{furthauer2012,liebchen2022,bowick2022,shankar2022,markovich2019} generate motion with preferred helical or rotational direction, generally due to the presence of torques at the microscale, which are fueled by local energy consumption (e.g., ATP hydrolysis). These materials are abundant in nature, with numerous examples such as the cytoskeleton that is twisted by motor proteins~\cite{naganathan2014,ramaiya2017,novak2018,afroze2021,meissner2024}, rotary clusters~\cite{grauer2021,vincenti2019} formed by torques due to bacteria flagella~\cite{diluzio2005,lauga2006,xing2006,mandadapu2015}, and self-spinning starfish embryos \cite{tan2022}. {Artificial chiral active materials were also realized recently via torque-driven colloids~\cite{snezhko2016,soni2019,shelke2019,schmidt2019,zhang2020,massana-cid2021,alvarez2021} or active granular spinners~\cite{scholz2018,lopez-castano2022,caprini2025c}, where external magnetic field or light serves as the energy (activity) source.} 



Due to their broken time-reversal and mirror-image (parity) symmetries, chiral active materials exhibit remarkable mechanical properties, which received a lot of attention recently. Specifically, they show a new type of viscosity dubbed odd viscosity~\cite{avron1998,banerjee2017,souslov2019,han2021,markovich2021,markovich2024,hosaka2023a,hosaka2024,banerjee2025,marconi2026} that is nondissipative and a new type of elastic response that is related to a new `odd' elastic modulus~\cite{scheibner2020,braverman2021,fossati2024,fruchart2023,banerjee2025}. Odd elasticity (viscosity) refers to the antisymmetric part of the elasticity (viscosity) tensor~\cite{fruchart2023}. 
In two-dimensional (2D) systems, odd elasticity/viscosity non-reciprocally couples the two shear deformations, such that a pure shear stress will result in a simple strain (strain rate), while simple shear stress results in a negative pure shear strain (strain rate). 
Among the unique mechanical properties of these materials are tilting under compression of an odd elastic material~\cite{scheibner2020}, growing wave modes~\cite{scheibner2020,markovich2021,markovich2024}, and unidirectional surface waves~\cite{abanov2018,souslov2019,soni2019,fossati2024,veenstra2025a,gao2022,caprini2025a,lee2026}.

The origins of odd viscosity in active matter have been extensively studied~\cite{banerjee2017,souslov2019,markovich2021,markovich2024,han2021,franca2025}, while progress in studying odd elasticity was mainly focused on ordered structures such as lattice designs of metamaterials with nonreciprocal springs \cite{scheibner2020,zhou2020,chen2021,veenstra2025a,shaat2023,veenstra2024,caprini2025,nemeth2026} or anisotropic prestress \cite{engstrom2025}.
Recently, some progress was achieved in understanding less ordered structures such as liquid crystals with nontrivial active chiral forces~\cite{kole2021,kole2024}. 
It remains unclear how, and whether, odd elasticity can emerge in structurally disordered elastic materials frequently-seen in biological and synthetic systems~\cite{naganathan2014,ramaiya2017,novak2018,afroze2021,meissner2024,snezhko2016,howard2019}.

In this Letter, we propose a minimal generic model for `odd solids' that reveals the required ingredients for odd elasticity to appear in structurally disordered elastic materials.
Essentially, all that is needed is injection of local active torques, and odd elasticity emerges from leading-order geometric nonlinearity (i.e., nonlinear strain), while a linear stress-strain relation suffices.

%
%
Using Poisson-bracket formalism we find the elasticity tensor $C_{ijkl}$ associated with Cauchy stress, which is symmetric with respect to $i\leftrightarrow j$ (as required by balance of angular momentum) and contains one `odd' modulus (usually referred to as odd elasticity~\cite{scheibner2020}) proportional to the active torque density.  
We further study viscoelasticity (Kelvin-Voigt model) of chiral active materials by immersing an odd solid in an odd fluid and find both 
unstable regions in which the solid is no longer homogeneous, and regions where `odd' mechanical waves propagate even in an overdamped solid.
%



\begin{figure}[t]
	\centering
	\includegraphics[width=8 cm]{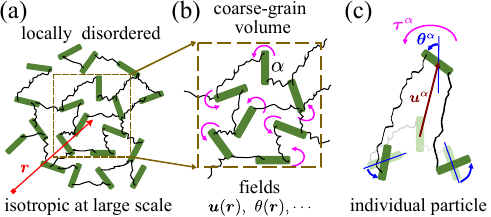}
	\caption{(a) Illustration of an elastic material composed of rigid  rod-like particles. Importantly, our model applies for any other complex rigid particles (granules, colloids, fiber composites, etc.).   
    (b) Coarse-graining at position $\bm{r}$ in the \textit{undeformed/Lagrangian} space. In this work we consider a locally-disordered, isotropic elastic material, in the presence of local active torque $\tau^\alpha$ ($\alpha$ being the particle index). The various fields ${\bm X}({\bm r})$ are the average of the particle's ${\bm X}^\alpha$ within the coarse-graining volume.
    (c) Particle displacement $\bm{u}^\alpha$ and internal rotation $\theta^\alpha$ away from the rest position and rest orientation (blue line) of the undeformed state. The internal rotation is measured with respect to the \textit{individual} rest orientation (not a universal direction).
 } 	\label{fig:CG scheme}
\end{figure}

We describe the 2D elastic material as made of identical complex particles (namely, not point-like) such that torques are applied at the particle level, see Fig.~\ref{fig:CG scheme}. Then, to incorporate local torques,  we consider the effect of local \textit{internal} rotation of material particles in the spirit of the well-known Cosserat elasticity~\cite{eringen1966,eringen1999,eremeyev2013}. In doing so, particles are treated as rigid, allowing only translations of center of mass (CM) and rotations around the CM.
The material is further assumed to be isotropic and homogeneous on large scales (but can be locally disordered). This would be the case in biological gels such as the cytoskeleton~\cite{Broedersz2014}. Importantly, deformations of such material are caused not only by displacement of particles CM, $\bm{u}^\alpha$, but also by their internal rotation $\theta^\alpha${, which, due to the finite particle size, effectively gives rise to  transverse forces} (see Fig.~\ref{fig:CG scheme}). Hence, another degree of freedom is introduced per particle, resulting in another deformation field and strain~\cite{eringen1999}, compared to classical elasticity~\cite{LLelastity}.

%


The local disorder is manifested by the random orientation of the particles (rods in Fig.~\ref{fig:CG scheme}) in the undeformed state, thus no local order is formed (i.e., nematic). Accordingly, the  $\alpha$'s particle \textit{internal} rotation $\theta^\alpha$ is defined as the rotation angle away from its rest orientation in the undeformed state~\footnote{$\theta^\alpha$ is different from the angle used in magnetic spin systems, which is the deviation from a universal direction defined by an external magnetic field.}, see Fig.~\ref{fig:CG scheme}(c). 
To describe material deformations on large-scales we use the coarse-grained (CG) fields $\theta(\bm{r})$ and $\bm{u}(\bm{r})$, defined via particle-averaging ${\bm X}(\bm{r})\equiv \sum_{\alpha \in \Delta V^\circ} {\bm X}^\alpha \delta(\bm{r}-\bm{r}^\alpha)/n^\circ(\bm{r})$. Here $\delta(\bm{r})$ is the Dirac delta function, $\bm{r}^\alpha$ is particle $\alpha$ position in the \textit{undeformed/Lagrangian} space, $\Delta V^\circ$ is the coarse-graining volume, and the particle number density in the undeformed space is $n^\circ(\bm{r}) = \sum_{\alpha \in \Delta V^\circ}\delta(\bm{r}-\bm{r}^\alpha)$. 
%


Chirality is introduced via injection of active \textit{internal} torques $\tau^\alpha$ acting on individual particles. These create nonzero internal rotation of particles in the solid. CG these internal active torques introduces a linear term in the elastic energy, $V_{\rm{torque}} = -\sum_{\alpha \in \Delta V^\circ} \tau^\alpha \theta^\alpha \delta(\bm{r}-\bm{r}^\alpha) \approx -\tau^\circ(\bm{r})\theta(\bm{r})$, where the CG torque density is $\tau^\circ(\bm{r})=\sum_{\alpha \in \Delta V^\circ}\tau^\alpha\delta(\bm{r}-\bm{r}^\alpha)$. Despite the distinct physical mechanism, this term is similar to an external field, although no global alignment is present~\footnote{This CG potential is obtained using the approximation~\cite{markovich2024} $\sum_{\alpha \in \Delta V^\circ} \tau^\alpha \theta^\alpha \delta(\bm{r}-\bm{r}^\alpha) \approx \sum_{\alpha \in \Delta V^\circ} \tau^\alpha \delta(\bm{r}-\bm{r}^\alpha)\sum_{\beta \in \Delta V^\circ}  \theta^\beta \delta(\bm{r}-\bm{r}^\beta) / n^\circ(\bm{r})=\tau^\circ(\bm{r})\theta(\bm{r})$. The CG potential due to an external field is: 
$\tilde\tau(\bm{r})\sum_{\alpha \in \Delta V^\circ}  \theta^\alpha \delta(\bm{r}-\bm{r}^\alpha)=\tilde\tau(\bm{r}) n^\circ(\bm{r}) \theta(\bm{r}) =\tau(\bm{r}) \theta(\bm{r})$, where $\tilde\tau$ is the torque due to the external field.}.

%

To continue we write the Hamiltonian
$H$ that is composed of the kinetic energy density, the active torque potential, and an elastic potential density $V$: 
\begin{align}\label{eq:H}
H=\int d\bm{r} \
\bigg(
\frac{\left({\bm g}^{\circ}\right)^2}{2\rho^\circ} 
+ \frac{\left(\ell^{\circ}\right)^2}{2 I^\circ}
- \tau^\circ\theta   
+ V 
\bigg)
\, .
\end{align}
The first two terms are the CM and rotational kinetic energies~\cite{markovich2024,Goldstein_book}, where the CM momentum density $\bm{g}^{\circ}$, angular momentum density $\bm\ell^\circ$,  mass density $\rho^\circ$, and the density of the moment of inertia $I^\circ$, are defined as
${\bm X}^\circ(\bm{r})\equiv \sum_{\alpha \in \Delta V^\circ} {\bm X}^\alpha \delta(\bm{r}-\bm{r}^\alpha)$.
%
%
Since we assume all particles are identical, these density fields $\bm{X}^\circ \propto n^\circ$.
The superscript in $\bm{X}^\circ$ refers to values of density-related fields in the \textit{undeformed/Lagrangian} space, which is different from their values in the \textit{deformed/Eulerian} space (namely, the real space), due to the local volume change. For the displacement field $\bm{u}$ and internal rotation field $\theta$, there is no such difference as they are defined via particle-averaging.

To write the elastic potential, one must first find the strain. Following ideas from micropolar elasticity by  Eringen~\cite{eringen1999},  we calculate the  distance change between two points in the material. Unlike classical elasticity, here both CM displacement $\bm{u}$ and internal rotation angle $\theta$ are required to describe length change. We find that, for small particles (compared with the CG length-scale), there are two strain measures (see details in SI Sec.~\ref{app:micropolar V}): 
the Green-Lagrange strain $u_{ij}$ and another strain due to internal rotation, $e_{ij}$~\cite{eringen1999,eremeyev2013}:
%
%
%
%
\begin{equation}
\label{eq:geometric strain def}
u_{ij}=
\frac{1}{2}(F_{ki}F_{kj}-\delta_{ij})
\approx u^s_{ij}
\,\,\, ; \,\,\, 
e_{ij}=O_{ki}F_{kj}-\delta_{ij}
\, ,
\end{equation}
where  $F_{kj} = \nabla^\circ_j R_k$
is the deformation gradient tensor with ${\bm R}({\bm r}) = {\bm r} + {\bm u}({\bm r})$ being the position in the deformed space, and $O_{ik}=\delta_{ik}
- \varepsilon_{ik} \sin\theta
- (1-\cos\theta)\delta_{ik}$ is the 2D rotation matrix. Here $\nabla^\circ_i \equiv \partial/\partial r_i$.
Both strains are rotation-invariant, see SI Sec.~\ref{app:micropolar V}. 

In principle, the strains can also depend on $\nabla^\circ_j\theta$~\cite{chaikin1995}, however, in the hydrodynamic limit these terms are negligible and ignored hereafter {(see SI Sec.~\ref{app:micropolar V})}.
We further assume small deformations, but allow for geometric nonlinearities due to active torques. Specifically, we keep terms to linear order in $\nabla^\circ_j u_k$ and quadratic order in $\theta$. 
In this approximation $u_{ij}$ is the regular linear symmetric strain $u^s_{ij}\equiv(\nabla^\circ_j u_i+\nabla^\circ_i u_j)/2$ 
and $e_{ij} \approx e_{ij}^{(1)} + e_{ij}^{(2)}$ where 
$e_{ij}^{(1)}=u_{ij}^s
+\varepsilon_{ij}
[\theta-(\nabla^\circ\times\bm{u}/2)
]$ is the linear Cosserat elastic strain~\cite{eringen1966,eringen1999,eremeyev2013} 
that captures the strain due to mismatch between orbital and internal rotations. The second-order correction $e_{ij}^{(2)}=\varepsilon_{ik}\theta
\nabla^\circ_j u_k 
-(\delta_{ij}\theta^2/2)$ is required to get odd elasticity as explained below. 
%



Now that the strain is known, the elastic potential $V$ is obtained by assuming a linear stress-strain relation
\begin{equation}
\label{eq:V_linear}
V
=
\frac{1}{2}
\big[
u_{ij}\bar{E}_{ijkl}u_{kl}
+
e_{ij}C'_{ijkl}e_{kl}
+
2 u_{ij}\hat{E}_{ijkl}e_{kl}
\big] \, ,
\end{equation}
which has the well-known quadratic form of classical elasticity, but now with two strain measures (Eq.~\eqref{eq:geometric strain def}). The form of the elastic tensors $\bar{\bm E}$, ${\bm C}'$, and $\hat{\bm E}$ is found from symmetry arguments. Together with the approximations of the strain measures (SI Sec.~\ref{app:micropolar V}),
we have 
%
%
%
\begin{align}
V
&\approx
\frac{1}{2}
E_{ijkl}
u^s_{ij} u^s_{kl} 
+ 
\kappa_c\big(\theta-\frac{1}{2}\nabla^\circ\times\bm{u}\big)^2 
\notag\\
&\hspace{10pt}+
(C'_{ijkl}+\hat{E}_{ijkl})e_{ij}^{(1)}e^{(2)}_{kl}
\, ,
\label{eq:V}
\end{align}
where $E_{ijkl}=\bar{E}_{ijkl}+E'_{ijkl}+2\hat{E}_{ijkl}$, and  $E_{ijkl}=\lambda\delta_{ij}\delta_{kl} + \mu (\delta_{ik}\delta_{jl}+\delta_{il}\delta_{jk})$
with the Lam\'{e} coefficients $\lambda$ and $\mu$ of linear isotropic elasticity. $\bar{E}_{ijkl}$, $E'_{ijkl}$ and  $\hat{E}_{ijkl}$ have the same form as $E_{ijkl}$, but with different Lam\'{e} coefficients. 
Here $C'_{ijkl} = E'_{ijkl} +(\kappa_c\varepsilon_{ij}\varepsilon_{kl}/2)$ where $\kappa_c$ is the coefficient accounting for the energetic cost of rotational mismatch.
The first two terms in Eq.~\eqref{eq:V} give the linear Cosserat elasticity. The full explicit expression of $V$ is given in SI Sec.~\ref{app:micropolar V}.

Using the Poisson-bracket  formalism~\cite{hohenberg1977,stark2003,stenull2004,stark2005,markovich2021,markovich2024} with the Hamiltonian of Eq.~\eqref{eq:H} we write the 
dynamics of the hydrodynamic fields and obtain the elastic stress.
%
%
The canonical conjugate pairs in our 2D model are the angular momentum $\ell^\alpha$ with internal rotation $\theta^\alpha$, and the CM momentum $\bm{P}^\alpha$ with the position $\bm{R}^\alpha$ in the \textit{deformed} space. 
%
The nonvanishing Poisson brackets are then $\lbrace \ell^\circ(\bm{r}),\theta(\bm{r}') \rbrace
=\delta(\bm{r}-\bm{r}')$ and $\lbrace g_i^{\circ}(\bm{r}), u_j(\bm{r}') \rbrace = \delta_{ij}\delta(\bm{r}-\bm{r}')$ (SI Sec.~\ref{app:PB dynamics})
and the dynamic equations read: 
\begin{equation}
\label{eq:dynamics}
\frac{{\rm d} \ell^\circ}{{\rm d}t} = -\frac{\delta H}{\delta \theta} \quad ; \quad 
\frac{{\rm d} g^{\circ}_i}{\rm{d}t} = -\frac{\delta H}{\delta u_i} = \nabla^\circ_j P_{ij} \, ,
\end{equation}
%
where ${\rm d}\bm{X}/{\rm d}t$ is the total time derivative of $\bm{X}$, and $P_{ij}$ is the first Piola-Kirchhoff (1st PK) stress. Note that the active torque potential $-\tau^\circ\theta$ of Eq.~\eqref{eq:H}  naturally introduces $\tau^\circ$ in ${\rm d}\ell^\circ/{\rm d}t$.




Since generically $\ell^\circ$ relaxes fast compared to the velocity field 
(i.e., it is nonhydrodynamic -- it has a finite relaxation time when dissipation is introduced)~\cite{markovich2024,markovich2025,chaikin1995,maitra2019,surowka2023}, we eliminate $\theta$ by applying $\delta H/\delta \theta=0$ (SI Sec.~\ref{app:PB dynamics}),
yielding
%
%
%
%
\begin{align}\label{eq:angle relax}
\theta-\frac{1}{2}\nabla^\circ\times\bm{u}  
=
\frac{\tau^\circ}{2\kappa_c}
\bigg[
1+\frac{\tilde\lambda+\tilde\mu-\kappa_c}{\kappa_c}\nabla^\circ\cdot\bm{u}
\bigg] \, ,
\end{align}
{where $\tilde\lambda\equiv \lambda'+\hat\lambda$ and $\tilde\mu\equiv\mu'+\hat\mu$ (see definition below Eq.~\eqref{eq:V}).}
The torque density $\tau^\circ$ drives the rotation mismatch $\theta-(\nabla^\circ\times\bm{u}/2)$ as expected, while the factor $\nabla^\circ\cdot\bm{u}$ is a nonlinear contribution. {Microscopically, this rotational mismatch generates transverse forces on the particles that are essential for the emergence of odd elasticity~\cite{scheibner2020,fruchart2023,caprini2025}.}

Substituting Eq.~\eqref{eq:angle relax} in Eq.~\eqref{eq:dynamics} gives the 1st PK stress after angle relaxation. However, we are interested in the Cauchy stress, which is the stress in real (deformed) space where balance laws must hold (e.g., balance of angular momentum) and that is usually what is measured in experiments~\cite{veenstra2025a,tan2022}.
%
%
%
Following the standard transformation ${\bm \sigma} = J^{-1}{\bm P}\cdot {\bm F}^{\rm T}$ ($J\equiv\det{\bm F}$ and ${\bm F}^{\rm T}$ is the transpose of ${\bm F}$), 
the Cauchy stress is (SI Sec.~\ref{app:transform Cauchy})
\begin{align}
\sigma_{ij} 
=
\frac{\tau}{2}\varepsilon_{ij}
&-\frac{\tau^2}{4\kappa_c^2} (\tilde\lambda+\tilde\mu-\kappa_c)\delta_{ij}
\notag\\
&+
\big[
E_{ijkl}
+\frac{\tau}{4} (\varepsilon_{ik}\delta_{jl}+\varepsilon_{jl}\delta_{ik})
\big]
\nabla_l u_k \, ,
\label{eq:angle-relaxed cauchy stress}
\end{align}
where $\tau=\tau^\circ(1-\nabla\cdot\bm{u})$ is the torque density in the deformed space  and the $\tau/4$ in the second line is the odd elastic modulus~\cite{scheibner2020,braverman2021,fossati2024,fruchart2023}. 
%
%
%
Importantly, there is no restriction on the active torque $\tau$. Therefore, in contrast to previous work on metamaterials with designed lattice structures~\cite{scheibner2020,zhou2020,chen2021,veenstra2025a, gao2022, shaat2023,veenstra2024}, it can be inhomogeneous (and time-dependent).
%

As a result of the elimination of the angle variable, ${\bm\sigma}$ is also the stress of the `total' momentum~\cite{markovich2024} (SI Secs.~\ref{app:Eulerian dynamics} and \ref{app:total momentum dynamics}).
Therefore, to obey balance of `total' angular momentum the elasticity tensor is, and must be, symmetric under the exchange $i\leftrightarrow j$. To facilitate discussion, we follow the stress-strain representation used in Refs.~\cite{scheibner2020,braverman2021,fruchart2023} and write the elasticity tensor in the basis of independent deformations~\footnote{Writing the elasticity tensor 
$C_{ijkl} = \big[
B\delta_{ij}\delta_{kl}
+
\mu(\delta_{ik}\delta_{jl}+\delta_{il}\delta_{jk}-\delta_{ij}\delta_{kl})
- A\varepsilon_{ij}\delta_{kl}
+K^o(\varepsilon_{ik}\delta_{jl}+\varepsilon_{jl}\delta_{ik})
\big]$, which is the square-bracket term in Eq.~\eqref{eq:angle-relaxed cauchy stress}, in the 2D irreducible basis gives the matrix ${\bm C}$~\cite{scheibner2020} of Eq.~\eqref{eq:elasticity tensor}\label{note:Cijkl decomp}}. 
Equation~\eqref{eq:angle-relaxed cauchy stress} is then written as:
\begin{equation}\label{eq:elasticity tensor}
\begin{matrix}
\includegraphics[width=.35\textwidth]{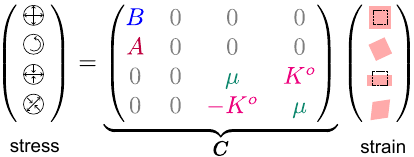}
\end{matrix}
\ \ ,
\end{equation}
where the bulk modulus $B=\lambda+\mu$, $K^o=\tau/4$ is the odd elastic modulus, and the $A$ modulus couples torque with compression. Notably, we find that $A=0$ as required from balance of total angular momentum.
%

For practical reasons, it is useful to write $\bm C$ in terms of the torque density in the undeformed space $\tau^\circ$, as it may be easier to control experimentally.
%
In doing so, $\bm C$ is written in a mixed representation (namely, mixing deformed and undeformed coordinates) such that $A$ may attain a nonzero value. 
%
%
Then, we find that $K^o=A/2=\tau^\circ/4$. {To progress analytically and compare with previous results on constant odd elastic moduli, we adopt in what follows a constant $\tau^\circ$ and use the mixed representation to
analyze the elastic and viscoelastic response. Studying the effects of inhomogeneous $\tau^\circ$, which results in inhomogeneous odd elasticity, is deferred to future work.} 




\begin{figure*}[t]
	\centering	\includegraphics[width=16 cm]{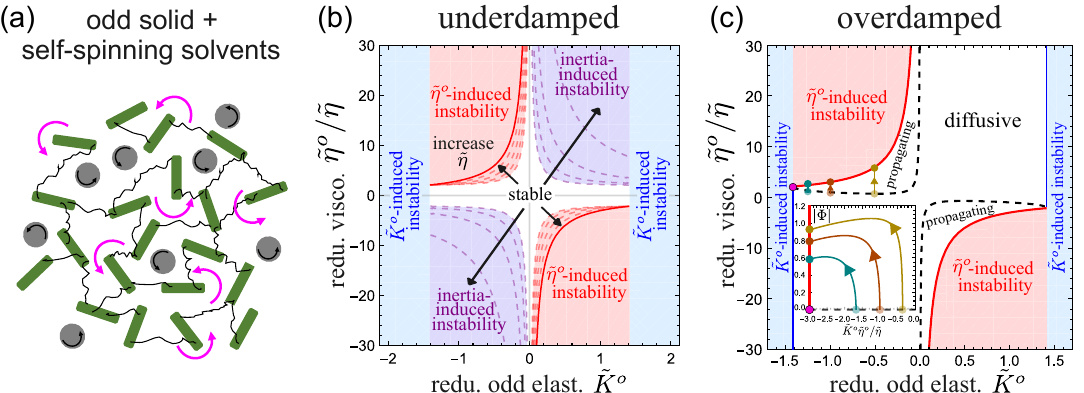}
	\caption{
Regions of dynamic instabilities and propagating displacement waves in an odd viscoelastic solid. 
(a) Illustration of the odd solid immersed in an odd solvent.
(b) The underdamped case. Arrows indicate increasing $\tilde\eta$ approaching the overdamped limit, where only $\tilde\eta^o$- and $\tilde{K}^o$-induced instabilities remain. 
(c) The overdamped case with focus on the nature of mechanical waves. 
Inset: Absolute value of the relative phase $|\Phi|$ between longitudinal and transverse modes as a function of the coupling $\tilde{\eta}^o\tilde{K}^o$.
Color coding matches the arrows in the main figure, which indicate increasing $\tilde{\eta}^o$ for various $\tilde{K}^o$. $|\Phi|$ shows non-monotonic behavior and vanishes at the tripoint (purple circle). 
%
 } 	\label{fig:instability}
\end{figure*}

\textit{Elastic response.} 
Under uniaxial compression, our odd solid exhibits Poisson ratio of $\nu
=
[\mu(B-\mu)-3\tau^{\circ^2}/16]/[\mu(B+\mu)-\tau^{\circ^2}/16]$. 
In odd solids there is another elastic response that is captured by the odd ratio, which measures the tilting due to compression~\cite{scheibner2020,fruchart2023} (SI Sec.~\ref{app:odd ratio}).
Here the odd ratio is  $\nu^o= [4\tau^\circ(B+2\mu)]/[16\mu(B+\mu)-\tau^{\circ^2}]$. 
These results are consistent with Refs.~\cite{scheibner2020,braverman2021}. 
For small $\tau^\circ$, $(\nu-\nu_\mathrm{pass}) \propto \tau^{\circ^2}$ and $\nu^o \propto \tau^\circ$, where the Poisson ratio of a passive elastic solid is $\nu_\mathrm{pass}=(B-\mu)/(B+\mu)$. This shows that the tilting encompassed in the odd ratio is a more pronounced effect.




\textit{Odd viscoelasticity and displacement waves.} Consider now a two-fluid model~\cite{levine2001,levine2009} where the odd solid is immersed in a solvent of odd active fluid~\cite{markovich2024}, Fig.~\ref{fig:instability}(a). {Assuming a strong viscous solid-fluid coupling}, we get a Kelvin-Voigt viscoelastic model (SI Sec.~\ref{app:odd viscoelas})
%
%
%
\begin{align} \label{eq:displacement dynamics}
\dot{g}_i^c 
&\approx 
\rho \ddot{u}_i 
= 
-\Gamma \dot{u}_i
+
\nabla_j(C_{ijkl}\nabla_l u_k +\eta_{ijkl}\nabla_l \dot{u}_k)
\, ,
\end{align}
where $\dot{\bm X}\equiv\partial{\bm X}/\partial t$ and $\Gamma$ is the friction coefficient.
The elasticity tensor $C_{ijkl}$ has both odd elastic moduli $K^o=A/2=\tau^\circ/4$ with uniform $\tau^\circ$~\cite{Note3} and the viscosity tensor
%
%
%
$\eta_{ijkl}=\eta(\delta_{ik}\delta_{jl}+\delta_{il}\delta_{jk}-\delta_{ij}\delta_{kl})+\eta^o(\varepsilon_{jl}\delta_{ik}+\varepsilon_{ik}\delta_{jl})+\eta^o\varepsilon_{kl}\delta_{ij}$,
where $\eta$ is the shear viscosity. Here we take, for simplicity, an `ideal' solvent such that there are no solvent-solvent interactions~\cite{markovich2024,markovich2025} with spatially-uniform angular momentum (for a more general discussion see SI Sec.~\ref{app:dynamic instability}). 
The terms $\propto\eta^o$, which are the odd pressure (pressure-vorticity coupling) and odd viscosity, are proportional to the solvent angular momentum density (but with opposite sign) and come from interaction of the elastic network with the surrounding fluid.

We continue with analysis of the mode structure of an odd viscoelastic solid, and specifically its dynamic instabilities. 
%
%
Fourier-transform of Eq.~\eqref{eq:displacement dynamics} with $u_i=\bar{u}_i\exp[i(\bm{k}\cdot\bm{R}-\omega t)]$ gives the following eigenvalue equation:
\begin{align} \label{eq:secular eq}
\big(1+\tilde{B}-\tilde{K}^{o^2}
&\big)
-i\tilde\omega
\big[
(2+\tilde{B})\tilde\eta
+\tilde\eta^o\tilde{K}^o
\big]
\\\notag
&-\tilde\omega^2
\big[
2+\tilde{B}+\tilde\eta^2
\big]
+2 i\tilde\eta \tilde\omega^3
+\tilde\omega^4
=0
\, ,
\end{align}
%
where we use $k\equiv|\bm{k}|$ and the dimensionless numbers $\tilde{\omega} \equiv  \omega\sqrt{\rho}/k\sqrt{\mu }$, $\{\tilde{B},\tilde{K^o}\} \equiv\{B,K^o\}/\mu$, $\tilde\eta^o \equiv \eta^o k/\sqrt{\mu\rho}$ and $\tilde\eta \equiv [k\eta+(\Gamma/k)]/\sqrt{\mu\rho}$ (see SI Sec.~\ref{app:dynamic matrix})
%
%
%

%
A generic scheme for deriving the instability boundaries is detailed in SI Sec.~\ref{app:instability lines}.
When odd elasticity increases beyond the threshold $1+\tilde{B}-{\tilde{K}^o}^2\leq 0$, there is always dynamic instability ($\tilde{K}^o$-induced, blue, Fig.~\ref{fig:instability}(b)).
This is similar to a pure odd elastic material with $\tilde{A}=2\tilde{K}^o$~\cite{scheibner2020}. However, due to the odd solid-fluid coupling $\tilde\eta^o\tilde{K}^o$, two previously-unreported types of instability arise. 
For $\tilde\eta^o\tilde{K}^o<0$, this coupling mainly acts as a destabilizing source, eventually causing an `$\tilde\eta^o$-induced instability' (red, Fig.~\ref{fig:instability}(b)). While for $\tilde\eta^o\tilde{K}^o>0$, the destabilization is a result of the $\tilde\eta^o\tilde{K}^o$ coupling together with inertia (purple, Fig.~\ref{fig:instability}(b)). 

Interpretation of these two instabilities becomes clear when considering an increasing $\tilde\eta$.
For small $\tilde\eta$, these instability boundaries differ only in sign: $\tilde\eta^o/\tilde\eta=\pm\tilde{B}/\tilde{K}^o$. 
Yet, as $\tilde\eta$ becomes large, the `inertia-induced' instability boundary becomes $\tilde\eta^o/\tilde\eta=2\tilde\eta^2/\tilde{K}^o$, moving significantly further than those of the `$\tilde\eta^o$-induced' instability. In fact, in the overdamped limit, only the `$\tilde\eta^o$-induced' instability  remains (red solid curves: $\tilde\eta^o/\tilde\eta = -(2+\tilde{B})/\tilde{K}^o$,  Fig.~\ref{fig:instability}(b-c))%
%
, indicating that the other instability is related to inertial effects (see SI Sec.~\ref{app:dynamic instability} for details).
%


The odd solid-fluid coupling $\tilde\eta^o\tilde{K}^o$ also changes the nature of displacement waves in the stable region, coupling longitudinal and transverse modes. 
In the underdamped case the change in the eigenvalues is only quantitative where the stabilizing term in Eq.~\eqref{eq:secular eq}$, (2+\tilde{B})\tilde\eta$, is modified by the $\tilde\eta^o\tilde{K}^o$ coupling.
The remarkable consequence of this odd solid-fluid coupling is that in the overdamped limit ($\rho\rightarrow 0$ in Eq.~\eqref{eq:secular eq}) propagating displacement waves are allowed (Fig.~\ref{fig:instability}(c)). This is in stark contrast with the passive case, where waves cannot propagate in an overdamped solid.

%

%
%
%

In the overdamped limit Eq.~\eqref{eq:secular eq} becomes ${\omega'}^2 + i a_1 \omega' -a_0 = 0$ 
with $\omega'\equiv\omega (\eta+\Gamma/k^2) / \mu$, $a_0\equiv 1+\tilde{B}-\tilde{K}^{o^2}$, and $a_1\equiv 2+\tilde{B}+\tilde{K}^o \tilde\eta^o / \tilde\eta$.
For  $-a_1^2+4a_0<0$ the waves are diffusive (not propagating). The line $-a_1^2+4a_0=0$ (dashed line, Fig.~\ref{fig:instability}(c)) is an exceptional line in which the two diffusive modes coalesce. Beyond this line ($-a_1^2+4a_0>0$) waves propagate. 
The eigenmode amplitudes, $\{ \bar{u}_\mathrm{L}, \bar{u}_\mathrm{T}\}=\mathcal{N}\{\tilde{K}^o,  1+\tilde{B}-i\omega' \}$, show that longitudinal and transverse modes are always coupled, {where $\mathcal{N}$ is the normalization factor}. Yet, at the onset of propagation,  a phase-shift $\Phi$ between $\bar{u}_\mathrm{L}$ and $\bar{u}_\mathrm{T}$ emerges [inset in Fig.~\ref{fig:instability}(c)], indicating an elliptical polarization in the shear deformation space. This phase-shift is a signature of odd wave propagation~\cite{scheibner2020}, see details in SI Sec.~\ref{app:overdamped}.
%
Importantly, waves can propagate in the overdamped limit only when $\tilde\eta^o\tilde{K}^o <0$, physically requiring the same direction for both the active torque in the solid and the spinning of fluid particles. Therefore, in the absence of coupling to an odd fluid (namely, $\tilde\eta^o/\tilde\eta=0$), our overdamped odd disordered solid does not have propagating waves (unlike some engineered ordered active solids~\cite{scheibner2020}). 
As the coupling $|\tilde\eta^o\tilde{K}^o|$ increases and approaches the $\tilde\eta^o$-induced instability, waves propagate faster and reach their maximum velocity $\sim a_0$, suggesting that the destabilization effect of $\tilde\eta^o\tilde{K}^o$ is also responsible for wave propagation. 

\textit{Conclusion.} We developed a minimal generic model for disordered chiral active solids.
{
Our central result is that odd elasticity $K^o$ emerges as a nonlinear elastic effect due to the presence of active torques $\tau$.
Importantly, these active torques drive internal rotations, which results in transverse forces required for odd elasticity to ensue~\cite{scheibner2020,fruchart2023,caprini2025}, hence $K^o\sim\tau$.
}
The elasticity tensor, when measured in {\it real} (deformed) space, is symmetric as it must in order to obey balance of angular momentum. Thus, the sole odd modulus is $K^o$.
Yet, when writing the elasticity tensor in terms of the torque density in the \textit{undeformed} space $\tau^\circ$ (which might be easier to control experimentally), the torque-compression elastic modulus $A$ also appears.
%
%

{Notably, in an odd viscoelastic solid, the odd solid-fluid coupling, together with inertia, drives new dynamic instabilities of bulk waves. In the overdamped limit, the same coupling enables bulk wave propagation, which is absent in `passive' viscoelastic solids.    
We point out that the predicted odd-inertial dynamics requires going beyond biological systems, where overdamped behavior typically dominates. Potential candidate systems include interconnected chiral spinners~\cite{scholz2018,lopez-castano2022,caprini2025c} interacting with smaller rotating elements~\cite{boriskovsky2024,boriskovsky2026}, as well as elastomers with embedded millimeter-scale magnetic particles~\cite{lucarini2022,zhai2025} immersed in a rotating solvent. 
}

In light of this work, we expect to find signatures of odd elasticity in a variety of living and synthetic materials, essentially whenever local torques are present.  
{
We hope this work will motivate further exploration of `odd' properties, especially in biological materials, where the active torque density generated by motor proteins in biological gels can be comparable to the elastic modulus (see SI Sec.~\ref{app:estimate ko} for details).}

\begin{acknowledgments}
\textit{Acknowledgment}.  This research was supported in part by Grant No. 2022/369 from the United States-Israel Binational Science Foundation (BSF).
T.M. acknowledges funding from the Israel Science Foundation (Grant No.~1356/22). T.C.L. acknowledges funding from the NSF Materials Research Science
and Engineering Center (MRSEC) at University of Pennsylvania (Grant No.~DMR-1720530).
\end{acknowledgments}

\bibliography{references.bib}

\begin{thebibliography}{100}%
\makeatletter
\providecommand \@ifxundefined [1]{%
 \@ifx{#1\undefined}
}%
\providecommand \@ifnum [1]{%
 \ifnum #1\expandafter \@firstoftwo
 \else \expandafter \@secondoftwo
 \fi
}%
\providecommand \@ifx [1]{%
 \ifx #1\expandafter \@firstoftwo
 \else \expandafter \@secondoftwo
 \fi
}%
\providecommand \natexlab [1]{#1}%
\providecommand \enquote  [1]{``#1''}%
\providecommand \bibnamefont  [1]{#1}%
\providecommand \bibfnamefont [1]{#1}%
\providecommand \citenamefont [1]{#1}%
\providecommand \href@noop [0]{\@secondoftwo}%
\providecommand \href [0]{\begingroup \@sanitize@url \@href}%
\providecommand \@href[1]{\@@startlink{#1}\@@href}%
\providecommand \@@href[1]{\endgroup#1\@@endlink}%
\providecommand \@sanitize@url [0]{\catcode `\\12\catcode `\$12\catcode `\&12\catcode `\#12\catcode `\^12\catcode `\_12\catcode `\%12\relax}%
\providecommand \@@startlink[1]{}%
\providecommand \@@endlink[0]{}%
\providecommand \url  [0]{\begingroup\@sanitize@url \@url }%
\providecommand \@url [1]{\endgroup\@href {#1}{\urlprefix }}%
\providecommand \urlprefix  [0]{URL }%
\providecommand \Eprint [0]{\href }%
\providecommand \doibase [0]{https://doi.org/}%
\providecommand \selectlanguage [0]{\@gobble}%
\providecommand \bibinfo  [0]{\@secondoftwo}%
\providecommand \bibfield  [0]{\@secondoftwo}%
\providecommand \translation [1]{[#1]}%
\providecommand \BibitemOpen [0]{}%
\providecommand \bibitemStop [0]{}%
\providecommand \bibitemNoStop [0]{.\EOS\space}%
\providecommand \EOS [0]{\spacefactor3000\relax}%
\providecommand \BibitemShut  [1]{\csname bibitem#1\endcsname}%
\let\auto@bib@innerbib\@empty
\bibitem [{\citenamefont {F{\"u}rthauer}\ \emph {et~al.}(2012)\citenamefont {F{\"u}rthauer}, \citenamefont {Strempel}, \citenamefont {Grill},\ and\ \citenamefont {J{\"u}licher}}]{furthauer2012}%
  \BibitemOpen
  \bibfield  {author} {\bibinfo {author} {\bibfnamefont {S.}~\bibnamefont {F{\"u}rthauer}}, \bibinfo {author} {\bibfnamefont {M.}~\bibnamefont {Strempel}}, \bibinfo {author} {\bibfnamefont {S.~W.}\ \bibnamefont {Grill}},\ and\ \bibinfo {author} {\bibfnamefont {F.}~\bibnamefont {J{\"u}licher}},\ }\href {https://doi.org/10.1140/epje/i2012-12089-6} {\bibfield  {journal} {\bibinfo  {journal} {Eur. Phys. J. E}\ }\textbf {\bibinfo {volume} {35}},\ \bibinfo {pages} {89} (\bibinfo {year} {2012})}\BibitemShut {NoStop}%
\bibitem [{\citenamefont {Liebchen}\ and\ \citenamefont {Levis}(2022)}]{liebchen2022}%
  \BibitemOpen
  \bibfield  {author} {\bibinfo {author} {\bibfnamefont {B.}~\bibnamefont {Liebchen}}\ and\ \bibinfo {author} {\bibfnamefont {D.}~\bibnamefont {Levis}},\ }\href {https://doi.org/10.1209/0295-5075/ac8f69} {\bibfield  {journal} {\bibinfo  {journal} {Europhys. Lett.}\ }\textbf {\bibinfo {volume} {139}},\ \bibinfo {pages} {67001} (\bibinfo {year} {2022})}\BibitemShut {NoStop}%
\bibitem [{\citenamefont {Bowick}\ \emph {et~al.}(2022)\citenamefont {Bowick}, \citenamefont {Fakhri}, \citenamefont {Marchetti},\ and\ \citenamefont {Ramaswamy}}]{bowick2022}%
  \BibitemOpen
  \bibfield  {author} {\bibinfo {author} {\bibfnamefont {M.~J.}\ \bibnamefont {Bowick}}, \bibinfo {author} {\bibfnamefont {N.}~\bibnamefont {Fakhri}}, \bibinfo {author} {\bibfnamefont {M.~C.}\ \bibnamefont {Marchetti}},\ and\ \bibinfo {author} {\bibfnamefont {S.}~\bibnamefont {Ramaswamy}},\ }\href {https://doi.org/10.1103/PhysRevX.12.010501} {\bibfield  {journal} {\bibinfo  {journal} {Phys. Rev. X}\ }\textbf {\bibinfo {volume} {12}},\ \bibinfo {pages} {010501} (\bibinfo {year} {2022})}\BibitemShut {NoStop}%
\bibitem [{\citenamefont {Shankar}\ \emph {et~al.}(2022)\citenamefont {Shankar}, \citenamefont {Souslov}, \citenamefont {Bowick}, \citenamefont {Marchetti},\ and\ \citenamefont {Vitelli}}]{shankar2022}%
  \BibitemOpen
  \bibfield  {author} {\bibinfo {author} {\bibfnamefont {S.}~\bibnamefont {Shankar}}, \bibinfo {author} {\bibfnamefont {A.}~\bibnamefont {Souslov}}, \bibinfo {author} {\bibfnamefont {M.~J.}\ \bibnamefont {Bowick}}, \bibinfo {author} {\bibfnamefont {M.~C.}\ \bibnamefont {Marchetti}},\ and\ \bibinfo {author} {\bibfnamefont {V.}~\bibnamefont {Vitelli}},\ }\href {https://doi.org/10.1038/s42254-022-00445-3} {\bibfield  {journal} {\bibinfo  {journal} {Nat. Rev. Phys.}\ }\textbf {\bibinfo {volume} {4}},\ \bibinfo {pages} {380} (\bibinfo {year} {2022})}\BibitemShut {NoStop}%
\bibitem [{\citenamefont {Markovich}\ \emph {et~al.}(2019)\citenamefont {Markovich}, \citenamefont {Tjhung},\ and\ \citenamefont {Cates}}]{markovich2019}%
  \BibitemOpen
  \bibfield  {author} {\bibinfo {author} {\bibfnamefont {T.}~\bibnamefont {Markovich}}, \bibinfo {author} {\bibfnamefont {E.}~\bibnamefont {Tjhung}},\ and\ \bibinfo {author} {\bibfnamefont {M.~E.}\ \bibnamefont {Cates}},\ }\href {https://doi.org/10.1088/1367-2630/ab54af} {\bibfield  {journal} {\bibinfo  {journal} {New J. Phys.}\ }\textbf {\bibinfo {volume} {21}},\ \bibinfo {pages} {112001} (\bibinfo {year} {2019})}\BibitemShut {NoStop}%
\bibitem [{\citenamefont {Naganathan}\ \emph {et~al.}(2014)\citenamefont {Naganathan}, \citenamefont {F{\"u}rthauer}, \citenamefont {Nishikawa}, \citenamefont {J{\"u}licher},\ and\ \citenamefont {Grill}}]{naganathan2014}%
  \BibitemOpen
  \bibfield  {author} {\bibinfo {author} {\bibfnamefont {S.~R.}\ \bibnamefont {Naganathan}}, \bibinfo {author} {\bibfnamefont {S.}~\bibnamefont {F{\"u}rthauer}}, \bibinfo {author} {\bibfnamefont {M.}~\bibnamefont {Nishikawa}}, \bibinfo {author} {\bibfnamefont {F.}~\bibnamefont {J{\"u}licher}},\ and\ \bibinfo {author} {\bibfnamefont {S.~W.}\ \bibnamefont {Grill}},\ }\href {https://doi.org/10.7554/eLife.04165} {\bibfield  {journal} {\bibinfo  {journal} {eLife}\ }\textbf {\bibinfo {volume} {3}},\ \bibinfo {pages} {e04165} (\bibinfo {year} {2014})}\BibitemShut {NoStop}%
\bibitem [{\citenamefont {Ramaiya}\ \emph {et~al.}(2017)\citenamefont {Ramaiya}, \citenamefont {Roy}, \citenamefont {Bugiel},\ and\ \citenamefont {Sch{\"a}ffer}}]{ramaiya2017}%
  \BibitemOpen
  \bibfield  {author} {\bibinfo {author} {\bibfnamefont {A.}~\bibnamefont {Ramaiya}}, \bibinfo {author} {\bibfnamefont {B.}~\bibnamefont {Roy}}, \bibinfo {author} {\bibfnamefont {M.}~\bibnamefont {Bugiel}},\ and\ \bibinfo {author} {\bibfnamefont {E.}~\bibnamefont {Sch{\"a}ffer}},\ }\href {https://doi.org/10.1073/pnas.1706985114} {\bibfield  {journal} {\bibinfo  {journal} {Proc. Natl. Acad. Sci. U.S.A.}\ }\textbf {\bibinfo {volume} {114}},\ \bibinfo {pages} {10894} (\bibinfo {year} {2017})}\BibitemShut {NoStop}%
\bibitem [{\citenamefont {Novak}\ \emph {et~al.}(2018)\citenamefont {Novak}, \citenamefont {Polak}, \citenamefont {Simuni{\'c}}, \citenamefont {Boban}, \citenamefont {Kuzmi{\'c}}, \citenamefont {Thomae}, \citenamefont {Toli{\'c}},\ and\ \citenamefont {Pavin}}]{novak2018}%
  \BibitemOpen
  \bibfield  {author} {\bibinfo {author} {\bibfnamefont {M.}~\bibnamefont {Novak}}, \bibinfo {author} {\bibfnamefont {B.}~\bibnamefont {Polak}}, \bibinfo {author} {\bibfnamefont {J.}~\bibnamefont {Simuni{\'c}}}, \bibinfo {author} {\bibfnamefont {Z.}~\bibnamefont {Boban}}, \bibinfo {author} {\bibfnamefont {B.}~\bibnamefont {Kuzmi{\'c}}}, \bibinfo {author} {\bibfnamefont {A.~W.}\ \bibnamefont {Thomae}}, \bibinfo {author} {\bibfnamefont {I.~M.}\ \bibnamefont {Toli{\'c}}},\ and\ \bibinfo {author} {\bibfnamefont {N.}~\bibnamefont {Pavin}},\ }\href {https://doi.org/10.1038/s41467-018-06005-7} {\bibfield  {journal} {\bibinfo  {journal} {Nat. Commun.}\ }\textbf {\bibinfo {volume} {9}},\ \bibinfo {pages} {3571} (\bibinfo {year} {2018})}\BibitemShut {NoStop}%
\bibitem [{\citenamefont {Afroze}\ \emph {et~al.}(2021)\citenamefont {Afroze}, \citenamefont {Inoue}, \citenamefont {Farhana}, \citenamefont {Hiraiwa}, \citenamefont {Akiyama}, \citenamefont {Kabir}, \citenamefont {Sada},\ and\ \citenamefont {Kakugo}}]{afroze2021}%
  \BibitemOpen
  \bibfield  {author} {\bibinfo {author} {\bibfnamefont {F.}~\bibnamefont {Afroze}}, \bibinfo {author} {\bibfnamefont {D.}~\bibnamefont {Inoue}}, \bibinfo {author} {\bibfnamefont {T.~I.}\ \bibnamefont {Farhana}}, \bibinfo {author} {\bibfnamefont {T.}~\bibnamefont {Hiraiwa}}, \bibinfo {author} {\bibfnamefont {R.}~\bibnamefont {Akiyama}}, \bibinfo {author} {\bibfnamefont {A.~M.~R.}\ \bibnamefont {Kabir}}, \bibinfo {author} {\bibfnamefont {K.}~\bibnamefont {Sada}},\ and\ \bibinfo {author} {\bibfnamefont {A.}~\bibnamefont {Kakugo}},\ }\href {https://doi.org/10.1016/j.bbrc.2021.05.037} {\bibfield  {journal} {\bibinfo  {journal} {Biochem. Biophys. Res. Commun.}\ }\textbf {\bibinfo {volume} {563}},\ \bibinfo {pages} {73} (\bibinfo {year} {2021})}\BibitemShut {NoStop}%
\bibitem [{\citenamefont {Mei{\ss}ner}\ \emph {et~al.}(2024)\citenamefont {Mei{\ss}ner}, \citenamefont {Niese},\ and\ \citenamefont {Diez}}]{meissner2024}%
  \BibitemOpen
  \bibfield  {author} {\bibinfo {author} {\bibfnamefont {L.}~\bibnamefont {Mei{\ss}ner}}, \bibinfo {author} {\bibfnamefont {L.}~\bibnamefont {Niese}},\ and\ \bibinfo {author} {\bibfnamefont {S.}~\bibnamefont {Diez}},\ }\href {https://doi.org/10.1016/j.ceb.2024.102367} {\bibfield  {journal} {\bibinfo  {journal} {Curr. Opin. Cell Biol.}\ }\textbf {\bibinfo {volume} {88}},\ \bibinfo {pages} {102367} (\bibinfo {year} {2024})}\BibitemShut {NoStop}%
\bibitem [{\citenamefont {Grauer}\ \emph {et~al.}(2021)\citenamefont {Grauer}, \citenamefont {Schmidt}, \citenamefont {Pineda}, \citenamefont {Midtvedt}, \citenamefont {L{\"o}wen}, \citenamefont {Volpe},\ and\ \citenamefont {Liebchen}}]{grauer2021}%
  \BibitemOpen
  \bibfield  {author} {\bibinfo {author} {\bibfnamefont {J.}~\bibnamefont {Grauer}}, \bibinfo {author} {\bibfnamefont {F.}~\bibnamefont {Schmidt}}, \bibinfo {author} {\bibfnamefont {J.}~\bibnamefont {Pineda}}, \bibinfo {author} {\bibfnamefont {B.}~\bibnamefont {Midtvedt}}, \bibinfo {author} {\bibfnamefont {H.}~\bibnamefont {L{\"o}wen}}, \bibinfo {author} {\bibfnamefont {G.}~\bibnamefont {Volpe}},\ and\ \bibinfo {author} {\bibfnamefont {B.}~\bibnamefont {Liebchen}},\ }\href {https://doi.org/10.1038/s41467-021-26319-3} {\bibfield  {journal} {\bibinfo  {journal} {Nat. Commun.}\ }\textbf {\bibinfo {volume} {12}},\ \bibinfo {pages} {6005} (\bibinfo {year} {2021})}\BibitemShut {NoStop}%
\bibitem [{\citenamefont {Vincenti}\ \emph {et~al.}(2019)\citenamefont {Vincenti}, \citenamefont {Ramos}, \citenamefont {Cordero}, \citenamefont {Douarche}, \citenamefont {Soto},\ and\ \citenamefont {Clement}}]{vincenti2019}%
  \BibitemOpen
  \bibfield  {author} {\bibinfo {author} {\bibfnamefont {B.}~\bibnamefont {Vincenti}}, \bibinfo {author} {\bibfnamefont {G.}~\bibnamefont {Ramos}}, \bibinfo {author} {\bibfnamefont {M.~L.}\ \bibnamefont {Cordero}}, \bibinfo {author} {\bibfnamefont {C.}~\bibnamefont {Douarche}}, \bibinfo {author} {\bibfnamefont {R.}~\bibnamefont {Soto}},\ and\ \bibinfo {author} {\bibfnamefont {E.}~\bibnamefont {Clement}},\ }\href {https://doi.org/10.1038/s41467-019-13031-6} {\bibfield  {journal} {\bibinfo  {journal} {Nat. Commun.}\ }\textbf {\bibinfo {volume} {10}},\ \bibinfo {pages} {5082} (\bibinfo {year} {2019})}\BibitemShut {NoStop}%
\bibitem [{\citenamefont {DiLuzio}\ \emph {et~al.}(2005)\citenamefont {DiLuzio}, \citenamefont {Turner}, \citenamefont {Mayer}, \citenamefont {Garstecki}, \citenamefont {Weibel}, \citenamefont {Berg},\ and\ \citenamefont {Whitesides}}]{diluzio2005}%
  \BibitemOpen
  \bibfield  {author} {\bibinfo {author} {\bibfnamefont {W.~R.}\ \bibnamefont {DiLuzio}}, \bibinfo {author} {\bibfnamefont {L.}~\bibnamefont {Turner}}, \bibinfo {author} {\bibfnamefont {M.}~\bibnamefont {Mayer}}, \bibinfo {author} {\bibfnamefont {P.}~\bibnamefont {Garstecki}}, \bibinfo {author} {\bibfnamefont {D.~B.}\ \bibnamefont {Weibel}}, \bibinfo {author} {\bibfnamefont {H.~C.}\ \bibnamefont {Berg}},\ and\ \bibinfo {author} {\bibfnamefont {G.~M.}\ \bibnamefont {Whitesides}},\ }\href {https://doi.org/10.1038/nature03660} {\bibfield  {journal} {\bibinfo  {journal} {Nature}\ }\textbf {\bibinfo {volume} {435}},\ \bibinfo {pages} {1271} (\bibinfo {year} {2005})}\BibitemShut {NoStop}%
\bibitem [{\citenamefont {Lauga}\ \emph {et~al.}(2006)\citenamefont {Lauga}, \citenamefont {DiLuzio}, \citenamefont {Whitesides},\ and\ \citenamefont {Stone}}]{lauga2006}%
  \BibitemOpen
  \bibfield  {author} {\bibinfo {author} {\bibfnamefont {E.}~\bibnamefont {Lauga}}, \bibinfo {author} {\bibfnamefont {W.~R.}\ \bibnamefont {DiLuzio}}, \bibinfo {author} {\bibfnamefont {G.~M.}\ \bibnamefont {Whitesides}},\ and\ \bibinfo {author} {\bibfnamefont {H.~A.}\ \bibnamefont {Stone}},\ }\href {https://doi.org/10.1529/biophysj.105.069401} {\bibfield  {journal} {\bibinfo  {journal} {Biophys. J.}\ }\textbf {\bibinfo {volume} {90}},\ \bibinfo {pages} {400} (\bibinfo {year} {2006})}\BibitemShut {NoStop}%
\bibitem [{\citenamefont {Xing}\ \emph {et~al.}(2006)\citenamefont {Xing}, \citenamefont {Bai}, \citenamefont {Berry},\ and\ \citenamefont {Oster}}]{xing2006}%
  \BibitemOpen
  \bibfield  {author} {\bibinfo {author} {\bibfnamefont {J.}~\bibnamefont {Xing}}, \bibinfo {author} {\bibfnamefont {F.}~\bibnamefont {Bai}}, \bibinfo {author} {\bibfnamefont {R.}~\bibnamefont {Berry}},\ and\ \bibinfo {author} {\bibfnamefont {G.}~\bibnamefont {Oster}},\ }\href {https://doi.org/10.1073/pnas.0507959103} {\bibfield  {journal} {\bibinfo  {journal} {Proc. Natl. Acad. Sci. U.S.A.}\ }\textbf {\bibinfo {volume} {103}},\ \bibinfo {pages} {1260} (\bibinfo {year} {2006})}\BibitemShut {NoStop}%
\bibitem [{\citenamefont {Mandadapu}\ \emph {et~al.}(2015)\citenamefont {Mandadapu}, \citenamefont {Nirody}, \citenamefont {Berry},\ and\ \citenamefont {Oster}}]{mandadapu2015}%
  \BibitemOpen
  \bibfield  {author} {\bibinfo {author} {\bibfnamefont {K.~K.}\ \bibnamefont {Mandadapu}}, \bibinfo {author} {\bibfnamefont {J.~A.}\ \bibnamefont {Nirody}}, \bibinfo {author} {\bibfnamefont {R.~M.}\ \bibnamefont {Berry}},\ and\ \bibinfo {author} {\bibfnamefont {G.}~\bibnamefont {Oster}},\ }\href {https://doi.org/10.1073/pnas.1501734112} {\bibfield  {journal} {\bibinfo  {journal} {Proc. Natl. Acad. Sci. U.S.A.}\ }\textbf {\bibinfo {volume} {112}},\ \bibinfo {pages} {E4381} (\bibinfo {year} {2015})}\BibitemShut {NoStop}%
\bibitem [{\citenamefont {Tan}\ \emph {et~al.}(2022)\citenamefont {Tan}, \citenamefont {Mietke}, \citenamefont {Li}, \citenamefont {Chen}, \citenamefont {Higinbotham}, \citenamefont {Foster}, \citenamefont {Gokhale}, \citenamefont {Dunkel},\ and\ \citenamefont {Fakhri}}]{tan2022}%
  \BibitemOpen
  \bibfield  {author} {\bibinfo {author} {\bibfnamefont {T.~H.}\ \bibnamefont {Tan}}, \bibinfo {author} {\bibfnamefont {A.}~\bibnamefont {Mietke}}, \bibinfo {author} {\bibfnamefont {J.}~\bibnamefont {Li}}, \bibinfo {author} {\bibfnamefont {Y.}~\bibnamefont {Chen}}, \bibinfo {author} {\bibfnamefont {H.}~\bibnamefont {Higinbotham}}, \bibinfo {author} {\bibfnamefont {P.~J.}\ \bibnamefont {Foster}}, \bibinfo {author} {\bibfnamefont {S.}~\bibnamefont {Gokhale}}, \bibinfo {author} {\bibfnamefont {J.}~\bibnamefont {Dunkel}},\ and\ \bibinfo {author} {\bibfnamefont {N.}~\bibnamefont {Fakhri}},\ }\href {https://doi.org/10.1038/s41586-022-04889-6} {\bibfield  {journal} {\bibinfo  {journal} {Nature}\ }\textbf {\bibinfo {volume} {607}},\ \bibinfo {pages} {287} (\bibinfo {year} {2022})}\BibitemShut {NoStop}%
\bibitem [{\citenamefont {Snezhko}(2016)}]{snezhko2016}%
  \BibitemOpen
  \bibfield  {author} {\bibinfo {author} {\bibfnamefont {A.}~\bibnamefont {Snezhko}},\ }\href {https://doi.org/10.1016/j.cocis.2015.11.010} {\bibfield  {journal} {\bibinfo  {journal} {Curr. Opin. Colloid Interface Sci.}\ }\textbf {\bibinfo {volume} {21}},\ \bibinfo {pages} {65} (\bibinfo {year} {2016})}\BibitemShut {NoStop}%
\bibitem [{\citenamefont {Soni}\ \emph {et~al.}(2019)\citenamefont {Soni}, \citenamefont {Bililign}, \citenamefont {Magkiriadou}, \citenamefont {Sacanna}, \citenamefont {Bartolo}, \citenamefont {Shelley},\ and\ \citenamefont {Irvine}}]{soni2019}%
  \BibitemOpen
  \bibfield  {author} {\bibinfo {author} {\bibfnamefont {V.}~\bibnamefont {Soni}}, \bibinfo {author} {\bibfnamefont {E.~S.}\ \bibnamefont {Bililign}}, \bibinfo {author} {\bibfnamefont {S.}~\bibnamefont {Magkiriadou}}, \bibinfo {author} {\bibfnamefont {S.}~\bibnamefont {Sacanna}}, \bibinfo {author} {\bibfnamefont {D.}~\bibnamefont {Bartolo}}, \bibinfo {author} {\bibfnamefont {M.~J.}\ \bibnamefont {Shelley}},\ and\ \bibinfo {author} {\bibfnamefont {W.~T.~M.}\ \bibnamefont {Irvine}},\ }\href {https://doi.org/10.1038/s41567-019-0603-8} {\bibfield  {journal} {\bibinfo  {journal} {Nat. Phys.}\ }\textbf {\bibinfo {volume} {15}},\ \bibinfo {pages} {1188} (\bibinfo {year} {2019})}\BibitemShut {NoStop}%
\bibitem [{\citenamefont {Shelke}\ \emph {et~al.}(2019)\citenamefont {Shelke}, \citenamefont {Srinivasan}, \citenamefont {Thampi},\ and\ \citenamefont {Mani}}]{shelke2019}%
  \BibitemOpen
  \bibfield  {author} {\bibinfo {author} {\bibfnamefont {Y.}~\bibnamefont {Shelke}}, \bibinfo {author} {\bibfnamefont {N.~R.}\ \bibnamefont {Srinivasan}}, \bibinfo {author} {\bibfnamefont {S.~P.}\ \bibnamefont {Thampi}},\ and\ \bibinfo {author} {\bibfnamefont {E.}~\bibnamefont {Mani}},\ }\href {https://doi.org/10.1021/acs.langmuir.9b00081} {\bibfield  {journal} {\bibinfo  {journal} {Langmuir}\ }\textbf {\bibinfo {volume} {35}},\ \bibinfo {pages} {4718} (\bibinfo {year} {2019})}\BibitemShut {NoStop}%
\bibitem [{\citenamefont {Schmidt}\ \emph {et~al.}(2019)\citenamefont {Schmidt}, \citenamefont {Liebchen}, \citenamefont {L{\"o}wen},\ and\ \citenamefont {Volpe}}]{schmidt2019}%
  \BibitemOpen
  \bibfield  {author} {\bibinfo {author} {\bibfnamefont {F.}~\bibnamefont {Schmidt}}, \bibinfo {author} {\bibfnamefont {B.}~\bibnamefont {Liebchen}}, \bibinfo {author} {\bibfnamefont {H.}~\bibnamefont {L{\"o}wen}},\ and\ \bibinfo {author} {\bibfnamefont {G.}~\bibnamefont {Volpe}},\ }\href {https://doi.org/10.1063/1.5079861} {\bibfield  {journal} {\bibinfo  {journal} {J. Chem. Phys.}\ }\textbf {\bibinfo {volume} {150}},\ \bibinfo {pages} {094905} (\bibinfo {year} {2019})}\BibitemShut {NoStop}%
\bibitem [{\citenamefont {Zhang}\ \emph {et~al.}(2020)\citenamefont {Zhang}, \citenamefont {Sokolov},\ and\ \citenamefont {Snezhko}}]{zhang2020}%
  \BibitemOpen
  \bibfield  {author} {\bibinfo {author} {\bibfnamefont {B.}~\bibnamefont {Zhang}}, \bibinfo {author} {\bibfnamefont {A.}~\bibnamefont {Sokolov}},\ and\ \bibinfo {author} {\bibfnamefont {A.}~\bibnamefont {Snezhko}},\ }\href {https://doi.org/10.1038/s41467-020-18209-x} {\bibfield  {journal} {\bibinfo  {journal} {Nat. Commun.}\ }\textbf {\bibinfo {volume} {11}},\ \bibinfo {pages} {4401} (\bibinfo {year} {2020})}\BibitemShut {NoStop}%
\bibitem [{\citenamefont {{Massana-Cid}}\ \emph {et~al.}(2021)\citenamefont {{Massana-Cid}}, \citenamefont {Levis}, \citenamefont {Hern{\'a}ndez}, \citenamefont {Pagonabarraga},\ and\ \citenamefont {Tierno}}]{massana-cid2021}%
  \BibitemOpen
  \bibfield  {author} {\bibinfo {author} {\bibfnamefont {H.}~\bibnamefont {{Massana-Cid}}}, \bibinfo {author} {\bibfnamefont {D.}~\bibnamefont {Levis}}, \bibinfo {author} {\bibfnamefont {R.~J.~H.}\ \bibnamefont {Hern{\'a}ndez}}, \bibinfo {author} {\bibfnamefont {I.}~\bibnamefont {Pagonabarraga}},\ and\ \bibinfo {author} {\bibfnamefont {P.}~\bibnamefont {Tierno}},\ }\href {https://doi.org/10.1103/PhysRevResearch.3.L042021} {\bibfield  {journal} {\bibinfo  {journal} {Phys. Rev. Res.}\ }\textbf {\bibinfo {volume} {3}},\ \bibinfo {pages} {L042021} (\bibinfo {year} {2021})}\BibitemShut {NoStop}%
\bibitem [{\citenamefont {Alvarez}\ \emph {et~al.}(2021)\citenamefont {Alvarez}, \citenamefont {{Fernandez-Rodriguez}}, \citenamefont {Alegria}, \citenamefont {{Arrese-Igor}}, \citenamefont {Zhao}, \citenamefont {Kr{\"o}ger},\ and\ \citenamefont {Isa}}]{alvarez2021}%
  \BibitemOpen
  \bibfield  {author} {\bibinfo {author} {\bibfnamefont {L.}~\bibnamefont {Alvarez}}, \bibinfo {author} {\bibfnamefont {M.~A.}\ \bibnamefont {{Fernandez-Rodriguez}}}, \bibinfo {author} {\bibfnamefont {A.}~\bibnamefont {Alegria}}, \bibinfo {author} {\bibfnamefont {S.}~\bibnamefont {{Arrese-Igor}}}, \bibinfo {author} {\bibfnamefont {K.}~\bibnamefont {Zhao}}, \bibinfo {author} {\bibfnamefont {M.}~\bibnamefont {Kr{\"o}ger}},\ and\ \bibinfo {author} {\bibfnamefont {L.}~\bibnamefont {Isa}},\ }\href {https://doi.org/10.1038/s41467-021-25108-2} {\bibfield  {journal} {\bibinfo  {journal} {Nat. Commun.}\ }\textbf {\bibinfo {volume} {12}},\ \bibinfo {pages} {4762} (\bibinfo {year} {2021})}\BibitemShut {NoStop}%
\bibitem [{\citenamefont {Scholz}\ \emph {et~al.}(2018)\citenamefont {Scholz}, \citenamefont {Engel},\ and\ \citenamefont {P{\"o}schel}}]{scholz2018}%
  \BibitemOpen
  \bibfield  {author} {\bibinfo {author} {\bibfnamefont {C.}~\bibnamefont {Scholz}}, \bibinfo {author} {\bibfnamefont {M.}~\bibnamefont {Engel}},\ and\ \bibinfo {author} {\bibfnamefont {T.}~\bibnamefont {P{\"o}schel}},\ }\href {https://doi.org/10.1038/s41467-018-03154-7} {\bibfield  {journal} {\bibinfo  {journal} {Nat. Commun.}\ }\textbf {\bibinfo {volume} {9}},\ \bibinfo {pages} {931} (\bibinfo {year} {2018})}\BibitemShut {NoStop}%
\bibitem [{\citenamefont {{L{\'o}pez-Casta{\~n}o}}\ \emph {et~al.}(2022)\citenamefont {{L{\'o}pez-Casta{\~n}o}}, \citenamefont {M{\'a}rquez~Seco}, \citenamefont {M{\'a}rquez~Seco}, \citenamefont {{Rodr{\'i}guez-Rivas}},\ and\ \citenamefont {Reyes}}]{lopez-castano2022}%
  \BibitemOpen
  \bibfield  {author} {\bibinfo {author} {\bibfnamefont {M.~A.}\ \bibnamefont {{L{\'o}pez-Casta{\~n}o}}}, \bibinfo {author} {\bibfnamefont {A.}~\bibnamefont {M{\'a}rquez~Seco}}, \bibinfo {author} {\bibfnamefont {A.}~\bibnamefont {M{\'a}rquez~Seco}}, \bibinfo {author} {\bibfnamefont {{\'A}.}~\bibnamefont {{Rodr{\'i}guez-Rivas}}},\ and\ \bibinfo {author} {\bibfnamefont {F.~V.}\ \bibnamefont {Reyes}},\ }\href {https://doi.org/10.1103/PhysRevResearch.4.033230} {\bibfield  {journal} {\bibinfo  {journal} {Phys. Rev. Res.}\ }\textbf {\bibinfo {volume} {4}},\ \bibinfo {pages} {033230} (\bibinfo {year} {2022})}\BibitemShut {NoStop}%
\bibitem [{\citenamefont {Caprini}\ \emph {et~al.}(2025)\citenamefont {Caprini}, \citenamefont {Abdoli}, \citenamefont {Marconi},\ and\ \citenamefont {L{\"o}wen}}]{caprini2025c}%
  \BibitemOpen
  \bibfield  {author} {\bibinfo {author} {\bibfnamefont {L.}~\bibnamefont {Caprini}}, \bibinfo {author} {\bibfnamefont {I.}~\bibnamefont {Abdoli}}, \bibinfo {author} {\bibfnamefont {U.~M.~B.}\ \bibnamefont {Marconi}},\ and\ \bibinfo {author} {\bibfnamefont {H.}~\bibnamefont {L{\"o}wen}},\ }\href {https://doi.org/10.1016/j.newton.2025.100253} {\bibfield  {journal} {\bibinfo  {journal} {Newton}\ }\textbf {\bibinfo {volume} {1}},\ \bibinfo {pages} {100253} (\bibinfo {year} {2025})}\BibitemShut {NoStop}%
\bibitem [{\citenamefont {Avron}(1998)}]{avron1998}%
  \BibitemOpen
  \bibfield  {author} {\bibinfo {author} {\bibfnamefont {J.~E.}\ \bibnamefont {Avron}},\ }\href {https://doi.org/10.1023/A:1023084404080} {\bibfield  {journal} {\bibinfo  {journal} {J. Stat. Phys.}\ }\textbf {\bibinfo {volume} {92}},\ \bibinfo {pages} {543} (\bibinfo {year} {1998})}\BibitemShut {NoStop}%
\bibitem [{\citenamefont {Banerjee}\ \emph {et~al.}(2017)\citenamefont {Banerjee}, \citenamefont {Souslov}, \citenamefont {Abanov},\ and\ \citenamefont {Vitelli}}]{banerjee2017}%
  \BibitemOpen
  \bibfield  {author} {\bibinfo {author} {\bibfnamefont {D.}~\bibnamefont {Banerjee}}, \bibinfo {author} {\bibfnamefont {A.}~\bibnamefont {Souslov}}, \bibinfo {author} {\bibfnamefont {A.~G.}\ \bibnamefont {Abanov}},\ and\ \bibinfo {author} {\bibfnamefont {V.}~\bibnamefont {Vitelli}},\ }\href {https://doi.org/10.1038/s41467-017-01378-7} {\bibfield  {journal} {\bibinfo  {journal} {Nat. Commun.}\ }\textbf {\bibinfo {volume} {8}},\ \bibinfo {pages} {1573} (\bibinfo {year} {2017})}\BibitemShut {NoStop}%
\bibitem [{\citenamefont {Souslov}\ \emph {et~al.}(2019)\citenamefont {Souslov}, \citenamefont {Dasbiswas}, \citenamefont {Fruchart}, \citenamefont {Vaikuntanathan},\ and\ \citenamefont {Vitelli}}]{souslov2019}%
  \BibitemOpen
  \bibfield  {author} {\bibinfo {author} {\bibfnamefont {A.}~\bibnamefont {Souslov}}, \bibinfo {author} {\bibfnamefont {K.}~\bibnamefont {Dasbiswas}}, \bibinfo {author} {\bibfnamefont {M.}~\bibnamefont {Fruchart}}, \bibinfo {author} {\bibfnamefont {S.}~\bibnamefont {Vaikuntanathan}},\ and\ \bibinfo {author} {\bibfnamefont {V.}~\bibnamefont {Vitelli}},\ }\href {https://doi.org/10.1103/PhysRevLett.122.128001} {\bibfield  {journal} {\bibinfo  {journal} {Phys. Rev. Lett.}\ }\textbf {\bibinfo {volume} {122}},\ \bibinfo {pages} {128001} (\bibinfo {year} {2019})}\BibitemShut {NoStop}%
\bibitem [{\citenamefont {Han}\ \emph {et~al.}(2021)\citenamefont {Han}, \citenamefont {Fruchart}, \citenamefont {Scheibner}, \citenamefont {Vaikuntanathan}, \citenamefont {{de Pablo}},\ and\ \citenamefont {Vitelli}}]{han2021}%
  \BibitemOpen
  \bibfield  {author} {\bibinfo {author} {\bibfnamefont {M.}~\bibnamefont {Han}}, \bibinfo {author} {\bibfnamefont {M.}~\bibnamefont {Fruchart}}, \bibinfo {author} {\bibfnamefont {C.}~\bibnamefont {Scheibner}}, \bibinfo {author} {\bibfnamefont {S.}~\bibnamefont {Vaikuntanathan}}, \bibinfo {author} {\bibfnamefont {J.~J.}\ \bibnamefont {{de Pablo}}},\ and\ \bibinfo {author} {\bibfnamefont {V.}~\bibnamefont {Vitelli}},\ }\href {https://doi.org/10.1038/s41567-021-01360-7} {\bibfield  {journal} {\bibinfo  {journal} {Nat. Phys.}\ }\textbf {\bibinfo {volume} {17}},\ \bibinfo {pages} {1260} (\bibinfo {year} {2021})}\BibitemShut {NoStop}%
\bibitem [{\citenamefont {Markovich}\ and\ \citenamefont {Lubensky}(2021)}]{markovich2021}%
  \BibitemOpen
  \bibfield  {author} {\bibinfo {author} {\bibfnamefont {T.}~\bibnamefont {Markovich}}\ and\ \bibinfo {author} {\bibfnamefont {T.~C.}\ \bibnamefont {Lubensky}},\ }\href {https://doi.org/10.1103/PhysRevLett.127.048001} {\bibfield  {journal} {\bibinfo  {journal} {Phys. Rev. Lett.}\ }\textbf {\bibinfo {volume} {127}},\ \bibinfo {pages} {048001} (\bibinfo {year} {2021})}\BibitemShut {NoStop}%
\bibitem [{\citenamefont {Markovich}\ and\ \citenamefont {Lubensky}(2024)}]{markovich2024}%
  \BibitemOpen
  \bibfield  {author} {\bibinfo {author} {\bibfnamefont {T.}~\bibnamefont {Markovich}}\ and\ \bibinfo {author} {\bibfnamefont {T.~C.}\ \bibnamefont {Lubensky}},\ }\href {https://doi.org/10.1073/pnas.2219385121} {\bibfield  {journal} {\bibinfo  {journal} {Proc. Natl. Acad. Sci. U.S.A.}\ }\textbf {\bibinfo {volume} {121}},\ \bibinfo {pages} {e2219385121} (\bibinfo {year} {2024})}\BibitemShut {NoStop}%
\bibitem [{\citenamefont {Hosaka}\ \emph {et~al.}(2023)\citenamefont {Hosaka}, \citenamefont {Golestanian},\ and\ \citenamefont {Vilfan}}]{hosaka2023a}%
  \BibitemOpen
  \bibfield  {author} {\bibinfo {author} {\bibfnamefont {Y.}~\bibnamefont {Hosaka}}, \bibinfo {author} {\bibfnamefont {R.}~\bibnamefont {Golestanian}},\ and\ \bibinfo {author} {\bibfnamefont {A.}~\bibnamefont {Vilfan}},\ }\href {https://doi.org/10.1103/PhysRevLett.131.178303} {\bibfield  {journal} {\bibinfo  {journal} {Phys. Rev. Lett.}\ }\textbf {\bibinfo {volume} {131}},\ \bibinfo {pages} {178303} (\bibinfo {year} {2023})}\BibitemShut {NoStop}%
\bibitem [{\citenamefont {Hosaka}\ \emph {et~al.}(2024)\citenamefont {Hosaka}, \citenamefont {Chatzittofi}, \citenamefont {Golestanian},\ and\ \citenamefont {Vilfan}}]{hosaka2024}%
  \BibitemOpen
  \bibfield  {author} {\bibinfo {author} {\bibfnamefont {Y.}~\bibnamefont {Hosaka}}, \bibinfo {author} {\bibfnamefont {M.}~\bibnamefont {Chatzittofi}}, \bibinfo {author} {\bibfnamefont {R.}~\bibnamefont {Golestanian}},\ and\ \bibinfo {author} {\bibfnamefont {A.}~\bibnamefont {Vilfan}},\ }\href {https://doi.org/10.1103/PhysRevResearch.6.L032044} {\bibfield  {journal} {\bibinfo  {journal} {Phys. Rev. Res.}\ }\textbf {\bibinfo {volume} {6}},\ \bibinfo {pages} {L032044} (\bibinfo {year} {2024})}\BibitemShut {NoStop}%
\bibitem [{\citenamefont {Banerjee}\ and\ \citenamefont {Sollich}(2025)}]{banerjee2025}%
  \BibitemOpen
  \bibfield  {author} {\bibinfo {author} {\bibfnamefont {D.}~\bibnamefont {Banerjee}}\ and\ \bibinfo {author} {\bibfnamefont {P.}~\bibnamefont {Sollich}},\ }\href {https://doi.org/10.48550/arXiv.2509.04693} {\bibinfo {title} {Emergent odd viscoelasticity in chiral soft glassy materials}} (\bibinfo {year} {2025}),\ \Eprint {https://arxiv.org/abs/2509.04693} {arXiv:2509.04693} \BibitemShut {NoStop}%
\bibitem [{\citenamefont {Marini Bettolo~Marconi}\ \emph {et~al.}(2026)\citenamefont {Marini Bettolo~Marconi}, \citenamefont {Petrini}, \citenamefont {Maire},\ and\ \citenamefont {Caprini}}]{marconi2026}%
  \BibitemOpen
  \bibfield  {author} {\bibinfo {author} {\bibfnamefont {U.}~\bibnamefont {Marini Bettolo~Marconi}}, \bibinfo {author} {\bibfnamefont {A.}~\bibnamefont {Petrini}}, \bibinfo {author} {\bibfnamefont {R.}~\bibnamefont {Maire}},\ and\ \bibinfo {author} {\bibfnamefont {L.}~\bibnamefont {Caprini}},\ }\href {https://doi.org/10.1088/1367-2630/ae6c55} {\bibfield  {journal} {\bibinfo  {journal} {New J. Phys.}\ }\textbf {\bibinfo {volume} {28}},\ \bibinfo {pages} {064401} (\bibinfo {year} {2026})}\BibitemShut {NoStop}%
\bibitem [{\citenamefont {Scheibner}\ \emph {et~al.}(2020)\citenamefont {Scheibner}, \citenamefont {Souslov}, \citenamefont {Banerjee}, \citenamefont {Sur{\'o}wka}, \citenamefont {Irvine},\ and\ \citenamefont {Vitelli}}]{scheibner2020}%
  \BibitemOpen
  \bibfield  {author} {\bibinfo {author} {\bibfnamefont {C.}~\bibnamefont {Scheibner}}, \bibinfo {author} {\bibfnamefont {A.}~\bibnamefont {Souslov}}, \bibinfo {author} {\bibfnamefont {D.}~\bibnamefont {Banerjee}}, \bibinfo {author} {\bibfnamefont {P.}~\bibnamefont {Sur{\'o}wka}}, \bibinfo {author} {\bibfnamefont {W.~T.~M.}\ \bibnamefont {Irvine}},\ and\ \bibinfo {author} {\bibfnamefont {V.}~\bibnamefont {Vitelli}},\ }\href {https://doi.org/10.1038/s41567-020-0795-y} {\bibfield  {journal} {\bibinfo  {journal} {Nat. Phys.}\ }\textbf {\bibinfo {volume} {16}},\ \bibinfo {pages} {475} (\bibinfo {year} {2020})}\BibitemShut {NoStop}%
\bibitem [{\citenamefont {Braverman}\ \emph {et~al.}(2021)\citenamefont {Braverman}, \citenamefont {Scheibner}, \citenamefont {VanSaders},\ and\ \citenamefont {Vitelli}}]{braverman2021}%
  \BibitemOpen
  \bibfield  {author} {\bibinfo {author} {\bibfnamefont {L.}~\bibnamefont {Braverman}}, \bibinfo {author} {\bibfnamefont {C.}~\bibnamefont {Scheibner}}, \bibinfo {author} {\bibfnamefont {B.}~\bibnamefont {VanSaders}},\ and\ \bibinfo {author} {\bibfnamefont {V.}~\bibnamefont {Vitelli}},\ }\href {https://doi.org/10.1103/PhysRevLett.127.268001} {\bibfield  {journal} {\bibinfo  {journal} {Phys. Rev. Lett.}\ }\textbf {\bibinfo {volume} {127}},\ \bibinfo {pages} {268001} (\bibinfo {year} {2021})}\BibitemShut {NoStop}%
\bibitem [{\citenamefont {Fossati}\ \emph {et~al.}(2024)\citenamefont {Fossati}, \citenamefont {Scheibner}, \citenamefont {Fruchart},\ and\ \citenamefont {Vitelli}}]{fossati2024}%
  \BibitemOpen
  \bibfield  {author} {\bibinfo {author} {\bibfnamefont {M.}~\bibnamefont {Fossati}}, \bibinfo {author} {\bibfnamefont {C.}~\bibnamefont {Scheibner}}, \bibinfo {author} {\bibfnamefont {M.}~\bibnamefont {Fruchart}},\ and\ \bibinfo {author} {\bibfnamefont {V.}~\bibnamefont {Vitelli}},\ }\href {https://doi.org/10.1103/PhysRevE.109.024608} {\bibfield  {journal} {\bibinfo  {journal} {Phys. Rev. E}\ }\textbf {\bibinfo {volume} {109}},\ \bibinfo {pages} {024608} (\bibinfo {year} {2024})}\BibitemShut {NoStop}%
\bibitem [{\citenamefont {Fruchart}\ \emph {et~al.}(2023)\citenamefont {Fruchart}, \citenamefont {Scheibner},\ and\ \citenamefont {Vitelli}}]{fruchart2023}%
  \BibitemOpen
  \bibfield  {author} {\bibinfo {author} {\bibfnamefont {M.}~\bibnamefont {Fruchart}}, \bibinfo {author} {\bibfnamefont {C.}~\bibnamefont {Scheibner}},\ and\ \bibinfo {author} {\bibfnamefont {V.}~\bibnamefont {Vitelli}},\ }\href {https://doi.org/10.1146/annurev-conmatphys-040821-125506} {\bibfield  {journal} {\bibinfo  {journal} {Annu. Rev. Condens. Matter Phys.}\ }\textbf {\bibinfo {volume} {14}},\ \bibinfo {pages} {471} (\bibinfo {year} {2023})}\BibitemShut {NoStop}%
\bibitem [{\citenamefont {Abanov}\ \emph {et~al.}(2018)\citenamefont {Abanov}, \citenamefont {Can},\ and\ \citenamefont {Ganeshan}}]{abanov2018}%
  \BibitemOpen
  \bibfield  {author} {\bibinfo {author} {\bibfnamefont {A.~G.}\ \bibnamefont {Abanov}}, \bibinfo {author} {\bibfnamefont {T.}~\bibnamefont {Can}},\ and\ \bibinfo {author} {\bibfnamefont {S.}~\bibnamefont {Ganeshan}},\ }\href {https://doi.org/10.21468/SciPostPhys.5.1.010} {\bibfield  {journal} {\bibinfo  {journal} {SciPost Phys.}\ }\textbf {\bibinfo {volume} {5}},\ \bibinfo {pages} {010} (\bibinfo {year} {2018})}\BibitemShut {NoStop}%
\bibitem [{\citenamefont {Veenstra}\ \emph {et~al.}(2025)\citenamefont {Veenstra}, \citenamefont {Scheibner}, \citenamefont {Brandenbourger}, \citenamefont {Binysh}, \citenamefont {Souslov}, \citenamefont {Vitelli},\ and\ \citenamefont {Coulais}}]{veenstra2025a}%
  \BibitemOpen
  \bibfield  {author} {\bibinfo {author} {\bibfnamefont {J.}~\bibnamefont {Veenstra}}, \bibinfo {author} {\bibfnamefont {C.}~\bibnamefont {Scheibner}}, \bibinfo {author} {\bibfnamefont {M.}~\bibnamefont {Brandenbourger}}, \bibinfo {author} {\bibfnamefont {J.}~\bibnamefont {Binysh}}, \bibinfo {author} {\bibfnamefont {A.}~\bibnamefont {Souslov}}, \bibinfo {author} {\bibfnamefont {V.}~\bibnamefont {Vitelli}},\ and\ \bibinfo {author} {\bibfnamefont {C.}~\bibnamefont {Coulais}},\ }\href {https://doi.org/10.1038/s41586-025-08646-3} {\bibfield  {journal} {\bibinfo  {journal} {Nature}\ }\textbf {\bibinfo {volume} {639}},\ \bibinfo {pages} {935} (\bibinfo {year} {2025})}\BibitemShut {NoStop}%
\bibitem [{\citenamefont {Gao}\ \emph {et~al.}(2022)\citenamefont {Gao}, \citenamefont {Qu},\ and\ \citenamefont {Christensen}}]{gao2022}%
  \BibitemOpen
  \bibfield  {author} {\bibinfo {author} {\bibfnamefont {P.}~\bibnamefont {Gao}}, \bibinfo {author} {\bibfnamefont {Y.}~\bibnamefont {Qu}},\ and\ \bibinfo {author} {\bibfnamefont {J.}~\bibnamefont {Christensen}},\ }\href {https://doi.org/10.1038/s43246-022-00297-5} {\bibfield  {journal} {\bibinfo  {journal} {Commun. Mater.}\ }\textbf {\bibinfo {volume} {3}},\ \bibinfo {pages} {1} (\bibinfo {year} {2022})}\BibitemShut {NoStop}%
\bibitem [{\citenamefont {Caprini}\ and\ \citenamefont {Marini Bettolo~Marconi}(2025{\natexlab{a}})}]{caprini2025a}%
  \BibitemOpen
  \bibfield  {author} {\bibinfo {author} {\bibfnamefont {L.}~\bibnamefont {Caprini}}\ and\ \bibinfo {author} {\bibfnamefont {U.}~\bibnamefont {Marini Bettolo~Marconi}},\ }\href {https://doi.org/10.1063/5.0262594} {\bibfield  {journal} {\bibinfo  {journal} {J. Chem. Phys.}\ }\textbf {\bibinfo {volume} {162}},\ \bibinfo {pages} {161101} (\bibinfo {year} {2025}{\natexlab{a}})}\BibitemShut {NoStop}%
\bibitem [{\citenamefont {Lee}\ and\ \citenamefont {Markovich}(2026)}]{lee2026}%
  \BibitemOpen
  \bibfield  {author} {\bibinfo {author} {\bibfnamefont {C.-T.}\ \bibnamefont {Lee}}\ and\ \bibinfo {author} {\bibfnamefont {T.}~\bibnamefont {Markovich}},\ }\href {https://doi.org/10.48550/arXiv.2603.21312} {\bibinfo {title} {Non-{{Hermitian}} chiral surface waves in disordered odd solids}} (\bibinfo {year} {2026}),\ \Eprint {https://arxiv.org/abs/2603.21312} {arXiv:2603.21312} \BibitemShut {NoStop}%
\bibitem [{\citenamefont {Fran{\c c}a}\ and\ \citenamefont {Jalaal}(2025)}]{franca2025}%
  \BibitemOpen
  \bibfield  {author} {\bibinfo {author} {\bibfnamefont {H.}~\bibnamefont {Fran{\c c}a}}\ and\ \bibinfo {author} {\bibfnamefont {M.}~\bibnamefont {Jalaal}},\ }\href {https://doi.org/10.48550/arXiv.2503.21649} {\bibinfo {title} {Odd {{Droplets}}: {{Fluids}} with {{Odd Viscosity}} and {{Highly Deformable Interfaces}}}} (\bibinfo {year} {2025}),\ \Eprint {https://arxiv.org/abs/2503.21649} {arXiv:2503.21649} \BibitemShut {NoStop}%
\bibitem [{\citenamefont {Zhou}\ and\ \citenamefont {Zhang}(2020)}]{zhou2020}%
  \BibitemOpen
  \bibfield  {author} {\bibinfo {author} {\bibfnamefont {D.}~\bibnamefont {Zhou}}\ and\ \bibinfo {author} {\bibfnamefont {J.}~\bibnamefont {Zhang}},\ }\href {https://doi.org/10.1103/PhysRevResearch.2.023173} {\bibfield  {journal} {\bibinfo  {journal} {Phys. Rev. Res.}\ }\textbf {\bibinfo {volume} {2}},\ \bibinfo {pages} {023173} (\bibinfo {year} {2020})}\BibitemShut {NoStop}%
\bibitem [{\citenamefont {Chen}\ \emph {et~al.}(2021)\citenamefont {Chen}, \citenamefont {Li}, \citenamefont {Scheibner}, \citenamefont {Vitelli},\ and\ \citenamefont {Huang}}]{chen2021}%
  \BibitemOpen
  \bibfield  {author} {\bibinfo {author} {\bibfnamefont {Y.}~\bibnamefont {Chen}}, \bibinfo {author} {\bibfnamefont {X.}~\bibnamefont {Li}}, \bibinfo {author} {\bibfnamefont {C.}~\bibnamefont {Scheibner}}, \bibinfo {author} {\bibfnamefont {V.}~\bibnamefont {Vitelli}},\ and\ \bibinfo {author} {\bibfnamefont {G.}~\bibnamefont {Huang}},\ }\href {https://doi.org/10.1038/s41467-021-26034-z} {\bibfield  {journal} {\bibinfo  {journal} {Nat. Commun.}\ }\textbf {\bibinfo {volume} {12}},\ \bibinfo {pages} {5935} (\bibinfo {year} {2021})}\BibitemShut {NoStop}%
\bibitem [{\citenamefont {Shaat}\ and\ \citenamefont {Park}(2023)}]{shaat2023}%
  \BibitemOpen
  \bibfield  {author} {\bibinfo {author} {\bibfnamefont {M.}~\bibnamefont {Shaat}}\ and\ \bibinfo {author} {\bibfnamefont {H.~S.}\ \bibnamefont {Park}},\ }\href {https://doi.org/10.1016/j.jmps.2022.105163} {\bibfield  {journal} {\bibinfo  {journal} {J. Mech. Phys. Solids}\ }\textbf {\bibinfo {volume} {171}},\ \bibinfo {pages} {105163} (\bibinfo {year} {2023})}\BibitemShut {NoStop}%
\bibitem [{\citenamefont {Veenstra}\ \emph {et~al.}(2024)\citenamefont {Veenstra}, \citenamefont {Gamayun}, \citenamefont {Guo}, \citenamefont {Sarvi}, \citenamefont {Meinersen},\ and\ \citenamefont {Coulais}}]{veenstra2024}%
  \BibitemOpen
  \bibfield  {author} {\bibinfo {author} {\bibfnamefont {J.}~\bibnamefont {Veenstra}}, \bibinfo {author} {\bibfnamefont {O.}~\bibnamefont {Gamayun}}, \bibinfo {author} {\bibfnamefont {X.}~\bibnamefont {Guo}}, \bibinfo {author} {\bibfnamefont {A.}~\bibnamefont {Sarvi}}, \bibinfo {author} {\bibfnamefont {C.~V.}\ \bibnamefont {Meinersen}},\ and\ \bibinfo {author} {\bibfnamefont {C.}~\bibnamefont {Coulais}},\ }\href {https://doi.org/10.1038/s41586-024-07097-6} {\bibfield  {journal} {\bibinfo  {journal} {Nature}\ }\textbf {\bibinfo {volume} {627}},\ \bibinfo {pages} {528} (\bibinfo {year} {2024})}\BibitemShut {NoStop}%
\bibitem [{\citenamefont {Caprini}\ and\ \citenamefont {Marini Bettolo~Marconi}(2025{\natexlab{b}})}]{caprini2025}%
  \BibitemOpen
  \bibfield  {author} {\bibinfo {author} {\bibfnamefont {L.}~\bibnamefont {Caprini}}\ and\ \bibinfo {author} {\bibfnamefont {U.}~\bibnamefont {Marini Bettolo~Marconi}},\ }\href {https://doi.org/10.1088/1367-2630/add366} {\bibfield  {journal} {\bibinfo  {journal} {New J. Phys.}\ }\textbf {\bibinfo {volume} {27}},\ \bibinfo {pages} {054401} (\bibinfo {year} {2025}{\natexlab{b}})}\BibitemShut {NoStop}%
\bibitem [{\citenamefont {N{\'e}meth}\ \emph {et~al.}(2026)\citenamefont {N{\'e}meth}, \citenamefont {Kobayashi},\ and\ \citenamefont {Adhikari}}]{nemeth2026}%
  \BibitemOpen
  \bibfield  {author} {\bibinfo {author} {\bibfnamefont {B.}~\bibnamefont {N{\'e}meth}}, \bibinfo {author} {\bibfnamefont {T.}~\bibnamefont {Kobayashi}},\ and\ \bibinfo {author} {\bibfnamefont {R.}~\bibnamefont {Adhikari}},\ }\href {https://doi.org/10.1088/1367-2630/ae4f19} {\bibfield  {journal} {\bibinfo  {journal} {New J. Phys.}\ }\textbf {\bibinfo {volume} {28}},\ \bibinfo {pages} {034401} (\bibinfo {year} {2026})}\BibitemShut {NoStop}%
\bibitem [{\citenamefont {Engstrom}\ and\ \citenamefont {Sussman}(2025)}]{engstrom2025}%
  \BibitemOpen
  \bibfield  {author} {\bibinfo {author} {\bibfnamefont {T.~A.}\ \bibnamefont {Engstrom}}\ and\ \bibinfo {author} {\bibfnamefont {D.~M.}\ \bibnamefont {Sussman}},\ }\href {https://doi.org/10.48550/arXiv.2509.19560} {\bibinfo {title} {Conservative yet constitutively odd elasticity in prestressed metamaterials}} (\bibinfo {year} {2025}),\ \Eprint {https://arxiv.org/abs/2509.19560} {arXiv:2509.19560} \BibitemShut {NoStop}%
\bibitem [{\citenamefont {Kole}\ \emph {et~al.}(2021)\citenamefont {Kole}, \citenamefont {Alexander}, \citenamefont {Ramaswamy},\ and\ \citenamefont {Maitra}}]{kole2021}%
  \BibitemOpen
  \bibfield  {author} {\bibinfo {author} {\bibfnamefont {S.~J.}\ \bibnamefont {Kole}}, \bibinfo {author} {\bibfnamefont {G.~P.}\ \bibnamefont {Alexander}}, \bibinfo {author} {\bibfnamefont {S.}~\bibnamefont {Ramaswamy}},\ and\ \bibinfo {author} {\bibfnamefont {A.}~\bibnamefont {Maitra}},\ }\href {https://doi.org/10.1103/PhysRevLett.126.248001} {\bibfield  {journal} {\bibinfo  {journal} {Phys. Rev. Lett.}\ }\textbf {\bibinfo {volume} {126}},\ \bibinfo {pages} {248001} (\bibinfo {year} {2021})}\BibitemShut {NoStop}%
\bibitem [{\citenamefont {Kole}\ \emph {et~al.}(2024)\citenamefont {Kole}, \citenamefont {Alexander}, \citenamefont {Maitra},\ and\ \citenamefont {Ramaswamy}}]{kole2024}%
  \BibitemOpen
  \bibfield  {author} {\bibinfo {author} {\bibfnamefont {S.~J.}\ \bibnamefont {Kole}}, \bibinfo {author} {\bibfnamefont {G.~P.}\ \bibnamefont {Alexander}}, \bibinfo {author} {\bibfnamefont {A.}~\bibnamefont {Maitra}},\ and\ \bibinfo {author} {\bibfnamefont {S.}~\bibnamefont {Ramaswamy}},\ }\href {https://doi.org/10.1093/pnasnexus/pgae398} {\bibfield  {journal} {\bibinfo  {journal} {PNAS Nexus}\ }\textbf {\bibinfo {volume} {3}},\ \bibinfo {pages} {398} (\bibinfo {year} {2024})}\BibitemShut {NoStop}%
\bibitem [{\citenamefont {Howard}\ \emph {et~al.}(2019)\citenamefont {Howard}, \citenamefont {Jadrich}, \citenamefont {Lindquist}, \citenamefont {Khabaz}, \citenamefont {Bonnecaze}, \citenamefont {Milliron},\ and\ \citenamefont {Truskett}}]{howard2019}%
  \BibitemOpen
  \bibfield  {author} {\bibinfo {author} {\bibfnamefont {M.~P.}\ \bibnamefont {Howard}}, \bibinfo {author} {\bibfnamefont {R.~B.}\ \bibnamefont {Jadrich}}, \bibinfo {author} {\bibfnamefont {B.~A.}\ \bibnamefont {Lindquist}}, \bibinfo {author} {\bibfnamefont {F.}~\bibnamefont {Khabaz}}, \bibinfo {author} {\bibfnamefont {R.~T.}\ \bibnamefont {Bonnecaze}}, \bibinfo {author} {\bibfnamefont {D.~J.}\ \bibnamefont {Milliron}},\ and\ \bibinfo {author} {\bibfnamefont {T.~M.}\ \bibnamefont {Truskett}},\ }\href {https://doi.org/10.1063/1.5119359} {\bibfield  {journal} {\bibinfo  {journal} {J. Chem. Phys.}\ }\textbf {\bibinfo {volume} {151}},\ \bibinfo {pages} {124901} (\bibinfo {year} {2019})}\BibitemShut {NoStop}%
\bibitem [{\citenamefont {Eringen}(1966)}]{eringen1966}%
  \BibitemOpen
  \bibfield  {author} {\bibinfo {author} {\bibfnamefont {A.~C.}\ \bibnamefont {Eringen}},\ }\href {https://doi.org/10.1512/iumj.1966.15.15060} {\bibfield  {journal} {\bibinfo  {journal} {J. Appl. Math. Mech.}\ }\textbf {\bibinfo {volume} {15}},\ \bibinfo {pages} {909} (\bibinfo {year} {1966})}\BibitemShut {NoStop}%
\bibitem [{\citenamefont {Eringen}(1999)}]{eringen1999}%
  \BibitemOpen
  \bibfield  {author} {\bibinfo {author} {\bibfnamefont {A.~C.}\ \bibnamefont {Eringen}},\ }\href {https://doi.org/10.1007/978-1-4612-0555-5} {\emph {\bibinfo {title} {Microcontinuum {{Field Theories}}}}}\ (\bibinfo  {publisher} {Springer},\ \bibinfo {address} {New York},\ \bibinfo {year} {1999})\BibitemShut {NoStop}%
\bibitem [{\citenamefont {Eremeyev}\ \emph {et~al.}(2013)\citenamefont {Eremeyev}, \citenamefont {Lebedev},\ and\ \citenamefont {Altenbach}}]{eremeyev2013}%
  \BibitemOpen
  \bibfield  {author} {\bibinfo {author} {\bibfnamefont {V.~A.}\ \bibnamefont {Eremeyev}}, \bibinfo {author} {\bibfnamefont {L.~P.}\ \bibnamefont {Lebedev}},\ and\ \bibinfo {author} {\bibfnamefont {H.}~\bibnamefont {Altenbach}},\ }\href@noop {} {\emph {\bibinfo {title} {Foundations of {{Micropolar Mechanics}}}}}\ (\bibinfo  {publisher} {Springer},\ \bibinfo {address} {Berlin},\ \bibinfo {year} {2013})\BibitemShut {NoStop}%
\bibitem [{\citenamefont {Broedersz}\ and\ \citenamefont {MacKintosh}(2014)}]{Broedersz2014}%
  \BibitemOpen
  \bibfield  {author} {\bibinfo {author} {\bibfnamefont {C.~P.}\ \bibnamefont {Broedersz}}\ and\ \bibinfo {author} {\bibfnamefont {F.~C.}\ \bibnamefont {MacKintosh}},\ }\href {https://doi.org/10.1103/RevModPhys.86.995} {\bibfield  {journal} {\bibinfo  {journal} {Rev. Mod. Phys.}\ }\textbf {\bibinfo {volume} {86}},\ \bibinfo {pages} {995} (\bibinfo {year} {2014})}\BibitemShut {NoStop}%
\bibitem [{\citenamefont {Landau}\ \emph {et~al.}(1986)\citenamefont {Landau}, \citenamefont {Pitaevskii}, \citenamefont {Kosevich},\ and\ \citenamefont {Lifshitz}}]{LLelastity}%
  \BibitemOpen
  \bibfield  {author} {\bibinfo {author} {\bibfnamefont {L.~D.}\ \bibnamefont {Landau}}, \bibinfo {author} {\bibfnamefont {L.~P.}\ \bibnamefont {Pitaevskii}}, \bibinfo {author} {\bibfnamefont {A.~M.}\ \bibnamefont {Kosevich}},\ and\ \bibinfo {author} {\bibfnamefont {E.~M.}\ \bibnamefont {Lifshitz}},\ }\href@noop {} {\emph {\bibinfo {title} {Theory of Elasticity}}},\ \bibinfo {edition} {3rd}\ ed.\ (\bibinfo  {publisher} {Pergamon Press},\ \bibinfo {address} {New York},\ \bibinfo {year} {1986})\BibitemShut {NoStop}%
\bibitem [{Note1()}]{Note1}%
  \BibitemOpen
  \bibinfo {note} {$\theta ^\alpha $ is different from the angle used in magnetic spin systems, which is the deviation from a universal direction defined by an external magnetic field.}\BibitemShut {Stop}%
\bibitem [{Note2()}]{Note2}%
  \BibitemOpen
  \bibinfo {note} {This CG potential is obtained using the approximation~\cite {markovich2024} $\DOTSB \sum@ \slimits@ _{\alpha \in \Delta V^\circ } \tau ^\alpha \theta ^\alpha \delta (\protect \bm {r}-\protect \bm {r}^\alpha ) \approx \DOTSB \sum@ \slimits@ _{\alpha \in \Delta V^\circ } \tau ^\alpha \delta (\protect \bm {r}-\protect \bm {r}^\alpha )\DOTSB \sum@ \slimits@ _{\beta \in \Delta V^\circ } \theta ^\beta \delta (\protect \bm {r}-\protect \bm {r}^\beta ) / n^\circ (\protect \bm {r})=\tau ^\circ (\protect \bm {r})\theta (\protect \bm {r})$. The CG potential due to an external field is: $\protect \cc@accent {"707E}\tau (\protect \bm {r})\DOTSB \sum@ \slimits@ _{\alpha \in \Delta V^\circ } \theta ^\alpha \delta (\protect \bm {r}-\protect \bm {r}^\alpha )=\protect \cc@accent {"707E}\tau (\protect \bm {r}) n^\circ (\protect \bm {r}) \theta (\protect \bm {r}) =\tau (\protect \bm {r}) \theta (\protect \bm {r})$, where $\protect \cc@accent {"707E}\tau $ is the torque due to the external field.}\BibitemShut
  {Stop}%
\bibitem [{\citenamefont {Goldstein}\ \emph {et~al.}(2002)\citenamefont {Goldstein}, \citenamefont {Poole},\ and\ \citenamefont {Safko}}]{Goldstein_book}%
  \BibitemOpen
  \bibfield  {author} {\bibinfo {author} {\bibfnamefont {H.}~\bibnamefont {Goldstein}}, \bibinfo {author} {\bibfnamefont {C.}~\bibnamefont {Poole}},\ and\ \bibinfo {author} {\bibfnamefont {J.}~\bibnamefont {Safko}},\ }\href@noop {} {\emph {\bibinfo {title} {Classical Mechanics}}}\ (\bibinfo  {publisher} {Addison-Wesley},\ \bibinfo {address} {Boston},\ \bibinfo {year} {2002})\BibitemShut {NoStop}%
\bibitem [{\citenamefont {Chaikin}\ and\ \citenamefont {Lubensky}(1995)}]{chaikin1995}%
  \BibitemOpen
  \bibfield  {author} {\bibinfo {author} {\bibfnamefont {P.~M.}\ \bibnamefont {Chaikin}}\ and\ \bibinfo {author} {\bibfnamefont {T.~C.}\ \bibnamefont {Lubensky}},\ }\href@noop {} {\emph {\bibinfo {title} {Principles of Condensed Matter Physics}}}\ (\bibinfo  {publisher} {Cambridge University Press},\ \bibinfo {address} {New York},\ \bibinfo {year} {1995})\BibitemShut {NoStop}%
\bibitem [{\citenamefont {Hohenberg}\ and\ \citenamefont {Halperin}(1977)}]{hohenberg1977}%
  \BibitemOpen
  \bibfield  {author} {\bibinfo {author} {\bibfnamefont {P.~C.}\ \bibnamefont {Hohenberg}}\ and\ \bibinfo {author} {\bibfnamefont {B.~I.}\ \bibnamefont {Halperin}},\ }\href {https://doi.org/10.1103/RevModPhys.49.435} {\bibfield  {journal} {\bibinfo  {journal} {Rev. Mod. Phys.}\ }\textbf {\bibinfo {volume} {49}},\ \bibinfo {pages} {435} (\bibinfo {year} {1977})}\BibitemShut {NoStop}%
\bibitem [{\citenamefont {Stark}\ and\ \citenamefont {Lubensky}(2003)}]{stark2003}%
  \BibitemOpen
  \bibfield  {author} {\bibinfo {author} {\bibfnamefont {H.}~\bibnamefont {Stark}}\ and\ \bibinfo {author} {\bibfnamefont {T.~C.}\ \bibnamefont {Lubensky}},\ }\href {https://doi.org/10.1103/PhysRevE.67.061709} {\bibfield  {journal} {\bibinfo  {journal} {Phys. Rev. E}\ }\textbf {\bibinfo {volume} {67}},\ \bibinfo {pages} {061709} (\bibinfo {year} {2003})}\BibitemShut {NoStop}%
\bibitem [{\citenamefont {Stenull}\ and\ \citenamefont {Lubensky}(2004)}]{stenull2004}%
  \BibitemOpen
  \bibfield  {author} {\bibinfo {author} {\bibfnamefont {O.}~\bibnamefont {Stenull}}\ and\ \bibinfo {author} {\bibfnamefont {T.~C.}\ \bibnamefont {Lubensky}},\ }\href {https://doi.org/10.1103/PhysRevE.69.051801} {\bibfield  {journal} {\bibinfo  {journal} {Phys. Rev. E}\ }\textbf {\bibinfo {volume} {69}},\ \bibinfo {pages} {051801} (\bibinfo {year} {2004})}\BibitemShut {NoStop}%
\bibitem [{\citenamefont {Stark}\ and\ \citenamefont {Lubensky}(2005)}]{stark2005}%
  \BibitemOpen
  \bibfield  {author} {\bibinfo {author} {\bibfnamefont {H.}~\bibnamefont {Stark}}\ and\ \bibinfo {author} {\bibfnamefont {T.~C.}\ \bibnamefont {Lubensky}},\ }\href {https://doi.org/10.1103/PhysRevE.72.051714} {\bibfield  {journal} {\bibinfo  {journal} {Phys. Rev. E}\ }\textbf {\bibinfo {volume} {72}},\ \bibinfo {pages} {051714} (\bibinfo {year} {2005})}\BibitemShut {NoStop}%
\bibitem [{\citenamefont {Markovich}\ and\ \citenamefont {Lubensky}(2025)}]{markovich2025}%
  \BibitemOpen
  \bibfield  {author} {\bibinfo {author} {\bibfnamefont {T.}~\bibnamefont {Markovich}}\ and\ \bibinfo {author} {\bibfnamefont {T.~C.}\ \bibnamefont {Lubensky}},\ }\href {https://doi.org/10.1103/tkbx-75zt} {\bibfield  {journal} {\bibinfo  {journal} {Phys. Rev. E}\ }\textbf {\bibinfo {volume} {112}},\ \bibinfo {pages} {035409} (\bibinfo {year} {2025})}\BibitemShut {NoStop}%
\bibitem [{\citenamefont {Maitra}\ and\ \citenamefont {Ramaswamy}(2019)}]{maitra2019}%
  \BibitemOpen
  \bibfield  {author} {\bibinfo {author} {\bibfnamefont {A.}~\bibnamefont {Maitra}}\ and\ \bibinfo {author} {\bibfnamefont {S.}~\bibnamefont {Ramaswamy}},\ }\href {https://doi.org/10.1103/PhysRevLett.123.238001} {\bibfield  {journal} {\bibinfo  {journal} {Phys. Rev. Lett.}\ }\textbf {\bibinfo {volume} {123}},\ \bibinfo {pages} {238001} (\bibinfo {year} {2019})}\BibitemShut {NoStop}%
\bibitem [{\citenamefont {Sur{\'o}wka}\ \emph {et~al.}(2023)\citenamefont {Sur{\'o}wka}, \citenamefont {Souslov}, \citenamefont {J{\"u}licher},\ and\ \citenamefont {Banerjee}}]{surowka2023}%
  \BibitemOpen
  \bibfield  {author} {\bibinfo {author} {\bibfnamefont {P.}~\bibnamefont {Sur{\'o}wka}}, \bibinfo {author} {\bibfnamefont {A.}~\bibnamefont {Souslov}}, \bibinfo {author} {\bibfnamefont {F.}~\bibnamefont {J{\"u}licher}},\ and\ \bibinfo {author} {\bibfnamefont {D.}~\bibnamefont {Banerjee}},\ }\href {https://doi.org/10.1103/PhysRevE.108.064609} {\bibfield  {journal} {\bibinfo  {journal} {Phys. Rev. E}\ }\textbf {\bibinfo {volume} {108}},\ \bibinfo {pages} {064609} (\bibinfo {year} {2023})}\BibitemShut {NoStop}%
\bibitem [{Note3()}]{Note3}%
  \BibitemOpen
  \bibinfo {note} {Writing the elasticity tensor $C_{ijkl} = \protect \big [ B\delta _{ij}\delta _{kl} + \mu (\delta _{ik}\delta _{jl}+\delta _{il}\delta _{jk}-\delta _{ij}\delta _{kl}) - A\varepsilon _{ij}\delta _{kl} +K^o(\varepsilon _{ik}\delta _{jl}+\varepsilon _{jl}\delta _{ik}) \protect \big ]$, which is the square-bracket term in Eq.~\protect \eqref {eq:angle-relaxed cauchy stress}, in the 2D irreducible basis gives the matrix ${\protect \bm {C}}$~\cite {scheibner2020} of Eq.~\protect \eqref {eq:elasticity tensor}\label {note:Cijkl decomp}}\BibitemShut {NoStop}%
\bibitem [{\citenamefont {Levine}\ and\ \citenamefont {Lubensky}(2001)}]{levine2001}%
  \BibitemOpen
  \bibfield  {author} {\bibinfo {author} {\bibfnamefont {A.~J.}\ \bibnamefont {Levine}}\ and\ \bibinfo {author} {\bibfnamefont {T.~C.}\ \bibnamefont {Lubensky}},\ }\href {https://doi.org/10.1103/PhysRevE.63.041510} {\bibfield  {journal} {\bibinfo  {journal} {Phys. Rev. E}\ }\textbf {\bibinfo {volume} {63}},\ \bibinfo {pages} {041510} (\bibinfo {year} {2001})}\BibitemShut {NoStop}%
\bibitem [{\citenamefont {Levine}\ and\ \citenamefont {MacKintosh}(2009)}]{levine2009}%
  \BibitemOpen
  \bibfield  {author} {\bibinfo {author} {\bibfnamefont {A.~J.}\ \bibnamefont {Levine}}\ and\ \bibinfo {author} {\bibfnamefont {F.~C.}\ \bibnamefont {MacKintosh}},\ }\href {https://doi.org/10.1021/jp808192w} {\bibfield  {journal} {\bibinfo  {journal} {J. Phys. Chem. B}\ }\textbf {\bibinfo {volume} {113}},\ \bibinfo {pages} {3820} (\bibinfo {year} {2009})}\BibitemShut {NoStop}%
\bibitem [{\citenamefont {Boriskovsky}\ \emph {et~al.}(2024)\citenamefont {Boriskovsky}, \citenamefont {Lindner},\ and\ \citenamefont {Roichman}}]{boriskovsky2024}%
  \BibitemOpen
  \bibfield  {author} {\bibinfo {author} {\bibfnamefont {D.}~\bibnamefont {Boriskovsky}}, \bibinfo {author} {\bibfnamefont {B.}~\bibnamefont {Lindner}},\ and\ \bibinfo {author} {\bibfnamefont {Y.}~\bibnamefont {Roichman}},\ }\href {https://doi.org/10.1039/D4SM00808A} {\bibfield  {journal} {\bibinfo  {journal} {Soft Matter}\ }\textbf {\bibinfo {volume} {20}},\ \bibinfo {pages} {8017} (\bibinfo {year} {2024})}\BibitemShut {NoStop}%
\bibitem [{\citenamefont {Boriskovsky}\ \emph {et~al.}(2026)\citenamefont {Boriskovsky}, \citenamefont {Goerlich}, \citenamefont {Lindner},\ and\ \citenamefont {Roichman}}]{boriskovsky2026}%
  \BibitemOpen
  \bibfield  {author} {\bibinfo {author} {\bibfnamefont {D.}~\bibnamefont {Boriskovsky}}, \bibinfo {author} {\bibfnamefont {R.}~\bibnamefont {Goerlich}}, \bibinfo {author} {\bibfnamefont {B.}~\bibnamefont {Lindner}},\ and\ \bibinfo {author} {\bibfnamefont {Y.}~\bibnamefont {Roichman}},\ }\href {https://doi.org/10.1039/D5SM00840A} {\bibfield  {journal} {\bibinfo  {journal} {Soft Matter}\ }\textbf {\bibinfo {volume} {22}},\ \bibinfo {pages} {297} (\bibinfo {year} {2026})}\BibitemShut {NoStop}%
\bibitem [{\citenamefont {Lucarini}\ \emph {et~al.}(2022)\citenamefont {Lucarini}, \citenamefont {Hossain},\ and\ \citenamefont {{Garcia-Gonzalez}}}]{lucarini2022}%
  \BibitemOpen
  \bibfield  {author} {\bibinfo {author} {\bibfnamefont {S.}~\bibnamefont {Lucarini}}, \bibinfo {author} {\bibfnamefont {M.}~\bibnamefont {Hossain}},\ and\ \bibinfo {author} {\bibfnamefont {D.}~\bibnamefont {{Garcia-Gonzalez}}},\ }\href {https://doi.org/10.1016/j.compstruct.2021.114800} {\bibfield  {journal} {\bibinfo  {journal} {Compos. Struct.}\ }\textbf {\bibinfo {volume} {279}},\ \bibinfo {pages} {114800} (\bibinfo {year} {2022})}\BibitemShut {NoStop}%
\bibitem [{\citenamefont {Zhai}\ \emph {et~al.}(2025)\citenamefont {Zhai}, \citenamefont {Li}, \citenamefont {Yu}, \citenamefont {Wang}, \citenamefont {Chang}, \citenamefont {Li}, \citenamefont {Cheng}, \citenamefont {Zhou}, \citenamefont {Fang}, \citenamefont {Liu}, \citenamefont {Yu}, \citenamefont {Zhu}, \citenamefont {Li},\ and\ \citenamefont {Li}}]{zhai2025}%
  \BibitemOpen
  \bibfield  {author} {\bibinfo {author} {\bibfnamefont {H.}~\bibnamefont {Zhai}}, \bibinfo {author} {\bibfnamefont {X.}~\bibnamefont {Li}}, \bibinfo {author} {\bibfnamefont {S.}~\bibnamefont {Yu}}, \bibinfo {author} {\bibfnamefont {J.}~\bibnamefont {Wang}}, \bibinfo {author} {\bibfnamefont {Y.}~\bibnamefont {Chang}}, \bibinfo {author} {\bibfnamefont {J.}~\bibnamefont {Li}}, \bibinfo {author} {\bibfnamefont {X.}~\bibnamefont {Cheng}}, \bibinfo {author} {\bibfnamefont {L.}~\bibnamefont {Zhou}}, \bibinfo {author} {\bibfnamefont {Y.}~\bibnamefont {Fang}}, \bibinfo {author} {\bibfnamefont {T.}~\bibnamefont {Liu}}, \bibinfo {author} {\bibfnamefont {X.}~\bibnamefont {Yu}}, \bibinfo {author} {\bibfnamefont {M.}~\bibnamefont {Zhu}}, \bibinfo {author} {\bibfnamefont {B.}~\bibnamefont {Li}},\ and\ \bibinfo {author} {\bibfnamefont {W.}~\bibnamefont {Li}},\ }\href {https://doi.org/10.1016/j.compositesb.2025.112387} {\bibfield  {journal} {\bibinfo  {journal} {Compos. B: Eng.}\ }\textbf {\bibinfo {volume} {298}},\ \bibinfo
  {pages} {112387} (\bibinfo {year} {2025})}\BibitemShut {NoStop}%
\bibitem [{\citenamefont {Bauchau}\ and\ \citenamefont {Trainelli}(2003)}]{bauchau2003}%
  \BibitemOpen
  \bibfield  {author} {\bibinfo {author} {\bibfnamefont {O.~A.}\ \bibnamefont {Bauchau}}\ and\ \bibinfo {author} {\bibfnamefont {L.}~\bibnamefont {Trainelli}},\ }\href {https://doi.org/10.1023/A:1024265401576} {\bibfield  {journal} {\bibinfo  {journal} {Nonlinear Dyn.}\ }\textbf {\bibinfo {volume} {32}},\ \bibinfo {pages} {71} (\bibinfo {year} {2003})}\BibitemShut {NoStop}%
\bibitem [{\citenamefont {Brauns}\ and\ \citenamefont {Marchetti}(2024)}]{brauns2024}%
  \BibitemOpen
  \bibfield  {author} {\bibinfo {author} {\bibfnamefont {F.}~\bibnamefont {Brauns}}\ and\ \bibinfo {author} {\bibfnamefont {M.~C.}\ \bibnamefont {Marchetti}},\ }\href {https://doi.org/10.1103/PhysRevX.14.021014} {\bibfield  {journal} {\bibinfo  {journal} {Phys. Rev. X}\ }\textbf {\bibinfo {volume} {14}},\ \bibinfo {pages} {021014} (\bibinfo {year} {2024})}\BibitemShut {NoStop}%
\bibitem [{\citenamefont {Grekova}(2019)}]{grekova2019}%
  \BibitemOpen
  \bibfield  {author} {\bibinfo {author} {\bibfnamefont {E.~F.}\ \bibnamefont {Grekova}},\ }\href {https://doi.org/10.1007/s00161-019-00829-4} {\bibfield  {journal} {\bibinfo  {journal} {Continuum Mech. Thermodyn.}\ }\textbf {\bibinfo {volume} {31}},\ \bibinfo {pages} {1805} (\bibinfo {year} {2019})}\BibitemShut {NoStop}%
\bibitem [{\citenamefont {Lee}\ and\ \citenamefont {Markovich}()}]{lee2026Poisson}%
  \BibitemOpen
  \bibfield  {author} {\bibinfo {author} {\bibfnamefont {C.-T.}\ \bibnamefont {Lee}}\ and\ \bibinfo {author} {\bibfnamefont {T.}~\bibnamefont {Markovich}},\ }\href@noop {} {}\bibinfo {note} {Eulerian Poisson-bracket formalism for elasticity: application to chiral odd solids (to be published)}\BibitemShut {NoStop}%
\bibitem [{\citenamefont {Robinson}\ \emph {et~al.}(2002)\citenamefont {Robinson}, \citenamefont {Cavet}, \citenamefont {Warrick},\ and\ \citenamefont {Spudich}}]{robinson2002}%
  \BibitemOpen
  \bibfield  {author} {\bibinfo {author} {\bibfnamefont {D.~N.}\ \bibnamefont {Robinson}}, \bibinfo {author} {\bibfnamefont {G.}~\bibnamefont {Cavet}}, \bibinfo {author} {\bibfnamefont {H.~M.}\ \bibnamefont {Warrick}},\ and\ \bibinfo {author} {\bibfnamefont {J.~A.}\ \bibnamefont {Spudich}},\ }\href {https://doi.org/10.1186/1471-2121-3-4} {\bibfield  {journal} {\bibinfo  {journal} {BMC Cell Biol}\ }\textbf {\bibinfo {volume} {3}},\ \bibinfo {pages} {4} (\bibinfo {year} {2002})}\BibitemShut {NoStop}%
\bibitem [{\citenamefont {Ali}\ \emph {et~al.}(2002)\citenamefont {Ali}, \citenamefont {Uemura}, \citenamefont {Adachi}, \citenamefont {Itoh}, \citenamefont {Kinosita},\ and\ \citenamefont {Ishiwata}}]{ali2002}%
  \BibitemOpen
  \bibfield  {author} {\bibinfo {author} {\bibfnamefont {M.~Y.}\ \bibnamefont {Ali}}, \bibinfo {author} {\bibfnamefont {S.}~\bibnamefont {Uemura}}, \bibinfo {author} {\bibfnamefont {K.}~\bibnamefont {Adachi}}, \bibinfo {author} {\bibfnamefont {H.}~\bibnamefont {Itoh}}, \bibinfo {author} {\bibfnamefont {K.}~\bibnamefont {Kinosita}},\ and\ \bibinfo {author} {\bibfnamefont {S.}~\bibnamefont {Ishiwata}},\ }\href {https://doi.org/10.1038/nsb803} {\bibfield  {journal} {\bibinfo  {journal} {Nat. Struct. Mol. Biol.}\ }\textbf {\bibinfo {volume} {9}},\ \bibinfo {pages} {464} (\bibinfo {year} {2002})}\BibitemShut {NoStop}%
\bibitem [{\citenamefont {Karplus}\ and\ \citenamefont {Gao}(2004)}]{karplus2004}%
  \BibitemOpen
  \bibfield  {author} {\bibinfo {author} {\bibfnamefont {M.}~\bibnamefont {Karplus}}\ and\ \bibinfo {author} {\bibfnamefont {Y.~Q.}\ \bibnamefont {Gao}},\ }\href {https://doi.org/10.1016/j.sbi.2004.03.012} {\bibfield  {journal} {\bibinfo  {journal} {Curr. Opin. Struct. Biol.}\ }\textbf {\bibinfo {volume} {14}},\ \bibinfo {pages} {250} (\bibinfo {year} {2004})}\BibitemShut {NoStop}%
\bibitem [{\citenamefont {Spetzler}\ \emph {et~al.}(2009)\citenamefont {Spetzler}, \citenamefont {Ishmukhametov}, \citenamefont {Hornung}, \citenamefont {Day}, \citenamefont {Martin},\ and\ \citenamefont {Frasch}}]{spetzler2009}%
  \BibitemOpen
  \bibfield  {author} {\bibinfo {author} {\bibfnamefont {D.}~\bibnamefont {Spetzler}}, \bibinfo {author} {\bibfnamefont {R.}~\bibnamefont {Ishmukhametov}}, \bibinfo {author} {\bibfnamefont {T.}~\bibnamefont {Hornung}}, \bibinfo {author} {\bibfnamefont {L.~J.}\ \bibnamefont {Day}}, \bibinfo {author} {\bibfnamefont {J.}~\bibnamefont {Martin}},\ and\ \bibinfo {author} {\bibfnamefont {W.~D.}\ \bibnamefont {Frasch}},\ }\href {https://doi.org/10.1021/bi9008215} {\bibfield  {journal} {\bibinfo  {journal} {Biochemistry}\ }\textbf {\bibinfo {volume} {48}},\ \bibinfo {pages} {7979} (\bibinfo {year} {2009})}\BibitemShut {NoStop}%
\bibitem [{\citenamefont {Schnurr}\ \emph {et~al.}(1997)\citenamefont {Schnurr}, \citenamefont {Gittes}, \citenamefont {MacKintosh},\ and\ \citenamefont {Schmidt}}]{schnurr1997}%
  \BibitemOpen
  \bibfield  {author} {\bibinfo {author} {\bibfnamefont {B.}~\bibnamefont {Schnurr}}, \bibinfo {author} {\bibfnamefont {F.}~\bibnamefont {Gittes}}, \bibinfo {author} {\bibfnamefont {F.~C.}\ \bibnamefont {MacKintosh}},\ and\ \bibinfo {author} {\bibfnamefont {C.~F.}\ \bibnamefont {Schmidt}},\ }\href {https://doi.org/10.1021/ma970555n} {\bibfield  {journal} {\bibinfo  {journal} {Macromolecules}\ }\textbf {\bibinfo {volume} {30}},\ \bibinfo {pages} {7781} (\bibinfo {year} {1997})}\BibitemShut {NoStop}%
\bibitem [{\citenamefont {Shin}\ \emph {et~al.}(2004)\citenamefont {Shin}, \citenamefont {Gardel}, \citenamefont {Mahadevan}, \citenamefont {Matsudaira},\ and\ \citenamefont {Weitz}}]{shin2004}%
  \BibitemOpen
  \bibfield  {author} {\bibinfo {author} {\bibfnamefont {J.~H.}\ \bibnamefont {Shin}}, \bibinfo {author} {\bibfnamefont {M.~L.}\ \bibnamefont {Gardel}}, \bibinfo {author} {\bibfnamefont {L.}~\bibnamefont {Mahadevan}}, \bibinfo {author} {\bibfnamefont {P.}~\bibnamefont {Matsudaira}},\ and\ \bibinfo {author} {\bibfnamefont {D.~A.}\ \bibnamefont {Weitz}},\ }\href {https://doi.org/10.1073/pnas.0308733101} {\bibfield  {journal} {\bibinfo  {journal} {Proc. Natl. Acad. Sci.}\ }\textbf {\bibinfo {volume} {101}},\ \bibinfo {pages} {9636} (\bibinfo {year} {2004})}\BibitemShut {NoStop}%
\bibitem [{\citenamefont {Luan}\ \emph {et~al.}(2008)\citenamefont {Luan}, \citenamefont {Lieleg}, \citenamefont {Wagner},\ and\ \citenamefont {Bausch}}]{luan2008}%
  \BibitemOpen
  \bibfield  {author} {\bibinfo {author} {\bibfnamefont {Y.}~\bibnamefont {Luan}}, \bibinfo {author} {\bibfnamefont {O.}~\bibnamefont {Lieleg}}, \bibinfo {author} {\bibfnamefont {B.}~\bibnamefont {Wagner}},\ and\ \bibinfo {author} {\bibfnamefont {A.~R.}\ \bibnamefont {Bausch}},\ }\href {https://doi.org/10.1529/biophysj.107.112417} {\bibfield  {journal} {\bibinfo  {journal} {Biophys. J.}\ }\textbf {\bibinfo {volume} {94}},\ \bibinfo {pages} {688} (\bibinfo {year} {2008})}\BibitemShut {NoStop}%
\bibitem [{\citenamefont {Kasza}\ \emph {et~al.}(2010)\citenamefont {Kasza}, \citenamefont {Broedersz}, \citenamefont {Koenderink}, \citenamefont {Lin}, \citenamefont {Messner}, \citenamefont {Millman}, \citenamefont {Nakamura}, \citenamefont {Stossel}, \citenamefont {MacKintosh},\ and\ \citenamefont {Weitz}}]{kasza2010}%
  \BibitemOpen
  \bibfield  {author} {\bibinfo {author} {\bibfnamefont {K.~E.}\ \bibnamefont {Kasza}}, \bibinfo {author} {\bibfnamefont {C.~P.}\ \bibnamefont {Broedersz}}, \bibinfo {author} {\bibfnamefont {G.~H.}\ \bibnamefont {Koenderink}}, \bibinfo {author} {\bibfnamefont {Y.~C.}\ \bibnamefont {Lin}}, \bibinfo {author} {\bibfnamefont {W.}~\bibnamefont {Messner}}, \bibinfo {author} {\bibfnamefont {E.~A.}\ \bibnamefont {Millman}}, \bibinfo {author} {\bibfnamefont {F.}~\bibnamefont {Nakamura}}, \bibinfo {author} {\bibfnamefont {T.~P.}\ \bibnamefont {Stossel}}, \bibinfo {author} {\bibfnamefont {F.~C.}\ \bibnamefont {MacKintosh}},\ and\ \bibinfo {author} {\bibfnamefont {D.~A.}\ \bibnamefont {Weitz}},\ }\href {https://doi.org/10.1016/j.bpj.2010.06.025} {\bibfield  {journal} {\bibinfo  {journal} {Biophys. J.}\ }\textbf {\bibinfo {volume} {99}},\ \bibinfo {pages} {1091} (\bibinfo {year} {2010})}\BibitemShut {NoStop}%
\bibitem [{\citenamefont {Dwyer}\ \emph {et~al.}(2022)\citenamefont {Dwyer}, \citenamefont {{Robertson-Anderson}},\ and\ \citenamefont {Gurmessa}}]{dwyer2022}%
  \BibitemOpen
  \bibfield  {author} {\bibinfo {author} {\bibfnamefont {M.~E.}\ \bibnamefont {Dwyer}}, \bibinfo {author} {\bibfnamefont {R.~M.}\ \bibnamefont {{Robertson-Anderson}}},\ and\ \bibinfo {author} {\bibfnamefont {B.~J.}\ \bibnamefont {Gurmessa}},\ }\href {https://doi.org/10.3390/polym14224980} {\bibfield  {journal} {\bibinfo  {journal} {Polymers}\ }\textbf {\bibinfo {volume} {14}},\ \bibinfo {pages} {4980} (\bibinfo {year} {2022})}\BibitemShut {NoStop}%
\bibitem [{\citenamefont {{Moreno-Mateos}}\ \emph {et~al.}(2022)\citenamefont {{Moreno-Mateos}}, \citenamefont {Hossain}, \citenamefont {Steinmann},\ and\ \citenamefont {{Garcia-Gonzalez}}}]{moreno-mateos2022}%
  \BibitemOpen
  \bibfield  {author} {\bibinfo {author} {\bibfnamefont {M.~A.}\ \bibnamefont {{Moreno-Mateos}}}, \bibinfo {author} {\bibfnamefont {M.}~\bibnamefont {Hossain}}, \bibinfo {author} {\bibfnamefont {P.}~\bibnamefont {Steinmann}},\ and\ \bibinfo {author} {\bibfnamefont {D.}~\bibnamefont {{Garcia-Gonzalez}}},\ }\href {https://doi.org/10.1038/s41524-022-00844-1} {\bibfield  {journal} {\bibinfo  {journal} {npj Comput. Mater.}\ }\textbf {\bibinfo {volume} {8}},\ \bibinfo {pages} {162} (\bibinfo {year} {2022})}\BibitemShut {NoStop}%
\bibitem [{\citenamefont {Zhao}\ \emph {et~al.}(2019)\citenamefont {Zhao}, \citenamefont {Kim}, \citenamefont {Chester}, \citenamefont {Sharma},\ and\ \citenamefont {Zhao}}]{zhao2019}%
  \BibitemOpen
  \bibfield  {author} {\bibinfo {author} {\bibfnamefont {R.}~\bibnamefont {Zhao}}, \bibinfo {author} {\bibfnamefont {Y.}~\bibnamefont {Kim}}, \bibinfo {author} {\bibfnamefont {S.~A.}\ \bibnamefont {Chester}}, \bibinfo {author} {\bibfnamefont {P.}~\bibnamefont {Sharma}},\ and\ \bibinfo {author} {\bibfnamefont {X.}~\bibnamefont {Zhao}},\ }\href {https://doi.org/10.1016/j.jmps.2018.10.008} {\bibfield  {journal} {\bibinfo  {journal} {J. Mech. Phys. Solids.}\ }\textbf {\bibinfo {volume} {124}},\ \bibinfo {pages} {244} (\bibinfo {year} {2019})}\BibitemShut {NoStop}%
\bibitem [{\citenamefont {Landau}\ and\ \citenamefont {Lifshitz}(1987)}]{LLfluidMechanics}%
  \BibitemOpen
  \bibfield  {author} {\bibinfo {author} {\bibfnamefont {L.~D.}\ \bibnamefont {Landau}}\ and\ \bibinfo {author} {\bibfnamefont {E.~M.}\ \bibnamefont {Lifshitz}},\ }\href@noop {} {\emph {\bibinfo {title} {Fluid Mechanics}}},\ \bibinfo {edition} {2nd}\ ed.\ (\bibinfo  {publisher} {Pergamon Press},\ \bibinfo {address} {New York},\ \bibinfo {year} {1987})\BibitemShut {NoStop}%
\bibitem [{\citenamefont {Tagliazucchi}\ \emph {et~al.}(2014)\citenamefont {Tagliazucchi}, \citenamefont {Zou},\ and\ \citenamefont {Weiss}}]{tagliazucchi2014}%
  \BibitemOpen
  \bibfield  {author} {\bibinfo {author} {\bibfnamefont {M.}~\bibnamefont {Tagliazucchi}}, \bibinfo {author} {\bibfnamefont {F.}~\bibnamefont {Zou}},\ and\ \bibinfo {author} {\bibfnamefont {E.~A.}\ \bibnamefont {Weiss}},\ }\href {https://doi.org/10.1021/jz5013609} {\bibfield  {journal} {\bibinfo  {journal} {J. Phys. Chem. Lett.}\ }\textbf {\bibinfo {volume} {5}},\ \bibinfo {pages} {2775} (\bibinfo {year} {2014})}\BibitemShut {NoStop}%
\bibitem [{\citenamefont {Gittes}\ \emph {et~al.}(1997)\citenamefont {Gittes}, \citenamefont {Schnurr}, \citenamefont {Olmsted}, \citenamefont {MacKintosh},\ and\ \citenamefont {Schmidt}}]{gittes1997}%
  \BibitemOpen
  \bibfield  {author} {\bibinfo {author} {\bibfnamefont {F.}~\bibnamefont {Gittes}}, \bibinfo {author} {\bibfnamefont {B.}~\bibnamefont {Schnurr}}, \bibinfo {author} {\bibfnamefont {P.~D.}\ \bibnamefont {Olmsted}}, \bibinfo {author} {\bibfnamefont {F.~C.}\ \bibnamefont {MacKintosh}},\ and\ \bibinfo {author} {\bibfnamefont {C.~F.}\ \bibnamefont {Schmidt}},\ }\href {https://doi.org/10.1103/PhysRevLett.79.3286} {\bibfield  {journal} {\bibinfo  {journal} {Phys. Rev. Lett.}\ }\textbf {\bibinfo {volume} {79}},\ \bibinfo {pages} {3286} (\bibinfo {year} {1997})}\BibitemShut {NoStop}%
\bibitem [{\citenamefont {Fruchart}\ \emph {et~al.}(2021)\citenamefont {Fruchart}, \citenamefont {Hanai}, \citenamefont {Littlewood},\ and\ \citenamefont {Vitelli}}]{fruchart2021}%
  \BibitemOpen
  \bibfield  {author} {\bibinfo {author} {\bibfnamefont {M.}~\bibnamefont {Fruchart}}, \bibinfo {author} {\bibfnamefont {R.}~\bibnamefont {Hanai}}, \bibinfo {author} {\bibfnamefont {P.~B.}\ \bibnamefont {Littlewood}},\ and\ \bibinfo {author} {\bibfnamefont {V.}~\bibnamefont {Vitelli}},\ }\href {https://doi.org/10.1038/s41586-021-03375-9} {\bibfield  {journal} {\bibinfo  {journal} {Nature}\ }\textbf {\bibinfo {volume} {592}},\ \bibinfo {pages} {363} (\bibinfo {year} {2021})}\BibitemShut {NoStop}%
\bibitem [{\citenamefont {Heiss}(2012)}]{heiss2012}%
  \BibitemOpen
  \bibfield  {author} {\bibinfo {author} {\bibfnamefont {W.~D.}\ \bibnamefont {Heiss}},\ }\href {https://doi.org/10.1088/1751-8113/45/44/444016} {\bibfield  {journal} {\bibinfo  {journal} {J. Phys. A: Math. Theor.}\ }\textbf {\bibinfo {volume} {45}},\ \bibinfo {pages} {444016} (\bibinfo {year} {2012})}\BibitemShut {NoStop}%
\end{thebibliography}%

\onecolumngrid

\appendix

\newpage


\section{Micropolar elasticity}\label{app:micropolar V}

In this appendix we derive the elastic potential of a solid material composed of complex (namely, not pointlike) rigid particles. This is the so-called micropolar (Cosserat) elastic material~\cite{eringen1999}.
We first find the strains (unlike classical elasticity, here there are two strains), one of which also accounts for the particle's rotation, and show that they are rotational invariant as required. 
Then we assume linear stress-strain relation and write the elastic potential to include the first geometric nonlinear contribution (whose origin is nonlinear strain due to particle's rotation).
%

\subsection{Strains in micropolar elasticity}
\label{app:micropolar strain}

\begin{figure}[t]
	\centering
\includegraphics[width=12 cm]{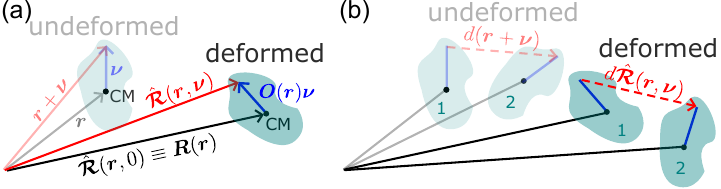}
	\caption{Illustration of the micropolar elasticity model~\cite{eringen1999}. (a) A single particle before deformation is described by the CM position $\bm{r}$ (gray) plus a rigid director $\bm\nu$ (light blue). Upon deformation, the endpoint position of the particle changes from $\bm{r}+\bm\nu$ (light red) to $\hat{\bm{\mathcal{R}}}(\bm{r},\bm\nu)$ (a function of $\bm{r}$ and $\bm\nu$, red). $\bm{R}(\bm{r})$ is the new CM position and $\bm{O}(\bm{r})$ is a rigid rotation operation on $\bm\nu$, both of which depend on $\bm{r}$. (b) Strains are accordingly defined via the distance change between the endpoints of two neighboring particles from $d(\bm{r}+\bm\nu)$ to $d\hat{\bm{\mathcal{R}}}(\bm{r},\bm\nu)$ (red dashed) upon deformation. Here $d$ denotes the small differential change.} 
	\label{appfig:geo strain}
\end{figure}

We follow conventional procedures~\cite{LLelastity,eringen1999} to derive the strains of a micropolar material.
%
%
Unlike classical elasticity, where deformation can be solely described by the center-of-mass (CM) displacement $\bm{u}$ for point-like particles, in micropolar elasticity one must consider the rigid particles direction, described by the vector $\bm{\nu}$ ($|\bm\nu|\sim$~particle size), see Fig.~\ref{appfig:geo strain}(a).
%
%
When the material is deformed the particle's CM is displaced, but the particle also rotates, namely $\bm\nu$ is changing.
Therefore, the position of a point mass within the particle before deformation (undeformed state) $\bm{r}+\bm\nu$ becomes $\hat{\bm{\mathcal{R}}}$ after deformation (deformed state): 
%
%
\begin{align}
\label{appeq:end dist vec}
\hat{\mathcal{R}}_i(\bm{r},\bm\nu)
\approx
\hat{\mathcal{R}}_i(\bm{r},0)
+
O_{ij} (\bm{r}) \nu_j
\equiv
R_i(\bm{r})
+
O_{ij} (\bm{r}) \nu_j
\, ,
\end{align}
where the point-mass position $\hat{\bm{\mathcal{R}}}(\bm{r},\bm\nu)$ is a function of both $\bm{r}$ and $\bm\nu$ of the undeformed state.
In the continuum coarse-grained (CG) description, the particles are assumed to be much smaller than typical lengths in the material,  $\hat{\bm{\mathcal{R}}}(\bm{r},\bm\nu)$ is linearly approximated at $\bm\nu=0$, hence, the difference between point-masses within the particle is discarded. Here $\hat{\bm{\mathcal{R}}}(\bm{r},\bm\nu=0)$ is  the deformation of the particle CM position $\bm{R}(\bm{r})$.
%
%
The difference from classical elasticity stems from the second term in~\eqref{appeq:end dist vec}, where $\bm{O}(\bm{r})$ is a rigid rotation matrix. In two  dimensions (2D) it assumes the following form:
%
%
%
\begin{align}\label{appeq: rotation matrix}
O_{ij} (\bm{r})
=  
\delta_{ij} 
- \varepsilon_{ij} \sin\theta(\bm{r})
- [1-\cos\theta(\bm{r})]\delta_{ij}
\, ,
\end{align}
with $\delta_{ij}$ and $\varepsilon_{ij}$ being the 2D Kronecker delta and Levi-Civita symbol, respectively. The $\bm{r}$-dependence of $\bm{O}(\bm{r})$ is embedded in the CG field describing the particles' internal rotation $\theta(\bm{r})$.
In 3D $\bm{O}$ can be expressed using Rodrigues' rotation formula in the axis-angle representation \cite{eringen1999,eremeyev2013,bauchau2003,brauns2024}. Moreover, when considering deformable particles $\bm{O}$ will also include stretching (see micromorphic theory~\cite{eremeyev2013}). We will not consider 3D or deformable particles in this work.

Rotation-invariant strains can be found via the distance change between two point-masses of two neighboring particles upon deformation, i.e., when $d(\bm{r}+\bm\nu)$ becomes $d\hat{\bm{\mathcal{R}}}(\bm{r},\bm\nu)$ after deformation, see Fig.~\ref{appfig:geo strain}(b). We then have:
\begin{align}\label{appeq:general strain}
(d \hat{\bm{\mathcal{R}}}(\bm{r},\bm\nu))^2
-(d\bm{r}+d\bm{\nu})^2
=&
\big\{
F_{ij}F_{il}
+
\big[
F_{ij}(\nabla^\circ_l O_{ik})
+F_{il}(\nabla^\circ_j O_{ik})
\big]\nu_k
+(\nabla^\circ_j O_{ik})(\nabla^\circ_l O_{im})\nu_k \nu_m
-\delta_{jl}
\big\}
d r_j d r_l
\notag\\
+&
2 \big\{
F_{ij}O_{il}
+O_{il}(\nabla^\circ_j O_{ik}) \nu_k
-\delta_{jl}
\big\} 
d r_j d \nu_l
\, ,
\end{align}
where $d\hat{\bm{\mathcal{R}}}(\bm{r},\bm\nu)$ is the differential form of Eq.~\eqref{appeq:end dist vec}, and the deformation gradient matrix $F_{ij}\equiv\partial R_i/\partial r_j$ with $\nabla^\circ_i \equiv \partial/\partial r_i$.
%
%
The introduction of an additional vector degree of freedom $\bm\nu$ expands the strain matrix into a $2\times2$ block matrix ($d\bm{r} \otimes d\bm{r}$, two symmetric blocks from $d\bm{r} \otimes d\bm\nu$, and $d\bm\nu \otimes d\bm\nu$). Therefore, there may be three different strains. In the case of rigid particles, the one related to $d\bm\nu \otimes d\bm\nu$ vanishes, such that  we are left with two strains (two curly-brackets in Eq.~\eqref{appeq:general strain}). 

We further consider the case in which the particle's size ($\sim |\bm\nu|$) 
is {much smaller than any other length-scale such that (see detailed discussion at the end of this subsection)} $\nu_k \nabla^\circ_j O_{ik}\ll F_{ij}$ and the effect of internal rotation gradient is negligible in Eq.~\eqref{appeq:general strain}:
\begin{align}\label{appeq:reduced cosserat diff in distance}
(d \hat{\bm{\mathcal{R}}}(\bm{r},\bm\nu))^2 
-(d\bm{r}+d\bm\nu)^2
\approx&
\
(F_{ij}F_{il}-\delta_{jl})d r_j d r_l
+
2
(F_{ij}O_{il}-\delta_{jl})
d r_j d \nu_l
\, .
\end{align}
The two strains in our elastic material (a reduced Cosserat media~\cite{eremeyev2013,grekova2019}) are:
\begin{align}\label{appeq:reduced cosserat two strains}
u_{ij}
\equiv
\frac{1}{2}\big(
F_{ki}F_{kj}-\delta_{ij} 
\big)
\ \, ,  \ \
e_{ij}
\equiv
O_{ki}F_{kj}-\delta_{ij} 
\, .
\end{align}
Here $u_{ij}$ is the classical elasticity Green-Lagrange strain~\cite{LLelastity}, while the other strain, $e_{ij}$,  accounts for the effect of internal rotation $\theta$ via $\bm{O}$. Note that when the particle's size ($\sim|\bm\nu|$) is no longer negligible in Eq.~\eqref{appeq:general strain}), the first-order correction adds a term $\sim\nabla_j^\circ\theta$ (from $\nabla^\circ_j O_{ik}$) such that the elasticity energy has a term $\sim (\nabla_i^\circ\theta)(\nabla_j^\circ\theta)$, similarly to a bond-angle elastic energy~\cite{chaikin1995} and the regular Cosserat elasticity~\cite{eringen1999,eremeyev2013}.



Importantly, for rigid body rotations $F_{ij}=O_{ij}$ (i.e., when internal particles' rotation matches  material volume rotation) we have $u_{ij}=e_{ij}=0$ indicating that the two strains, $u_{ij}$ and $e_{ij}$, are rotation-invariant.
Rotational invariance of the strains can also be proved directly by applying a global rotation ${\bm Q}$ to both the displacement $\bm{u}$ and the internal rotation $\theta$, namely, $u_i \rightarrow Q_{ij}u_{j}$ and $O_{ij} \rightarrow Q_{ik}O_{kj}$ in Eq.~\eqref{appeq:reduced cosserat two strains}, recovering the same strain measures.

{
In our work we neglect terms $\sim \nabla^\circ_j O_{ik}$, and, as we show, these terms are not crucial for odd elasticity. 
Nevertheless, these terms do play a role in micropolar elasticity~\cite{eringen1999}. 
Therefore, we now discuss the validity of such assumption in detail, which, to be precise is $\nu_k \nabla^\circ_j O_{ik}\ll F_{ij}$.
Considering typical length-scales in the system, this condition is expressed as $\nu \ll u\, L_\text{rot}/L_\text{dis}$, where $\nu\sim |\bm\nu|$ is the particle size, $L_\text{rot}$ denotes the characteristic length scale of variations in the internal rotation,  $L_\text{dis}$ is that of the displacement, and $u$ is the displacement magnitude. 
Since we consider the hydrodynamic limit (long-wavelength), one typically has $L_\text{rot},\, L_\text{dis},\,  u \gg \nu$. 
If internal rotations vary slow ($L_{\mathrm{rot}} \geq L_{\mathrm{dis}}$) our assumption is naturally satisfied. This is typically the case for homogeneous torque, or in the absence of torque, where the internal rotation varies only through the displacement. 
In the general case, we have $\nu/L_\text{rot}  \ll\,  u /L_\text{dis}$, which may be invalidated when internal rotations vary sharply. For example, in the case of inhomogeneous torque, $L_{\mathrm{rot}}$ is determined by the spatial variation of $\tau^\circ/\kappa_c$. Therefore, the validity of neglecting terms $\sim \nabla^\circ_j O_{ik}$ gives a limit on the inhomogeneity of $\tau^\circ$.
%
A practical realization of a system in which such terms are negligible, is a system of short rod-like particles interconnected over a relatively large separation length. Then, the rotation of an individual particle has only a weak effect on its neighbors.
}

\subsection{Micropolar elastic potential of isotropic materials}\label{app:V}

Assuming linear stress-strain relations, the elastic potential density $V$ is expressed in a general quadratic form:
\begin{align}\label{appeq:V_gen}
V 
=
\frac{1}{2}u_{ij}\bar{E}_{ijkl}u_{kl}
+
\frac{1}{2}e_{ij}C'_{ijkl}e_{kl}
+
\frac{1}{2}u_{ij} D_{ijkl}e_{kl}
+
\frac{1}{2}e_{ij} D'_{ijkl}u_{kl}
\, ,
\end{align}
where the strains $u_{ij}$ and $e_{ij}$ are given in Eq.~\eqref{appeq:reduced cosserat two strains}. For an isotropic material we further have $\bar{E}_{ijkl} =\bar\lambda (\delta_{ij}\delta_{kl})+ \bar\mu (\delta_{ik}\delta_{jl}+\delta_{il}\delta_{jk})$ with the Lam\'{e} coefficients $\bar\lambda$ and $\bar\mu$ as in classical elasticity.
Our goal in this subsection is to find the other elasticity tensors $\bm{C}'$, $\bm{D}$ and $\bm{D}'$ based on the isotropy assumption and the symmetry of $u_{ij}$ and $e_{ij}$. 

First, notice that, unlike $u_{ij}$ that is purely symmetric, $e_{ij}$ has both symmetric and antisymmetric parts due to the rotation matrix $O_{ij}$.  Therefore, antisymmetric coupling (for $i\leftrightarrow j$ or $k\leftrightarrow l$) is also allowed in $\bm{C}'$. The quadratic form of $V$ also requires $\bm{C}'$ to be symmetric under $(i,j) \leftrightarrow (k,l)$. We then find that 
$C'_{ijkl}=E'_{ijkl}+(\kappa_c\varepsilon_{ij}\varepsilon_{kl}/2)$,
%
%
where $\kappa_c$ is the antisymmetric coupling strength (sometimes termed the Cosserat elastic constant~\cite{eringen1999,eremeyev2013}).  $\bm{E}'$ has the same form as $\bar{\bm{E}}$, but with different Lam\'{e} coefficients. 

The quadratic form of the potential that requires all the elastic tensors to be symmetric under $(i,j) \leftrightarrow (k,l)$,
together with the symmetry of $u_{ij}$, requires that both $\bm{D}$ and $\bm{D}'$ in Eq.~\eqref{appeq:V_gen} have the same form as $\bar{\bm{E}}$ (again, with different Lam\'{e} coefficients).
To see why a symmetric-antisymmetric coupling (e.g., via $\delta_{ij}\varepsilon_{kl}$) is prohibited, consider the 3D isotropic case, in which we seek for the existence of a contraction $M_{nij}\varepsilon_{nkl}$ with $M_{nij}$  symmetric under $i\leftrightarrow j$. A systematic analysis from Ref.~\cite{markovich2019} implied that no such symmetric tensor $M_{nij}$ is allowed (by replacing $p_\alpha p_\beta$ with $\delta_{\alpha\beta}$ in their Eq.~(B.24) and taking all odd-order terms of $p_\alpha$ to zero). 
%
%
%
Taking advantage of the symmetric $u_{ij}$-$e_{kl}$ coupling, we can further merge the $\bm{D}$- and $\bm{D}'$-terms, yielding
\begin{align}\label{appeq:V_gen final}
V 
= 
\frac{1}{2}u_{ij}\bar{E}_{ijkl}u_{kl}
+
\frac{1}{2}e_{ij}C'_{ijkl}e_{kl}
+
u_{ij} \hat{E}_{ijkl}e_{kl}
,
\end{align}
where $\hat{\bm{E}}$ and $\bm{E}'$ (embedded in $\bm{C}'$) are defined similarly as $\bar{\bm{E}}$ but with two different pairs of the Lam\'{e} coefficients $(\hat\lambda, \hat\mu)$ and $(\lambda', \mu')$, respectively. This is Eq.~(3)
of the main text.

Note that in some of the micropolar elasticity literature~\cite{eringen1999,eremeyev2013} the strain $u_{ij}$ is treated as being a higher-order term of $e_{ij}$ because $2 u_{ij}=(e_{ki}+\delta_{ki})(e_{kj}+\delta_{kj})-\delta_{ij}$. Then, in many cases the only strain considered is $e_{ij}$ and the elastic energy taken as $V=e_{ij}C'_{ijkl}e_{kl}/2$. However, unlike the theory presented here, this does not recover the expected classical elasticity $u_{ij}\bar{E}_{ijkl}u_{kl}/2$ (that includes nonlinear contribution) when  particles cannot rotate (e.g., for point-like particles). 
%

\subsection{Geometric nonlinearity in the micropolar elastic potential}\label{app:V expansion}

In chiral active solids, the presence of active torques drives the internal rotation $\theta$ and induces a geometric nonlinearity in $V$, even in the regime of small displacements, namely, in the case $\nabla_j u_i \ll \theta \ll 1$, which we consider hereafter. Such nonlinearity is important for the emergence of odd elasticity (see Sec.~\ref{app:transform Cauchy} below).
Therefore, we expand $V$ (Eq.~\eqref{appeq:V_gen final}) to the first order beyond the quadratic approximation, $\sim \theta^2\nabla_j u_i$. To this order it is sufficient to use the linearized strain $u_{ij}\approx (\nabla^\circ_j u_i+\nabla^\circ_i u_j)/2 \equiv u^s_{ij}$ together with 
\begin{align}
e_{ij}
&\approx 
\underbrace{
\big[
u_{ij}^s
+
\varepsilon_{ij}
\big(\theta-\frac{1}
{2}\nabla^\circ\times\bm{u}
\big)
\big]
}_{e^{(1)}_{ij}}
+
\underbrace{
\varepsilon_{im}\theta\big(
\nabla^\circ_j u_m 
+\frac{1}{2}\varepsilon_{mj}\theta
\big)
}_{e^{(2)}_{ij}}
\, .
\label{appeq:eij}
\end{align}
Here $e^{(1)}_{ij}$ is the linear Cosserat strain~\cite{eringen1966,eremeyev2013} that encompasses the effect of mismatch between internal rotations and CM orbital rotation via $\theta-(\nabla^\circ\times\bm{u}/2)\equiv \phi^\circ$, which is absent in classical elasticity. The nonlinear quadratic part, despite its more complicated form, is proportional to this rotation mismatch, $e^{(2)}_{ij} \propto \theta \propto \phi^\circ$ (within our approximation scheme).

Using these strain expansions, $V$ is re-organized to separate the linear and geometric nonlinear contributions:
\begin{align}
V 
&\approx
\frac{1}{2}u^s_{ij}\bar{E}_{ijkl}u^s_{kl}
+
\frac{1}{2}\big[e^{(1)}_{ij}+e^{(2)}_{ij}\big]C'_{ijkl}\big[e^{(1)}_{kl}+e^{(2)}_{kl}\big]
+
u^s_{ij} \hat{E}_{ijkl}\big[e^{(1)}_{kl}+e^{(2)}_{kl}\big]
\notag\\
&\approx
\Big(
\frac{1}{2}
E_{ijkl}
u^s_{ij} u^s_{kl} 
+ 
\kappa_c\phi^{\circ^2} 
\Big)
+
(C'_{ijkl}+\hat{E}_{ijkl})e_{ij}^{(1)}e^{(2)}_{kl}
\label{appeq:V_gen expansion}
\\
&=
\underbrace{
\Big(
\frac{1}{2}u^s_{ij}E_{ijkl}u^s_{kl}
+\kappa_c\phi^{\circ^2}
\Big)
}_\text{linear Cosserat elasticity}
+
\underbrace{
\phi^\circ
\Big\{
\Big[
\tilde\mu
\big(
\varepsilon_{ij}\delta_{kl}
+
\varepsilon_{jk}\delta_{il}
\big)
-
\frac{\kappa_c}{2}\varepsilon_{ij}\delta_{kl}
\Big]
(\nabla^\circ_j u_i)(\nabla^\circ_l u_k)
-(\tilde\lambda+\tilde\mu-\kappa_c)\phi^\circ(\nabla^\circ\cdot\bm{u})
\Big\}
}_\text{geometric nonlinearity}
\label{appeq:V final}
\, ,
\end{align}
where $\tilde\lambda\equiv \lambda'+\hat\lambda$ and $\tilde\mu\equiv\mu'+\hat\mu$.
To derive the second line we abandoned a higher-order term $\sim e_{ij}^{(2)}e_{kl}^{(2)}$, wrote the antisymmetric coupling as $\kappa_c\varepsilon_{ij}\varepsilon_{kl}e_{ij}^{(1)}e^{(1)}_{kl}/2 = \kappa_c \phi^{\circ^2}$, and used symmetry to substitute $e^{(1)}_{ij} E'_{ijkl}u^s_{kl} = u^s_{ij}E'_{ijkl}u^s_{kl}$ (also for $\hat{E}_{ijkl}$) and $u^s_{ij}\hat{E}_{ijkl}e^{(2)}_{kl}=e^{(1)}_{ij}\hat{E}_{ijkl}e^{(2)}_{kl}$. 
%
%
The first two terms in Eq.~\eqref{appeq:V final} are the linear Cosserat elasticity~\cite{eringen1999,eremeyev2013} with $\bm{E}\equiv\bar{\bm{E}}+\bm{E}'+2\hat{\bm{E}}$, while $(C'_{ijkl}+\hat{E}_{ijkl})e_{ij}^{(1)}e^{(2)}_{kl}$ is the geometric nonlinear contribution.
%
%
Importantly, since $e_{ij}^{(1)}e^{(2)}_{kl} \propto (\varepsilon_{ij}\phi^\circ + u_{ij}^s) \phi^\circ$, a linear factor of $\phi^\circ$ can be extracted out, indicating its central role for geometric nonlinearity. Together with the fact that $\phi^\circ$ is the actual quantity driven by active torques (see Eq.~\eqref{appeq:ell dynamics} below), the elastic geometric nonlinearity is directly induced by active torque as stated  in the beginning of this subsection. 

Equation~\eqref{appeq:V final} is the elastic potential density $V$ used in our following analysis, which reduces to the linear isotropic elasticity $V =u^s_{ij} E_{ijkl} u^s_{kl}/2$ when there is no rotation mismatch $\phi^\circ=0$ (e.g., without active torque driving in our case). 

\section{Elastic stress via Lagrangian Poisson-bracket formalism}
\label{app:PB dynamics}

In the \textit{undeformed/Lagrangian} space, 
the Poisson bracket (PB) formalism gives the CG field dynamics using the system's Hamiltonian $H$~\cite{hohenberg1977,stark2003,stenull2004,stark2005,markovich2021,markovich2024}: 
\begin{align}\label{appeq:PB integral}
\frac{d \Phi_a(\bm{r})}{d t}
= 
-\int d\bm{r} \sum_b \lbrace \Phi_a(\bm{r}), \Phi_b(\bm{r}') \rbrace \frac{\delta H}{\delta \Phi_b(\bm{r}')},
\end{align} 
where $\Phi_a$ and $\Phi_b$ are the CG fields that can be expressed in terms of the microscopic generalized coordinates and momenta of the particles.
%
%
$d/dt$ is the \textit{total} time derivative as a pack of particles within the CG volume is traced in the Lagrangian formalism. The PB $\{\Phi_a(\bm{r}), \Phi_b(\bm{r}')\}$ dictates the dynamic coupling between the two CG fields, $\Phi_a$ and $\Phi_b$, at two different positions $\bm{r}$ and $\bm{r}'$ in the {undeformed/Lagrangian} space:  
\begin{align}\label{appeq:general PB}
\lbrace \Phi_a(\bm{r}), \Phi_b(\bm{r}') \rbrace = \sum_{\alpha}\left[ \frac{\partial\Phi_a(\bm{r})}{\partial \pi_k^\alpha}\frac{\partial\Phi_b(\bm{r}')}{\partial q_k^\alpha}-
\frac{\partial\Phi_a(\bm{r})}{\partial q_k^\alpha}\frac{\partial\Phi_b(\bm{r}')}{\partial \pi_k^\alpha}
\right]
\, .  
\end{align}
Here $\pi_k^\alpha$ and $q_k^\alpha$ are canonical conjugates of generalized microscopic momentum and position for particle 
$\alpha$, respectively. 
Importantly, the generalized coordinates and momenta are defined in real space, namely, in the \textit{deformed/Eulerian} space (and not in the undeformed space).

\subsection{Poisson brackets in the undeformed space} \label{app:PB formalism}

To calculate PBs using Eq.~\eqref{appeq:general PB}, we first find the canonical momenta for the position $\bm{R}^\alpha$ and  the internal rotation $\theta^\alpha$ of particle $\alpha$ in the deformed space.
These are defined via $\partial \mathcal{L}/\partial \dot{\bm{R}}^\alpha$ and $\partial \mathcal{L}/\partial \dot{\theta}^\alpha$,
with the particle time derivative $\dot{\bm X}^\alpha\equiv d{\bm X}^\alpha/dt$.
The Lagrangian $\mathcal{L}$ of our 2D system is:
\begin{align}\label{appeq:Lagrangian action}
\mathcal{L}=
\left[
\sum_\alpha 
\left(
\frac{{\bm{P}^\alpha}^2}{2 m^\alpha} 
+\frac{1}{2} I^\alpha \Omega^{\alpha^2}
\right)
\right]
- \mathcal{V}(\{ \theta^\alpha, \bm{u}^\alpha \})
\, ,   
\end{align}
where $\bm{P}^\alpha = m^\alpha \dot{\bm{R}}^\alpha$ is particle $\alpha$ linear momentum and $\Omega^\alpha=\dot{\theta}^\alpha$ its angular velocity. Here $m^\alpha$ and $I^\alpha$ are particle $\alpha$ mass and moment of inertia, respectively. 
The square-bracketed terms in Eq.~\eqref{appeq:Lagrangian action} are the kinetic energy, while $\mathcal{V}$ is the potential energy, which is only a function of  the particles' internal rotation $\theta^\alpha$ and displacement $\bm{u}^\alpha$. 
Hence, $\partial \mathcal{L}/\partial \dot{\bm{R}}^\alpha$ and $\partial \mathcal{L}/\partial \dot{\theta}^\alpha$ apply only to the kinetic part of $\mathcal{L}$, 
yielding the two canonical conjugate pairs: $(\bm{R}^\alpha,\bm{P}^\alpha)$ and $(\theta^\alpha,\ell^\alpha)$ where $\ell^\alpha = I^\alpha \Omega^\alpha$ is the particle's angular momentum. 

Our model has 6 independent CG fields in total: CM momentum density $\bm{g}^{\circ}(\bm{r})$, angular momentum density $\ell^\circ(\bm{r})$, mass density $\rho^\circ(\bm{r})$, active torque density $\tau^\circ(\bm{r})$, displacement $\bm{u}(\bm{r})$, and the internal rotation $\theta(\bm{r})$. The density-related fields are defined by ${\bm X}^\circ(\bm{r})\equiv \sum_{\alpha \in \Delta V^\circ} {\bm X}^\alpha \delta(\bm{r}-\bm{r}^\alpha)$, with the particle position ${\bm r}^\alpha$ and the CG volume $\Delta V^\circ$ in the undeformed space, while $\bm{u}(\bm{r})$ and $\theta(\bm{r})$ are defined by particle-averaging ${\bm X}(\bm{r})\equiv \sum_{\alpha \in \Delta V^\circ} {\bm X}^\alpha \delta(\bm{r}-\bm{r}^\alpha)/n^\circ(\bm{r})$, with the particle number density in the undeformed space $n^\circ(\bm{r}) = \sum_{\alpha \in \Delta V^\circ}\delta(\bm{r}-\bm{r}^\alpha)$. Note that due to our assumption of identical particles and isotropy, the field of the moment of inertia density $I(\bm{r})$, $\rho^\circ(\bm r)$ and $n^\circ(\bm r)$ are linearly-dependent $I \propto \rho^\circ \propto n^\circ$.
Using the conjugate pairs $(\bm{R}^\alpha,\bm{P}^\alpha)$ and $(\theta^\alpha,\ell^\alpha)$, Eq.~\eqref{appeq:general PB} has only two nonzero Lagrangian PBs within our 2D model:
\begin{align}
\{ u_i(\bm{r}) , g_j^{\circ}(\bm{r}')  \} 
&= -\sum_{\alpha} \frac{\partial u_i(\bm{r})}{\partial R_k^\alpha} \frac{\partial g_j^{\circ}(\bm{r}')}{\partial P_k^\alpha}
=
-\delta_{ij}\delta(\bm{r}-\bm{r}')
\quad ; \quad
\label{appeq: PB for u g}
\{\ell^\circ (\bm{r}) , \theta(\bm{r'})\} 
= \sum_{\alpha} 
\frac{\partial \ell^\circ(\bm{r})}{\partial\ell^\alpha}
\frac{\partial \theta(\bm{r}')}{\partial\theta^\alpha}
=
\delta(\bm{r}-\bm{r}')
\, ,
\end{align} 
where we used $\partial u_i^\alpha/ \partial R_j^\alpha = \delta_{ij}$ (the particle displacement is $\bm{u}^\alpha = \bm{R}^\alpha-\bm{r}^\alpha$) and the equality $\delta(\bm{r}-\bm{r}^\alpha) \delta(\bm{r}'-\bm{r}^\alpha) = \delta(\bm{r}-\bm{r}') \delta(\bm{r}'-\bm{r}^\alpha)$
which (after summation) cancels the $n^\circ(\bm{r}')$ dependence. 

We remark that many of the PBs vanish because $\delta(\bm{r}-\bm{r}^\alpha)$ and $\bm{R}^\alpha$ are decoupled, which is a result of tracing the same particle pack in the deformed space.
In contrast, PBs of Eulerian fields, where particle positions are described by $\delta(\bm{R}-\bm{R}^\alpha)$, are more complicated since $\partial \delta(\bm{R}-\bm{R}^\alpha)/\partial \bm{R}^\alpha$ does not vanish. The derivation of the Eulerian PBs will be done elsewhere~\cite{lee2026Poisson}.

\subsection{Dynamics in the undeformed/Lagrangian space and the first Piola-Kirchhoff stress}\label{app:Lagrangian dynamics and angle-relaxed Pij}


Using the PB formalism (Eqs.~\eqref{appeq:PB integral} and \eqref{appeq: PB for u g}) with the field Hamiltonian of Eq.~(1)
in the main text and $V$ from Eq.~\eqref{appeq:V final}, we find that
${d \bm{u}}/{d t}= \bm{g}^{\circ}/\rho^\circ\equiv \bm{v}^c$ with $\bm{v}^c$ being the CM velocity, ${d \theta}/{d t}=\Omega$, and $d\rho^\circ/dt=d\tau^\circ/dt=0$ as expected from the Lagrangian description.
The dynamics of the momentum and angular momentum densities then read 
\begin{align}
\frac{d\ell^\circ(\bm{r})}{dt}
&=-\frac{\delta H}{\delta \theta} 
=\tau^\circ-\frac{\partial V}{\partial \phi^\circ}
=
\tau^\circ
+
2\phi^\circ
\big[
-\kappa_c
+
(\tilde\lambda+\tilde\mu-\kappa_c)(\nabla^\circ\cdot\bm{u})
\big]
\, ,
\label{appeq:ell dynamics}
\\ 
\frac{d g_i^{\circ}(\bm{r})}{dt}
&=
-\frac{\delta H}{\delta u_i}
=
\nabla^\circ_j
\underbrace{
\bigg\lbrace
\kappa_c\varepsilon_{ij}\phi^\circ 
- (\tilde\lambda+\tilde\mu-\kappa_c)\phi^{\circ^2}\delta_{ij}
+ 
\bigg[
E_{ijkl}
-\frac{\phi^\circ}{2}
\big[
(2\tilde\lambda+2\tilde\mu-\kappa_c)\varepsilon_{ij}\delta_{kl}
+\kappa_c\varepsilon_{kl}\delta_{ij}
\big]
\bigg]
\nabla^\circ_l u_k
\bigg\rbrace
}_{\bm{P}}
\label{appeq:gc dynamics}
\, ,
\end{align}
where $\bm{P}$ is the first Piola-Kirchhoff (1st PK) stress tensor and $\phi^\circ \equiv \theta-(\nabla^\circ\times\bm{u}/2)$ is the mismatch between orbital and internal rotation. 
To obtain Eq.~\eqref{appeq:gc dynamics}, we applied the following decompositions: $\varepsilon_{il}\delta_{jk} = [(\varepsilon_{il}\delta_{jk}+\varepsilon_{ik}\delta_{jl}+\varepsilon_{jl}\delta_{ik}+\varepsilon_{jk}\delta_{il})/4]
+\varepsilon_{ij}\delta_{kl}/2
+\varepsilon_{kl}\delta_{ij}/2$ and $\varepsilon_{kj}\delta_{il}=
-[(\varepsilon_{il}\delta_{jk}+\varepsilon_{ik}\delta_{jl}+\varepsilon_{jl}\delta_{ik}+\varepsilon_{jk}\delta_{il})/4]
+(\varepsilon_{ij}\delta_{kl}/2)
+(\varepsilon_{kl}\delta_{ij}
/2)$. This is obtained by first symmetrizing under $k\leftrightarrow l$ and then under $i\leftrightarrow j$, together with the identities $\varepsilon_{il}\delta_{jk}-\varepsilon_{ik}\delta_{jl}
=\varepsilon_{im}(\delta_{ml}\delta_{jk}-\delta_{mk}\delta_{jl})
=(\varepsilon_{im}\varepsilon_{mj})\varepsilon_{lk}
=\varepsilon_{kl}\delta_{ij}$ and $\varepsilon_{il}\delta_{jk}-\varepsilon_{jl}\delta_{ik}
=\varepsilon_{ml}(\delta_{mi}\delta_{jk}-\delta_{mj}\delta_{ik})
=(\varepsilon_{ml}\varepsilon_{mk})\varepsilon_{ij}
=\varepsilon_{ij}\delta_{kl}$.

Because $\ell^\circ$ is not hydrodynamic {(see Sec.~\ref{sec:AM_est} below)},
we set $d\ell^\circ/dt =0$, which together with the small displacement approximation $\nabla^\circ\cdot\bm{u}\ll 1$ gives Eq.~(6)
of the main text:
$\phi^\circ = \tau^\circ \left[  1+\left(\tilde\lambda+\tilde\mu-\kappa_c\right)\nabla^\circ\cdot\bm{u}/\kappa_c \right]/(2\kappa_c)$, where the active torque density $\tau^\circ$ linearly drives $\phi^\circ$.

In the 1st PK stress ($\bm{P}$ in Eq.~\eqref{appeq:gc dynamics}), the pressure and additional elastic moduli (other than $\bm{E}$) are a consequence of the geometric nonlinearity in $V$, which vanishes in the linear Cosserat elasticity. 
Applying $\phi^\circ$ from the elimination of $\ell^\circ$ (see above and Eq.~(6)
of the main text) to Eq.~\eqref{appeq:gc dynamics}, we find the angle-relaxed $\bm{P}$, in which the rotation
mismatch $\phi^\circ$ is replaced by the active torque density $\tau^\circ$: 
\begin{align}
P_{ij}
=
\frac{\tau^\circ}{2}\varepsilon_{ij}
-\frac{\tau^{\circ^2}}{4\kappa_c^2}(\tilde\lambda+\tilde\mu-\kappa_c)\delta_{ij}
+
\underbrace{
\big[
E_{ijkl}
-\frac{\tau^\circ}{4}
(
\varepsilon_{ij}\delta_{kl}
+\varepsilon_{kl}\delta_{ij}
)
\big]
}_{\bm{C}_\text{PK}}
\nabla^\circ_l u_k 
\, .
\label{appeq:angle-relaxed Piola stress}
\end{align}
%
%


Note that despite angle-relaxation, the elasticity tensor in the {\it undeformed} space, $\bm{C}_\text{PK}$ of Eq.~\eqref{appeq:angle-relaxed Piola stress}, is not symmetric for $i \leftrightarrow j$, unlike what is expected from balance of angular momentum (see discussion in Sec.~\ref{app:total momentum dynamics}). This discrepancy comes from the geometric nonlinear contributions $\sim\tau^\circ\nabla_l^\circ u_k$. It reflects the fact that the undeformed space is not the natural `physical' space to examine conservation laws. 

{
\subsection{Estimation of $K^o$ magnitude}
\label{app:estimate ko}
To show that odd elasticity $K^o = \tau^\circ/4$ should have a significant effect, we compare its magnitude to other elastic moduli. As examples we take biological gels and synthetic elastomers with embedded magnetic beads.

In biological gels such as the cytoskeleton, the torques are generated by motor proteins such as myosin. 
The active torque density is then estimated as $\tau^\circ \sim n_s \tau_s$, where $n_s \sim 10^3\,(\mathrm{{\mu m})^{-3}}$ is the motor-protein number density~\cite{robinson2002} and the torque generated by a single motor protein is $\tau_s \sim 10^1\,\mathrm{pN}\!\cdot\!\mathrm{nm}$~\cite{ali2002,karplus2004,spetzler2009}. This gives $K^o \sim 1\,\mathrm{Pa}$, comparable to the shear modulus of biological gels, $\mu \sim 10^{-1}$--$100\,\mathrm{Pa}$~\cite{schnurr1997,shin2004,luan2008,kasza2010,dwyer2022}. 

For synthetic elastomers with magnetic particles with diameters of tens of microns~\cite{lucarini2022,zhai2025}, $\tau^\circ = M_r B_m$, where $M_\text{r} \sim  10-100\text{\ kA/m}$ is the remanent magnetization~\cite{moreno-mateos2022,zhao2019} and the commonly used magnetic fields are $B_m\sim 10 - 100 \text{\ mT}$~\cite{zhao2019,moreno-mateos2022}. In this case the odd elastic modulus is estimated to be $K^o \sim 0.1 - 10 \text{\ kPa}$.  This is  comparable to the shear modulus $\mu \sim 1-10 \text{\ kPa}$ of ultra-soft elastomer matrices of polydimethylsiloxane (PDMS)~\cite{moreno-mateos2022}.
}

{
\subsection{Estimation of the angular momentum relaxation time} \label{sec:AM_est}
In the main text and above we eliminated the angular momentum because it is not hydrodynamic. This means that it relaxes at some finite time that does not depend on the system size, unlike the linear momentum. Therefore, there is a length-scale (system size) above which it is appropriate to eliminate $\ell^\circ$.
To get a simple estimate of this length-scale, we keep only the essential ingredients needed to estimate the relaxation times of the linear and angular momenta. To this end, we use  the linear Cosserat elasticity in one dimension.
This reduces the rotational mismatch to the internal rotation, namely, $\theta-(\nabla^\circ\times\bm{u})/2\rightarrow\theta$, and the elasticity and viscosity tensors to single constants:  $E_{ijkl}\rightarrow E$ and $\eta_{ijkl}\rightarrow \eta$. The dynamics of the translational momentum and angular momentum then becomes:
\begin{align}
\rho^\circ\frac{d^2 u}{dt^2} 
=
\nabla^\circ \big[E(\nabla^\circ u)+
\eta\nabla^\circ \big(
\frac{d u}{dt}
\big)
\big]
-\Gamma \frac{d u}{dt}
\quad ; \quad
I^\circ \frac{d^2\theta}{dt^2} 
= \tau^\circ
-2\kappa_c
\theta
-\Gamma_\theta \frac{d }{dt}
\theta
\label{eq:theta}
\, ,
\end{align}
where we used $\ell^\circ = I^\circ d\theta/dt$ and $d{I}^\circ/dt=0$ in the reference (Lagrangian) description, and added an angular friction term with the coefficient $\Gamma_\theta$.

Solving Eqs.~\eqref{eq:theta} using the Fourier transform $u=\bar{u}e^{i(kx-\omega t)}$ and $\theta = \tau^\circ/2\kappa_c + \bar\theta e^{-i\omega t}$ yields the angular frequencies:
\begin{align}
\omega_u
=
\frac{- i (\Gamma+\eta k^2) \pm\sqrt{4 E  \rho^\circ k^2 -(\Gamma+\eta k^2)^2}}{2 \rho^\circ}
\quad ; \quad
\omega_\theta
=
\frac{- i \Gamma_\theta \pm\sqrt{8 
\kappa_c 
I^\circ -\Gamma^2_\theta}}{2 I^\circ}
\label{eq:omega_theta}
\, .
\end{align}
%
The relaxation times are defined as the inverse of the negative of the imaginary part of the respective frequencies (because $\{u,\theta\} \sim e^{\mathrm{Im}\{\omega\}t}$). For the estimation we take the longest relaxation time (out of the two solutions for each of $\{\omega_u,\omega_\theta\}$). 
In the hydrodynamic limit $k\rightarrow 0$, $\omega_u$ is always overdamped, unless $\Gamma =0$. Here `overdamped' means that the frequency does not have a real part.
From the structure of the Eq.~\eqref{eq:omega_theta} we have four cases based on whether $4 E  \rho^\circ k^2 \ll (\Gamma+\eta k^2)^2$ and $8 (\kappa_c + \kappa k^2) I^\circ \ll \Gamma^2_\theta$:
(i) overdamped $\omega_u$, underdamped $\omega_\theta$; (ii) overdamped $\omega_u$, overdamped $\omega_\theta$; (iii) underdamped $\omega_u$ (for $\Gamma =  0$), underdamped  $\omega_\theta$; (iv) underdamped $\omega_u$ (for $\Gamma = 0$), overdamped  $\omega_\theta$.
This leads to four different ratios between the relaxation times of the angle and displacement: 
\begin{align}
\frac{\tau_\theta^\text{under}}{\tau_u^\text{over}}
&= 
\bigg(\frac{2I^\circ E}{\Gamma\Gamma_\theta}\bigg)k^2
\, , \,
\frac{\tau_\theta^\text{over}}{\tau_u^\text{over}}
= 
\bigg(\frac{\Gamma_\theta E}{2\kappa_c \Gamma}\bigg)k^2
\,  \hspace{20 pt} \text{for\ $\Gamma\neq0$}\, ;
\label{eq:relax ratio}\\
\frac{\tau_\theta^\text{under}}{\tau_u^\text{under}}
&= 
\bigg(\frac{I^\circ\eta}{\rho^\circ\Gamma_\theta}\bigg)k^2
\, , \,
\frac{\tau_\theta^\text{over}}{\tau_u^\text{under}}
= 
\bigg(\frac{\Gamma_\theta\eta}{4\kappa \rho^\circ}\bigg)k^2
\hspace{20 pt} \text{for\ $\Gamma=0$}
\, .
\label{eq:relax ratio without friction}
\end{align}
Importantly, all of the relaxation-time ratios in Eqs.~\eqref{eq:relax ratio} and \eqref{eq:relax ratio without friction} scale quadratically with $k$. Therefore, in the long-wavelength limit (small $k$), the angle $\theta$ necessarily relaxes much faster than the displacement. This is the hydrodynamic limit.
The prefactors in Eqs.~\eqref{eq:relax ratio} and \eqref{eq:relax ratio without friction} set the length-scale for which the fast-relaxation assumption holds.
Because the overdamped relaxation time is shorter than the underdamped one, $\tau^\text{under}_\theta/\tau^\text{over}_u$ yields the largest prefactor. 
Therefore, $\tau^\text{under}_\theta/\tau^\text{over}_u\ll 1$ sets the minimal wavelength $\lambda_\text{min}$ required for the fast angle relaxation to be valid.

In the following analysis, we use interconnected spherical particles suspended in water as an illustrative example. The Stokes drags~\cite{LLfluidMechanics} for translation and rotation are
$\Gamma=6n\pi\eta r$ and $\Gamma_\theta=8n\pi\eta r^3$, respectively,
where $r$ is the particle radius, $n$ the particle number density, and $\eta$ the solvent viscosity. The moment of inertia density of spherical particles is $I^\circ=2nmr^2/5$, where $m$ is the mass of a single particle. 
We then find that $2I^\circ E/\left(\Gamma \Gamma_\theta\right) = 4E\rho_p r^4/\left(135\phi^2\eta^2\right)$.
%
%
%
where we also used $\rho_p = m n$ and $n=3\phi/(4\pi r^3)$, 
with $\phi$ being the volume fraction and $\rho_p$ the particle mass density. 
%
We now take $\rho_p \sim \mathrm{10^{3}\, kg/m^3}$, characteristic of water-rich soft particles such as hydrogels~\cite{tagliazucchi2014}, $\eta \sim 10^{-3}\,\mathrm{Pa\cdot s}$ for water at room temperature, $E \sim  \mathrm{100 \, Pa}$ which is within the typical range of shear moduli for soft networks~\cite{schnurr1997}, and $\phi \sim 10^{-1}$. This gives $\tau_\theta^\text{under}/\tau_u^\text{over} \sim \left(10^6 [\mathrm{m}^{-1}] r^2 / \lambda\right)^2
\ll 1$
%
where we used $k\equiv 2\pi/\lambda$, and $\lambda$ is the wavelength we seek. 
Therefore the fast-angle relaxation is valid for typical lengths of $\lambda \gg 10^6 r^2$. For $r=1 \, \mathrm{\mu m}$, this gives $\lambda_{\min} \sim 1 \, \mathrm{\mu m}$, indicating that the hydrodynamic limit can already be reached on the scale of only a few particles. However, $\lambda_{\min}$ increases sharply with particle size. For example, for $r= 1 \, \mathrm {mm}$, one finds $\lambda_{\min}\sim 1 \, \mathrm{m}$, so that more than a thousand particles are required.
}

\section{Transformation between the first Piola-Kirchhoff and Cauchy stress}\label{app:transform Cauchy}

We are interested in the real forces within the system, namely, the stresses in the \textit{real/deformed} space. These are related to the Cauchy stress $\bm{\sigma}$.
Using standard transformation~\cite{LLelastity}of the first PK stress $\bm{P}$ (Eq.~\eqref{appeq:angle-relaxed Piola stress}) we have:
\begin{align}
\sigma_{ij}
=
J^{-1}P_{il}F_{jl}
\approx
(1-\nabla_k u_k)P_{il}(\delta_{lj}+\nabla_l u_j)
=
P_{ij}- P_{ij} \delta_{lk} \nabla_l u_k+ P_{il}\delta_{jk}\nabla_l u_k
\, ,
\label{appeq:stress transform}
\end{align}
where the Jacobian $J\equiv\det{\bm F}$, and we expanded $J^{-1}$ and the deformation gradient tensor $F_{ij}$ to linear order in $\nabla_l u_k \equiv \partial u_k/\partial R_l$ (i.e., the displacement gradient in the deformed space). 
Due to the presence of active torque $\tau^\circ$, we keep terms up to order ${\cal O} \left(\phi^\circ\nabla_j u_i\right)$ and ${\cal O} \left(\phi^{\circ^2}\right)$ in $\bm{\sigma}$. 
The term $\varepsilon_{ij}\tau^\circ/2$ in Eq.~\eqref{appeq:angle-relaxed Piola stress} contributes two new additional terms $-\tau^\circ\varepsilon_{ij}\delta_{kl}\nabla_l u_k/2$ and $\tau^\circ\varepsilon_{il}\delta_{jk}\nabla_l u_k/2$ in Eq.~\eqref{appeq:stress transform}, while other contributions cancels off:
\begin{align}
\sigma_{ij}
&=
\frac{\tau^\circ}{2}\varepsilon_{ij}
-\frac{\tau^{\circ^2}}{4\kappa_c^2}(\tilde\lambda+\tilde\mu-\kappa_c)\delta_{ij}
+
\big[
E_{ijkl}
-\frac{\tau^\circ}{4}
(
3\varepsilon_{ij}\delta_{kl}
+\varepsilon_{kl}\delta_{ij}
-2\varepsilon_{il}\delta_{jk}
)
\big]
\nabla_l u_k
\notag\\
&=
\underbrace{
\frac{\tau^\circ}{2}\varepsilon_{ij}
-\frac{\tau^{\circ^2}}{4\kappa_c^2}(\tilde\lambda+\tilde\mu-\kappa_c)\delta_{ij}
}_{\bm\sigma^\text{pre}}
+
\underbrace{
\big\lbrace
E_{ijkl}
-\frac{\tau^\circ}{4}
\big[
2\varepsilon_{ij}\delta_{kl}
-(\varepsilon_{jl}\delta_{ik}+\varepsilon_{ik}\delta_{jl})
\big]
}_{\bm{C}}
\big\rbrace
\nabla_l u_k
\label{appeq:Cauchy stress Lagra tau}
\\
&\approx
\underbrace{
\frac{\tau}{2}\varepsilon_{ij}
-\frac{\tau^2}{4\kappa_c^2}(\tilde\lambda+\tilde\mu-\kappa_c)\delta_{ij}
}_{\bm\sigma^\text{pre}}
+
\underbrace{
\big[
E_{ijkl}
+\frac{\tau}{4}
(\varepsilon_{jl}\delta_{ik}+\varepsilon_{ik}\delta_{jl})
\big]
}_{\bm{C}}
\nabla_l u_k
\label{appeq:Cauchy stress Eule tau}
\ .
\end{align}
Here $\bm\sigma^\text{pre}$ is the prestress due to the active torques, even in the absence of deformation, and $\bm{C}$ is the elasticity tensor in {\it real} space.
To get Eq.~\eqref{appeq:Cauchy stress Lagra tau}, we used $\nabla^\circ_i = F_{ji}\nabla_j\approx (\delta_{ij}+\nabla_i u_j) \nabla_j$ and the decomposition $2\varepsilon_{il}\delta_{jk}=[(\varepsilon_{jk}\delta_{il}
+\varepsilon_{jl}\delta_{ik}
+\varepsilon_{ik}\delta_{jl}
+\varepsilon_{il}\delta_{jk})/2]+\varepsilon_{ij}\delta_{kl}+\varepsilon_{kl}\delta_{ij}$ (see below Eq.~\eqref{appeq:gc dynamics}) and also $(\varepsilon_{jk}\delta_{il}
+\varepsilon_{jl}\delta_{ik}
+\varepsilon_{ik}\delta_{jl}
+\varepsilon_{il}\delta_{jk})\nabla_l u_k 
=2 (\varepsilon_{jl}\delta_{ik}+\varepsilon_{ik}\delta_{jl})\nabla_l u_k$. 
Then, using the active torque in the deformed space,  $\tau^\circ=(1+\nabla\cdot\bm{u})\tau$, an additional term $\tau\varepsilon_{ij}\delta_{kl}\nabla_l u_k/2$ comes from $\tau^\circ\varepsilon_{ij}/2$ of Eq.~\eqref{appeq:Cauchy stress Lagra tau}, yielding Eq.~\eqref{appeq:Cauchy stress Eule tau}.

Remarkably, odd elasticity~\cite{scheibner2020} appears naturally (as a result of the geometric nonlinearity) in the Cauchy stress. It is the last term in Eqs.~\eqref{appeq:Cauchy stress Lagra tau} and \eqref{appeq:Cauchy stress Eule tau}, where the odd modulus here is $K^o=\tau/4$, such that it vanishes in the absence of active driving.
Importantly, within the Cauchy stress, the elasticity tensor $\bm{C}$  is symmetric under $i \leftrightarrow j$, thus obeying balance of angular momentum (see also Sec.~\ref{app:total momentum dynamics}) as required, see Eq.~\eqref{appeq:Cauchy stress Eule tau}. 
Indeed, the only asymmetric part of $\bm\sigma$ is $\varepsilon_{ij} \tau/2$~\cite{markovich2025}. 

In Eq.~\eqref{appeq:Cauchy stress Lagra tau} a mixed representation is used: the Cauchy stress $\bm\sigma$ (of the \textit{deformed} space) is written in terms of $\tau^\circ$ (of the \textit{undeformed} space). In such representation the symmetry for $i \leftrightarrow j$ is broken and an additional antisymmetric modulus that couples dilation with rotation (the coefficient of $\varepsilon_{ij}\delta_{kl}$) $A=\tau^\circ/2$  appears.
Clearly, balance of angular momentum is obeyed as seen from Eq.~\eqref{appeq:Cauchy stress Eule tau}, but in such mixed representation, which might be useful as explained in the main text, the stress may contain antisymmetric terms.

\section{Dynamics in the deformed/Eulerian space}\label{app:Eulerian dynamics}

%
In this section we write the dynamics of the fields in the \textit{deformed} (Eulerian) space $\bm{R}$. To this end, the Lagrangian dynamics in the \textit{undeformed} space $\bm{r}$ (Eqs.~\eqref{appeq:ell dynamics}-\eqref{appeq:gc dynamics}) is transformed by expanding the total time derivative to include the effect of the changing volume on density-related fields. 
For the particle-averaged fields $\bm{u}$ and $\theta$, the transformation is trivial: $d{\bm X}/dt= \dot{\bm X} + {\bm v}^c \cdot \nabla {\bm X}$ with $\dot{\bm X}\equiv \partial {\bm X}/\partial t$.  Therefore, ${d {\bm u}}/{dt} = \dot{\bm u} + {\bm v}^c \cdot \nabla {\bm u} = {\bm v}^c$ and ${d \theta}/{dt} =\dot\theta + {\bm v}^c \cdot \nabla\theta = \Omega$. 
For the density-related fields ($\rho^\circ$, $\tau^\circ$, $\ell^\circ$ and $\bm{g}^{\circ}$), we apply $d/dt$ on both sides of 
${\bm X}(\bm{R}) = J^{-1} X^\circ(\bm{r})$, where ${\bm X}(\bm{R})$ is a field in the deformed (Eulerian) space $\bm{R}$ and the volume change factor $J^{-1}\approx 1-\nabla\cdot\bm{u}$, to obtain 
\begin{align}
%
\dot{\bm X} +  \nabla_j(v^c_j {\bm X})
&=
(1-\nabla\cdot\bm{u})\frac{d {\bm X}^\circ}{dt}
\label{appeq:dynamic transform}
\, ,
\end{align}
where $\nabla_j(v^c_j {\bm X})$ is the streaming term expected in the Eulerian formulation, the term $\left(\nabla \cdot \bm{u}\right) dX^\circ/dt \sim \tau\nabla\cdot\bm{u}$ is a nonlinear contribution that is important in the presence of active torques, and we have discarded a term $\sim (\nabla \cdot \bm{u})\nabla_j( du_j/dt)$.

Using Eq.~\eqref{appeq:dynamic transform}, $d\rho^\circ/dt =0$ and $d\tau^\circ/dt =0$ give
$\dot\rho + \nabla_j ( v^c_j \rho)=0$ and $\dot\tau + \nabla_j ( v^c_j \tau)=0$, while the angular momentum dynamics, Eq.~\eqref{appeq:ell dynamics}, and CM momentum dynamics, Eq.~\eqref{appeq:gc dynamics}, become:
\begin{align}
\dot\ell + \nabla_j(v^c_j\ell)
&=
(1-\nabla\cdot\bm{u})(\tau^\circ-2\kappa_c\phi^\circ)
-2\phi^\circ(\kappa_c-\tilde\lambda-\tilde\mu)(\nabla^\circ\cdot\bm{u})
\approx\
\tau-2\kappa_c\phi
+2\phi(\tilde\lambda+\tilde\mu)(\nabla\cdot\bm{u})
\, ,
\label{appeq:ell Eul dynamics}
\\ \nonumber
\dot{g}^c_i + \nabla_j(v^c_j g^c_i)
&=
(1-\nabla\cdot\bm{u})\nabla^\circ_l P_{il}
=
(1-\nabla\cdot\bm{u})F^T_{lj}\nabla_j P_{il}
\approx
\nabla_j\big[
(1-\nabla\cdot\bm{u})P_{il}F^T_{lj} 
\big]
\\
&=
\nabla_j
\bigg\lbrace
\kappa_c\varepsilon_{ij}\phi
-\phi^2(\tilde\lambda+\tilde\mu-\kappa_c)\delta_{ij}
+ E_{ijkl} \nabla_l u_k
+\phi\bigg[
\frac{\kappa_c}{2}(\varepsilon_{ik}\delta_{jl}+\varepsilon_{jl}\delta_{ik})
-(\tilde\lambda+\tilde\mu)\varepsilon_{ij}\delta_{kl}
\bigg]\nabla_l u_k
\bigg\rbrace
\label{appeq:gc Eul dynamics}
\, ,
\end{align}
where $\phi\equiv\theta-(\nabla\times\bm{u})/2 \approx \phi^\circ$ (see $\phi^\circ$ definition in Eq.~\eqref{appeq:V_gen final}). 
As before, we kept up to order ${\cal O}\left(\phi\nabla_l u_k\right)$ and ${\cal O}\left( \phi^2\right)$.
%
Note that Eq.~\eqref{appeq:gc Eul dynamics} also recovers the stress transformation $(1-\nabla\cdot\bm{u})P_{il}F_{jl}$ of Eq.~\eqref{appeq:stress transform}. 

The elasticity tensor in Eq.~\eqref{appeq:gc Eul dynamics} contains an antisymmetric part (for $i \leftrightarrow j$) as it is only the elasticity of the CM dynamics $\dot{\bm{g}}^c$. 
In Section~\ref{app:total momentum dynamics} we show that such asymmetry is removed by using the \textit{total} momentum dynamics $\dot{\bm{g}}$. It is further shown that after angle-relaxation the elasticity tensor of CM dynamics coincides with that of the total momentum.
Finally, let us note that the dynamics in the deformed space can also be derived directly from an Eulerian PB formalism, which is out of the scope of this paper and is left to future work~\cite{Note1}.


\section{Symmetric elasticity tensor in total momentum dynamics}\label{app:total momentum dynamics}

Particles in our odd solid are non-point-like and can internally rotate to possess spin angular momentum density $\ell$. Hence, the CM momentum $\bm{g}^c$ alone is not sufficient to balance the angular momentum. We must examine the \textit{total} momentum ${\bm g} = {\bm g}^c +(\nabla\times\ell/2)$~\cite{markovich2021,markovich2024}, which takes into account the particle-spinning. Making use of Eqs.~\eqref{appeq:ell Eul dynamics} and \eqref{appeq:gc Eul dynamics} the dynamics of the total momentum reads:
\begin{align}
\dot{g}_i 
&=
-\nabla_j(g_i v_j)
+
\nabla_j\bigg\{
\frac{\tau}{2}\varepsilon_{ij}
-\phi^2(\tilde\lambda+\tilde\mu-\kappa_c)\delta_{ij}
+\eta^\mathrm{tot}_{ijkl}\nabla_l v_k
+\bigg[
E_{ijkl}
+\frac{\kappa_c\phi}{2}(\varepsilon_{ik}\delta_{jl}+\varepsilon_{jl}\delta_{ik})
\bigg]
\nabla_l u_k
\bigg\}
\, ,
\label{appeq:g total dynamics}
\end{align}
where we have used $\nabla_j(g_i^c v_j^c)+[\varepsilon_{ip}\nabla_j\nabla_p\left(\ell v_j^c\right)/2] = \nabla_j(v_j g_i)-\nabla_j(
\eta_{ijkl}^\mathrm{tot}\nabla_l v_k)$ with the total velocity $\bm{v} \equiv \bm{g}/\rho$. The viscosity tensor, $\eta^\mathrm{tot}_{ijkl}=
\ell\big[(\varepsilon_{ik}\delta_{jl}+\varepsilon_{il}\delta_{jk})+\varepsilon_{kl}\delta_{ij}\big]\nabla_l v_k/2$~\cite{markovich2021,markovich2024}, contains odd viscosity and odd pressure (pressure-vorticity coupling) due to the particle spinning. 

In Eq.~\eqref{appeq:g total dynamics}, the only source of change of the \textit{total} angular momentum is the antisymmetric term $\sim\tau$, giving $dL/dt\equiv\int (\bm{R}\times \dot{\bm{g}} ) d\bm{R}=\int \tau d\bm{R} +\ [\mathrm{boundary\ terms}]$ as expected. This indicates that the symmetric $\eta^\mathrm{tot}_{ijkl}$ and elasticity tensor (square-bracketed)
%
conserve angular momentum.
%

Moreover, in our case, angle-relaxation in the undeformed space $d\ell^\circ/dt =0$ implies that  $\ell^\circ =0$ (in our formulation dissipation was not introduced so far such that $\ell^\circ$ oscillates around zero, but with any dissipation it will vanish). Then, from the volume change we have $\ell=\ell^\circ(1-\nabla\cdot\bm{u})=0$ such that   $\nabla_j(\ell v_j^c)=0$, and we deduce $\dot\ell=0$ from Eq.~\eqref{appeq:dynamic transform}. Therefore, after angle-relaxation $\dot{\bm{g}}$ reduces to $\dot{\bm{g}}^c$ in the deformed space, which guarantees a symmetric elasticity tensor. Indeed, Eq.~\eqref{appeq:g total dynamics} after angle relaxation (with Eq.~(6)
inserted and $\eta^\mathrm{tot}_{ijkl}\propto\ell=0$) recovers the symmetric $\bm{C}$ of Eq.~\eqref{appeq:Cauchy stress Eule tau}.




\section{Odd elastostatics: Poisson ratio and odd ratio under a uniaxial stress}\label{app:odd ratio}

\begin{figure}[t]
	\centering
	\includegraphics[width=7 cm]{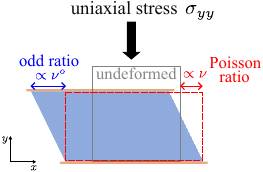}
	\caption{Odd solid under uniaxial stress $\sigma_{yy}$ from the top. The sides are stress-free $\sigma_{xx}=\sigma_{yx}=0$. Due to the presence of the odd elastic modulus $K^o$, the solid is tilted (blue shaded). This tilt is captured by the  odd ratio $\nu^o$. This tilt is in addition to the extension measured by the Poisson ratio $\nu$ (red dashed). The undeformed shape is shown in gray line.} 
	\label{appfig:odd ratio}
\end{figure}

In this section we study the elastic response of our odd solid to uniaxial stress.
Let us consider the odd solid with uniform local active torques $\tau^\circ$ (of the \textit{undeformed} space) placed between two paralleled plates at its top and bottom. 
Even without deformation, the plates balance the inherent surface torque and pressure that are present due to the active torques (first two terms of Eq.~(7)).
%
A small uniaxial stress $\sigma_{yy}=p \ll 1$ is then applied at the top plate, while both plates are maintained straight during the compression (namely, $\nabla_x u_y=0$), see Fig.~\ref{appfig:odd ratio}. The sides of the odd solid, initially perpendicular to the $x$-axis, are traction free, $\hat{\bm n}\cdot{\bm\sigma} =0$, which for small deflection means that $\sigma_{xx}=\sigma_{yx}=0$ (to leading order) on the sides of the solid. 
With these conditions, the static elastic response in the irreducible stress/strain decomposition~\cite{fruchart2023} is:
\begin{equation}
\begin{pmatrix}
\textcolor{blue}{B} & 
\textcolor{gray}{0}
& \textcolor{gray}{0} & \textcolor{gray}{0}\\
\textcolor{purple}{A} & \textcolor{gray}{0} & \textcolor{gray}{0} & \textcolor{gray}{0}\\
\textcolor{gray}{0} & \textcolor{gray}{0} & \textcolor{PineGreen}{\mu} & \textcolor{RubineRed}{K^o}\\
\textcolor{gray}{0} & \textcolor{gray}{0} & \textcolor{RubineRed}{-K^o} & \textcolor{PineGreen}{\mu}
\end{pmatrix}
\begin{pmatrix}
\nabla_x u_x +\nabla_y u_y\\
 - \nabla_y u_x \\
\nabla_x u_x - \nabla_y u_y \\
\nabla_y u_x 
\end{pmatrix}
=\frac{1}{2}
\begin{pmatrix}
\sigma_{xx} + \sigma_{yy}   \\
\sigma_{yx} - \sigma_{xy}    \\
\sigma_{xx} - \sigma_{yy}    \\
\sigma_{xy} + \sigma_{yx}    
\end{pmatrix}
=
\frac{1}{2}
\begin{pmatrix}
p   \\
-\sigma_{xy}  \\
-p    \\
\sigma_{xy}   
\end{pmatrix}
\, ,
\end{equation}
%
which gives: 
\begin{align}
\nonumber\frac{\nabla_x u_x}{p}
&=
\frac{(K^o+A)K^o -\mu(B-\mu)}{4B(K^{o^2}+\mu^2)}
 &;&   
& &\frac{\nabla_y u_y}{p}
= \frac{(K^o-A)K^o+\mu(B+\mu)}{4 B(K^{o^2}+\mu^2)} 
\, , \\
\frac{\nabla_y u_x}{p}
&=\frac{-(B K^o +A \mu)}{2B(K^{o^2}+\mu^2)} 
  &;& 
& &\frac{\sigma_{xy}}{p} 
= -\frac{A }{ B} \, .
\end{align}
The Poisson ratio $\nu\equiv-\nabla_x u_x/\nabla_y u_y$ and the odd ratio $\nu^o\equiv-\nabla_y u_x/2\nabla_y u_y$ are then 
\begin{align}
\nu
&=
\frac{(K^o+A)K^o -\mu(B-\mu)}{(A-K^o)K^o-\mu(B+\mu)}
=
\frac{16\mu(B-\mu)-3\tau^{\circ^2}}{16\mu(B+\mu)-\tau^{\circ^2}}
\quad ; \quad
\nu^o 
=
\frac{B K^o +A \mu}{(K^o-A)K^o+\mu(B+\mu)}
=
\frac{4\tau^\circ(B+2\mu)}{16\mu(B+\mu)-\tau^{\circ^2}}
\, ,
\end{align}
where  $A= 2 K^o=\tau^\circ/2$. To leading order in $\tau^o$ we find, $\nu - \nu_\mathrm{pass} \propto \tau^{\circ^2}$ and $\nu^o\propto \tau^\circ$, with $\nu_\mathrm{pass} \equiv (B-\mu)/(B+\mu)$. Clearly, the active torques have a more pronounced effect on the odd ratio $\nu^\circ$.

\section{Odd viscoelastic dynamics of displacement waves}\label{app:odd viscoelas}

This section explores the odd viscoelastic  model of the main text, where the odd solid is immersed in an odd chiral active fluid.   

\subsection{Kelvin-Voigt viscoelasticity from a `two-fluid' model}\label{app:two fluid model}

We introduce the `two-fluid model'~\cite{levine2001,levine2009} to incorporate the effects of viscosity by embedding our odd solid in a fluid of non-point-like particles. 
In particular, these fluid particles can spin due to chiral activity (e.g., active torques) and are of a much smaller size than the mesh size of the solid, such that the fluid particles' \textit{total} momentum $\bar{g}_i=\bar{g}^c_i+(\nabla\times\bar\ell/2)_i$~\cite{markovich2021,markovich2024}. 
Here $\bar{g}^c_i$ and $\bar\ell$ are the CG fields of the CM momentum density and spin angular momentum density for the solvent fluid, respectively.
%

%
%
Let us begin with a general formulation without specifying the elasticity and viscosity tensors. We assume small displacements $|\bm{u}|\ll 1$ and fast angle-relaxation for the solid, so we can linearize the dynamics: $\bm{v}=\bm{v}^c \approx \dot{\bm{u}}$, $\dot{\bm{g}}=\dot{\bm{g}}^c \approx \rho \ddot{\bm{u}}$ (see discussion in Sec.~\ref{app:total momentum dynamics}). 
A phenomenological solid-fluid coupling is then introduced via the velocity difference $(\dot{\bm{u}}-\bar{\bm{v}})$~\cite{levine2001,levine2009} where $\bar{\bm{v}}\equiv \bar{\bm{g}} /\bar\rho$ and $\bar\rho$ is the fluid mass density. The coupled dynamics, assuming the fluid relaxes faster than the solid network are:
%
%
%
%
%
%
%
\begin{align}
0 
&=
\bar\Gamma \big(\dot{u}_i-\bar{v}_i\big) + \nabla_j\big(
\bar\eta_{ijkl}\nabla_l \bar{v}_k
\big)
- \nabla_i P
\label{appeq:solvent dynamics}
\, ,
\\
\rho \ddot{u}_i 
&=
-\bar\Gamma\big(\dot{u}_i-\bar{v}_i\big)
+
\nabla_j 
\left(\sigma^\text{pre}_{ij}
+C_{ijkl}\nabla_l u_k\right)
- \Gamma \dot{u}_i 
\, .
\label{appeq:odd solid dynamics}
\end{align}
%
Equation~\eqref{appeq:solvent dynamics} is the force-balance (Stokes equation) for the solvent, {while $\bar\Gamma$ is the viscous coupling constant, arising from friction between the solid network and the fluid particles~\cite{gittes1997,schnurr1997,levine2001,levine2009}}, $P$ is the (thermodynamic) pressure, and $\bar\eta_{ijkl}$ is the solvent viscosity tensor, which may contain odd viscosity and odd pressure (vorticity-pressure coupling)~\cite{markovich2024}. Equation~\eqref{appeq:odd solid dynamics} is the linearized dynamics of the solid, which includes the elastic stress $\sigma^\text{pre}_{ij}
+C_{ijkl}\nabla_l u_k
$ and friction $-\Gamma\dot{u}_i$ with a substrate. The prestress $\bm\sigma^\text{pre}$ vanishes in the absence of active torques, whereas $\nabla\cdot\bm\sigma^\text{pre}=0$ when the  active torque density $\tau^\circ$ is constant, as in the main text (see Eq.~\eqref{appeq:Cauchy stress Lagra tau}). 

For strong coupling $\bar\Gamma$, the velocity difference $(\dot{\bm{u}}-\bar{\bm{v}})$ is expected to be small, namely $\bar{\bm{v}} = \dot{\bm{u}} + \bm\triangle$ with $\dot{\bm{u}} \gg \bm{\triangle}$, hence we can approximate $\nabla_j \bar{v}_i = \nabla_j (\dot{u}_i + \triangle_i) \approx \nabla_j \dot{u}_i$, neglecting terms of order ${\cal O}\left(\nabla_j \triangle_i\right)$. 
By eliminating $\bar\Gamma(\dot{u}_i - \bar{v}_i)$ of Eq.~\eqref{appeq:odd solid dynamics} using  Eq.~\eqref{appeq:solvent dynamics}, we get: 
%
%
%
%
%
%
%
%
\begin{align} \label{appeq:odd solid dynamics_step_2}
\rho \ddot{u}_i 
= 
&
- \Gamma \dot{u}_i
+\nabla_j 
\left(
\sigma^\text{pre}_{ij}
+
C_{ijkl}\nabla_l u_k
+
\bar\eta_{ijkl}
\nabla_l \dot{u}_k
\right)
-\nabla_i P
\, .
\end{align}
To further eliminate $\nabla_j P$, we note that $P$ is generally a function of $\bar\rho$, which can be written as $\bar\rho=\bar\rho_0+\delta\bar\rho$ where $\bar\rho_0$ is uniform and  $|\delta\bar\rho|\ll 1$. The pressure is then  $P=P_0\left(\bar\rho_0\right) + c^2 \delta\bar\rho$ with $c^2\equiv\partial P /\partial\bar\rho$ at $\bar\rho=\bar\rho_0$, and $c$ is the speed of sound. 
This approximation then yields $\nabla_i \dot{P} = - \nabla_i(c^2\delta\dot{\bar\rho}) = -\nabla_i\nabla_j(c^2\bar\rho_0 \bar{v}_j)$. Here we use the linearized continuity equation $\delta\dot{\bar\rho}=-\nabla_j(\bar\rho_0 \bar{v}^{c}_j)$ and the equality $\nabla_j(\bar\rho_0 \bar{v}_j) = \nabla_j(\bar\rho_0 \bar{v}^c_j)$ with $\bar{v}_i^c$ being the solvent CM velocity (for the later we utilize the definition of the total momentum $\bar\rho \bar{v}_i = \bar\rho \bar{v}^c_i+ (\nabla\times\bar\ell/2)_i$). 
Since $\nabla_j \bar{v}_i = \nabla_j \bar{v}^c_i \approx \nabla_j \dot{u}_i$, integrating over time gives $\nabla_i P \approx -\nabla_i \nabla_j(c^2\bar\rho_0 u_j)$, which only modifies the bulk modulus of the solid. 
%
Equation~\eqref{appeq:odd solid dynamics_step_2} becomes:
\begin{align} \label{appeq:two fluid model final eq}
\rho \ddot{u}_i 
= 
&
- \Gamma \dot{u}_i
+\nabla_j 
\big[
\sigma^\text{pre}_{ij}
+
\tilde{C}_{ijkl}
\nabla_l u_k
+
\bar\eta_{ijkl}
\nabla_l \dot{u}_k
\big]
\, ,
\end{align}
where the bulk modulus of $\tilde{\bm C}$ is $\tilde B = B + c^2\bar\rho_0$ while the rest of the elastic moduli are unchanged.

Equation~\eqref{appeq:two fluid model final eq} is a general result, displaying the effects of the solvent fluid interaction with the solid in the limit of strong solid-fluid coupling (large $\bar\Gamma$).
We thus have Kelvin-Voigt viscoelastic solid, where the response is fluid-like at short times and elastic at long times. 

{To verify the validity of the strong solid-fluid coupling assumption, we follow previous works and estimate 
$\bar\Gamma$~\cite{levine2001,levine2009}. 
A solid particle of size $\sim\zeta$ experiences Stokes drag force~\cite{LLfluidMechanics} of $\sim\bar\eta \zeta{\bf v}$ where $\bar\eta$ is the fluid shear viscosity. The drag force density is then $\sim\bar\eta \zeta{\bf v} n$ with $n\sim 1/\zeta^{3}$ being the number density of the solid. 
In the two-fluid model the drag force density is $\bar\Gamma{\bf v}$, and the solid network mesh size $\zeta_{\mathrm{mesh}}$ plays the role of a particle. The coupling strength is then estimated as $\bar\Gamma \sim \bar\eta/\zeta_{\mathrm{mesh}}^2$.
Comparing the drag and viscous terms in Eq.~\eqref{appeq:solvent dynamics} we find that drag dominates when $\bar\Gamma \gg \bar\eta k^2$, where $k$ is the wavenumber. Namely, we have strong coupling whenever $k \zeta_{\mathrm{mesh}} \ll 1$, which is essentially the hydrodynamic limit of long wavelengths.
We have assumed inertia is negligible for the fluid, but in principle one should also compare the drag and inertial terms. Following similar logic this leads to the condition $\bar\Gamma \gg \bar\rho\omega$, where $\bar\rho$ is the fluid density and $\omega$ the wave frequency. Then, we have strong solid-fluid coupling when $\bar\rho\omega\zeta_{\mathrm{mesh}}^2/\bar\eta \ll 1$, which is valid at low frequencies (the hydrodynamic limit) or large viscosity (negligible inertia), as expected.
In order to assess these conditions numerically we take $\bar\eta \simeq 10^{-3}\,\mathrm{Pa\cdot s}$ and $\bar\rho\simeq 1000 \, {\mathrm{kg}/\mathrm{m}^3}$ for water, and use a mesh size of $\zeta_\text{mesh}\simeq 0.1 \,\mu\mathrm{m}$ applicable for e.g., actin networks~\cite{schnurr1997,shin2004,luan2008,kasza2010,dwyer2022}.
This gives $\bar\Gamma \sim 10^{11} \, \mathrm{Pa\cdot s/m^2}$. 
The strong coupling assumption is then valid for wavelengths $\lambda \gg 0.1  \,\mu\mathrm{m} $ and frequencies $\omega \ll 100 \, \mathrm{MHz}$.
}

For a general solid-fluid coupling more thorough analysis of the dynamic spectrum is required~\cite{levine2001,levine2009}. This will not be further discussed in this work.



\subsection{Dynamic matrix and secular equation of odd viscoelasticity}
\label{app:dynamic matrix}

In this subsection we find the normal modes of the Kelvin-Voigt odd viscoelastic solid.
We begin with writing the elasticity  and viscosity  tensors ($\bm{C}$ and $\bm\eta$, respectively) of Eq.~\eqref{appeq:two fluid model final eq} in a more general form than what is used in the main text. This general form further allows for breaking of rotational invariance (the term $\sim\Lambda$ below)~\cite{fruchart2023}. We still assume the elastic moduli and viscosities are uniform time-independent, such that $\nabla\cdot\bm\sigma^\text{pre} = 0$.
Later, we will reduce the model to the one used in the main text.
%
%
%
We write
\begin{eqnarray}
\nonumber& &C_{ijkl}=B\delta_{ij}\delta_{kl}+\mu(\delta_{ik}\delta_{jl}+\delta_{il}\delta_{jk}-\delta_{ij}\delta_{kl})+K^o(\varepsilon_{jl}\delta_{ik}+\varepsilon_{ik}\delta_{jl})-A\varepsilon_{ij}\delta_{kl}-\Lambda\varepsilon_{kl}\delta_{ij} \, , \\
& &\eta_{ijkl}=\eta(\delta_{ik}\delta_{jl}+\delta_{il}\delta_{jk})+\eta^o(\varepsilon_{jl}\delta_{ik}+\varepsilon_{ik}\delta_{jl})-\eta^B\varepsilon_{kl}\delta_{ij} \, ,
\end{eqnarray}
where, without losing generality, the bulk modulus $B$ already contains the factor $c^2\bar\rho_0$ from the fluid (see after Eq.~\eqref{appeq:two fluid model final eq}), $\mu$ is the shear modulus, $K^o$ is the odd elastic modulus, $A$  couples torque with compression, and $\Lambda$ couples pressure with rotation.
In the viscosity tensor, $\eta$ is the shear viscosity, $\eta^o$ the odd viscosity, and $\eta^B$ the odd pressure that couples vorticity and pressure~\cite{markovich2024,han2021}.
%
%

Employing the Fourier transform $u_i =\bar{u}_i \exp[i(\bm{k}\cdot\bm{R}-\omega t)]$ in Eq.~\eqref{appeq:two fluid model final eq} and decomposing the amplitudes $\bar{u}_i$ into  longitudinal and transverse directions with $\bar{u}_\mathrm{L}\equiv k_i\bar{u}_i/k$ and $\bar{u}_\mathrm{T}\equiv\varepsilon_{ij}k_i\bar{u}_j/k$ ($k\equiv |\bm{k}|$) we obtain the dynamical matrix: 
\begin{equation}\label{appeq:general dynamic matrix}
\begin{pmatrix}
\rho\omega^2+i\Gamma\omega & 0
\\
0 & \rho\omega^2+i\Gamma\omega
\end{pmatrix}
\begin{pmatrix}
\bar{u}_\mathrm{L}
\\    
\bar{u}_\mathrm{T}
\end{pmatrix}
 =
 k^2
 \begin{pmatrix}
 B+\mu-i\omega\eta &
 K^o+\Lambda-i \omega (\eta^o + \eta^B)   
 \\
- K^o+A+i\omega \eta^o 
 &
 \mu-i\omega\eta
 \\
 \end{pmatrix}
\begin{pmatrix}
\bar{u}_\mathrm{L}
\\    
\bar{u}_\mathrm{T}
\end{pmatrix}
\, ,
\end{equation} 
with the amplitudes of the eigenmodes given by:
\begin{align}\label{appeq:general eigenmodes}
\{\bar{u}_\text{L},\bar{u}_\text{T} \}    
=
\mathcal{N}
\{
K^o+\Lambda-i \omega (\eta^o + \eta^B)  
\, ,
B+\mu - i\omega(\eta+\Gamma) -\rho \omega^2
\} 
\, ,
\end{align}
where $\mathcal{N}$ is the normalization factor such that $|\{\bar{u}_L,\bar{u}_T\}|=1$.
For nontrivial solutions of $\bar{u}_\mathrm{L}$ and $\bar{u}_\mathrm{T}$ in Eq.~\eqref{appeq:general dynamic matrix}, $\omega$ and $\bm{k}$ satisfy the secular equation for the dispersion relation:
\begin{align}\label{appeq:secular gen}
k^4 
\big\{\mu (B+\mu) 
-(A-K^o)&(K^o+\Lambda)\big]
-i k^2\omega
\big\{
(B+2 \mu)(\Gamma+k^2\eta)
- k^2
\big[
\eta^B(A-K^o)+\eta^o(A-2K^o-\Lambda)
\big]
\big\}  
\notag\\
&
-\omega^2
\big[
(\Gamma+k^2\eta)^2
+k^2\rho(B+2\mu)
+k^4\eta^o(\eta^o+\eta^B)
\big]
+2i\omega^3 \rho(\Gamma+k^2\eta)
+\rho ^2 \omega^4 =0
\, .   
\end{align}
In general, with the presence of odd moduli ($K^o$, $A$, and $\Lambda$) 
the dynamic matrix of Eq.~\eqref{appeq:general dynamic matrix} becomes asymmetric, indicating a nonreciprocal coupling between longitudinal and transverse modes, which is a signature of odd dynamics~\cite{markovich2021,markovich2024,scheibner2020,fruchart2023,nemeth2026}. (To be precise, when the dynamical matrix is non-Hermitian nonreciprocity emerges~\cite{markovich2021,markovich2024}).

In the simple example of the main text 
%
($A=2K^o \propto\tau^\circ$ and $\Lambda=0$), 
the elastic part of Eq.~\eqref{appeq:general dynamic matrix} becomes symmetric (with the off-diagonal terms being $K^o$). However, the matrix in Eq.~\eqref{appeq:general dynamic matrix} still remains non-Hermitian due to $\eta^B$, which is a result of the coupling with the odd chiral fluid. It is therefore expected that $\bar{u}_\text{L}$ and $\bar{u}_\text{T}$ in Eq.~\eqref{appeq:general eigenmodes} will be coupled. 
To write Eq.~\eqref{appeq:secular gen} in dimensionless form we define $\tilde{\omega}\equiv w\sqrt{\rho}/k\sqrt{\mu }$, $\{\tilde{B},\tilde{A},\tilde{\lambda},\tilde{K^o}\}\equiv\{B,A,\Lambda,K^o\}/\mu$, $\{\tilde\eta^o,\tilde\eta^B\}\equiv\{\eta^o,\eta^B\}k/\sqrt{\mu\rho}$, and $\tilde\eta\equiv[k\eta+(\Gamma/k)]/\sqrt{\mu\rho}$ and take $\tilde{A}=2\tilde{K}^o$ and $\tilde\Lambda=0$, yielding:
\begin{align}
\big(1+\tilde{B}-\tilde{K}^{o^2}
\big)
-i\tilde\omega
&\big[
(2+\tilde{B})\tilde\eta
-\tilde\eta^B\tilde{K}^o
\big]
-\tilde\omega^2
\big[
2+\tilde{B}+\tilde\eta^2
+\tilde\eta^o(\tilde\eta^o+\tilde\eta^B)
\big]
+2 i\tilde\eta \tilde\omega^3
+\tilde\omega^4
=0
\, .
\label{appeq:secular uniform torque dimensionless}
\end{align}

Solving Eq.~\eqref{appeq:secular uniform torque dimensionless} gives the eigenmodes of our viscoelastic solid. When one of these modes acquires a positive imaginary part the system becomes dynamically unstable. Therefore, Eq.~\eqref{appeq:secular uniform torque dimensionless} is central for exploring dynamic instabilities. 
%
%
Although Eq.~\eqref{appeq:secular uniform torque dimensionless} does not guarantee a simple analytical form for $\tilde\omega$-solutions, we provide a generic scheme 
to obtain analytical expressions for where instability occurs in Sec.~\ref{app:dynamic instability}. 
This scheme can also be extended to find  instabilities in more general settings such as the one in Eq.~\eqref{appeq:secular gen}.

Section~\ref{app:dynamic instability} will also show that the presence of $\tilde\eta^o$ and $\tilde\eta^B$ provides with new and diverse instability behavior, in particular, due to the odd solid-fluid coupling $\tilde\eta^B \tilde{K}^o$. This is extending the analysis done in Ref.~\cite{fruchart2023}, which included only the effects of $\tilde{K}^o$ and even viscosities.
%
%
Note that generally speaking $\tilde\eta^o \neq \tilde\eta^B$~\cite{han2021}. In our simple example of the main text, we consider a non-interacting odd active fluid such that $\tilde\eta^o=-\tilde\eta^B = -\bar\ell/2$. 
In this particular case, the effect that is unique to $\tilde\eta^o$ in Eq.~\eqref{appeq:secular uniform torque dimensionless} is canceled off, and only the coupling
$\tilde\eta^B\tilde{K}^o$ (or equivalently, $-\tilde\eta^o\tilde{K}^o$ in this case) governs the dynamic instabilities in Fig.~2.
%
We thus expect that even in the case where only $\tilde\eta^B$ is present (and $\tilde\eta^o=0$) the same instability will occur.

\section{Dynamic instability of odd viscoelasticity in the underdamped case}
\label{app:dynamic instability}

This section provides a generic method to find stability boundaries of a quartic eigenvalue equation. Beyond these boundaries dynamic instability occurs where at least one mode is growing, i.e., the imaginary part of $\tilde\omega$ is positive, denoted as $\mathrm{Im}[\tilde\omega]>0$.
On the instability boundary $\mathrm{Im}[\tilde\omega]=0$.
At the onset of instability at least one of the modes become unstable $\mathrm{Im}[\tilde\omega]=0$, while others remain stable $\mathrm{Im}[\tilde\omega]<0$.

Specifically, we use this generic method to derive the instability boundaries for Eq.~\eqref{appeq:secular uniform torque dimensionless}.
Instead of directly solving $\tilde\omega$ of Eq.~\eqref{appeq:secular uniform torque dimensionless}, which gives a complicated analytical form that is hard to analyze, we use an alternative scheme that allows simpler analytical expressions. 
First notice that the solutions of a quartic equation can be written as $(\tilde\omega-s_1)(\tilde\omega-s_2)(\tilde\omega-s_3)(\tilde\omega-s_4)=0$, where $\{s_i\}$ are complex numbers. Employing this on Eq.~\eqref{appeq:secular uniform torque dimensionless} imposes the conditions:
%
%
%
%
\begin{align}
s_1+s_2+s_3+s_4 
&= 
-2i\tilde\eta 
\, ,
\label{appeq:q linear}
\\
s_1 s_2+s_1 s_3+s_1 s_4+s_2 s_3+s_2 s_4+s_3 s_4 
& = 
- \big[2+\tilde{B}+\tilde{\eta}^2+\tilde\eta^o(\tilde\eta^o+\tilde\eta^B)\big] 
\, ,
\\
s_1 s_2 s_3 + s_1 s_2 s_4 +  s_1 s_3 s_4 + s_2 s_3 s_4
&=
[ (2+\tilde{B})\tilde\eta-\tilde{K}^o\tilde\eta^B ] i
\, ,\\
s_1 s_2 s_3 s_4 
&= 
1 +\tilde{B} -\tilde{K}^{o^2}
\, . \label{appeq:q const}
\end{align}
The solutions $\{s_i\}$ must also attain one of the following three patterns: (i) $\{s_1,s_2,s_3,s_4\} =\{a i,b i,c i,d i\}$, (ii) $\{s_1,s_2,s_3,s_4\} =\{a - b i,-a -b i, -c i, -d i\}$, and (iii) $\{s_1,s_2,s_3,s_4\} =\{a - b i,-a -b i,c- d i,-c -d i\}$, where $a$, $b$, $c$ and $d$ are real numbers. These patterns are derived by transforming Eq.~\eqref{appeq:secular uniform torque dimensionless} into a real equation via $z \equiv i \tilde{\omega}$ and using the complex conjugate root theorem.


Our goal is to find the condition that makes one of the solutions $\mathrm{Im}[s_i]=0$ obey Eqs.~\eqref{appeq:q linear} to \eqref{appeq:q const} with $a$, $b$, $c$, and $d$ being real numbers. 
%


\subsection{Instability in different \texorpdfstring{$\tilde{K}^o$}{odd}-regions}\label{app:instability lines}

To summarize our result, we find that in the region $1+\tilde{B}-\tilde{K}^{o^2}\geq 0$ the instability boundaries are $1+\tilde{B}-\tilde{K}^{o^2}=0$ and Eq.~\eqref{appeq:BD line express}, under the condition that $Q \equiv 2+\tilde{B}+\tilde{\eta}^2+\tilde\eta^o(\tilde\eta^o+\tilde\eta^B)>0$. 
In the region $1+\tilde{B}-\tilde{K}^{o^2}<0$ there are only unstable solutions. These results are numerically confirmed in Fig.~\ref{appfig:numInstability}.




\subsubsection{The case of \quad \texorpdfstring{$s_1 s_2 s_3 s_4=1 +\tilde{B} -\tilde{K}^{o^2}>0$}{positive cons}}\label{app:core bd region}

In this case pattern (i) $\{s_1,s_2,s_3,s_4\} =\{a i,b i,c i,d i\}$ is ruled out because it contradicts the premise $s_1 s_2 s_3 s_4> 0$ at the instability boundaries, where $\mathrm{Im}[s_i]=0$ implies $s_1 s_2 s_3 s_4=0$. 
For pattern (ii) $\{s_1,s_2,s_3,s_4\} =\{a - b i, -a -b i, -c i, -d i\}$, $\mathrm{Im}[s_i]=0$ imposes $b=0$ only, because the other two possibilities $c=0$ or $d=0$ contradict the premise $s_1 s_2 s_3 s_4 >0$. 
Pattern (iii) will later be shown to share the same properties as pattern (ii) after re-parametrization.  

Focusing on pattern (ii), our goal is to find the conditions to solve $a\in \mathrm{Real}$, $c>0$ and $d>0$, where the last two conditions are the requirement of stable solutions at the boundaries. To this end, we insert $\{s_1,s_2,s_3,s_4\} =\{a, -a, -c i, -d i\}$ into Eqs.~\eqref{appeq:q linear}-\eqref{appeq:q const}:
\begin{align}
a^2 cd 
&= 
1 +\tilde{B} -\tilde{K}^{o^2}
\, ,
\label{appeq:a2cd}\\
c + d 
&= 
2 \tilde{\eta}
\, , 
\label{appeq:c+d}\\
a^2+cd 
&= 2+\tilde{B}+\tilde{\eta}^2+\tilde\eta^o(\tilde\eta^o+\tilde\eta^B)
\, , 
\label{appeq:a2+cd}\\
a^2(c+d) 
&= 
(2+\tilde{B})\tilde\eta-\tilde{K}^o\tilde\eta^B 
\label{appeq:a2(c+d)}
\, .
\end{align}
The product $cd$ can be solved from Eqs.~\eqref{appeq:a2cd} and \eqref{appeq:a2+cd}: 
\begin{align}
cd 
&= \frac{1}{2}\bigg(
Q
\pm
\sqrt{Q^2-4(1+\tilde{B}-\tilde{K}^{o^2})}
\bigg)
\, ,
\label{appeq:cd solution}
\\
\mathrm{with\ } 
Q 
&\equiv
2+\tilde{B}+\tilde{\eta}^2+\tilde\eta^o(\tilde\eta^o+\tilde\eta^B)
\, .
\label{appeq:Q express}
\end{align}
Next, $a^2$ can be found from either dividing Eq.\eqref{appeq:a2(c+d)} by Eq.~\eqref{appeq:c+d} or by dividing Eq.~\eqref{appeq:a2cd} by Eq.~\eqref{appeq:cd solution}. This gives a constraint on the parameters:
\begin{align}
\frac{1+\tilde{B}-\tilde{K}^{o^2}}{\frac{1}{2}\bigg(
Q\pm\sqrt{Q^2-4(1+\tilde{B}-\tilde{K}^{o^2})}
\bigg)} 
= \frac{(2+\tilde{B})\tilde\eta-\tilde{K}^o\tilde\eta^B}{2\tilde\eta}   
\label{appeq:BD line express}.
\end{align}

Equation~\eqref{appeq:BD line express} is the expression of the instability boundaries in the region $1+\tilde{B}-\tilde{K}^{o^2}>0$. To apply Eq.~\eqref{appeq:BD line express}, it is also required that $Q^2\geq 4(1+\tilde{B}-\tilde{K}^{o^2})$ and $Q>0$ such that $cd >0$. 
The constraint $Q>0$ cuts off some of the branches when plotting Eq.~\eqref{appeq:BD line express} in the real plane.


At last, for pattern (iii) $\{s_1,s_2,s_3,s_4\} = \{a - b i,-a -b i,c- d i,-c -d i\}$, we take $b=0$ and insert $\{s_1,s_2,s_3,s_4\} = \{a,-a, c- d i,-c -d i\}$ into Eqs.~\eqref{appeq:q linear}-\eqref{appeq:q const}, which gives:
\begin{align}
a^2(c^2+d^2)
&=
1 +\tilde{B} -\tilde{K}^{o^2},
\\
d
&=\tilde{\eta},
\\
a^2+(c^2+d^2)
&=2+\tilde{B}+\tilde{\eta}^2+\tilde\eta^o(\tilde\eta^o+\tilde\eta^B),
\\
2 a^2 d 
&= (2+\tilde{B})\tilde\eta-\tilde{K}^o\tilde\eta^B \, .
\end{align}
These equations are, in fact, the same as Eqs.~\eqref{appeq:a2cd}-\eqref{appeq:a2(c+d)} via re-parametrization: $(c^2+d^2) \rightarrow c d$ and $d \rightarrow (c+d)/2$. Hence, Eq.~\eqref{appeq:BD line express} derived from pattern (ii) provides with all the possible instability boundaries.

\subsubsection{The case of \quad \texorpdfstring{$s_1 s_2 s_3 s_4=1 +\tilde{B} -\tilde{K}^{o^2}\leq 0$}{negative cons}}\label{app:ko negative region}

\begin{figure}[t]
	\centering
	\includegraphics[width=13 cm]{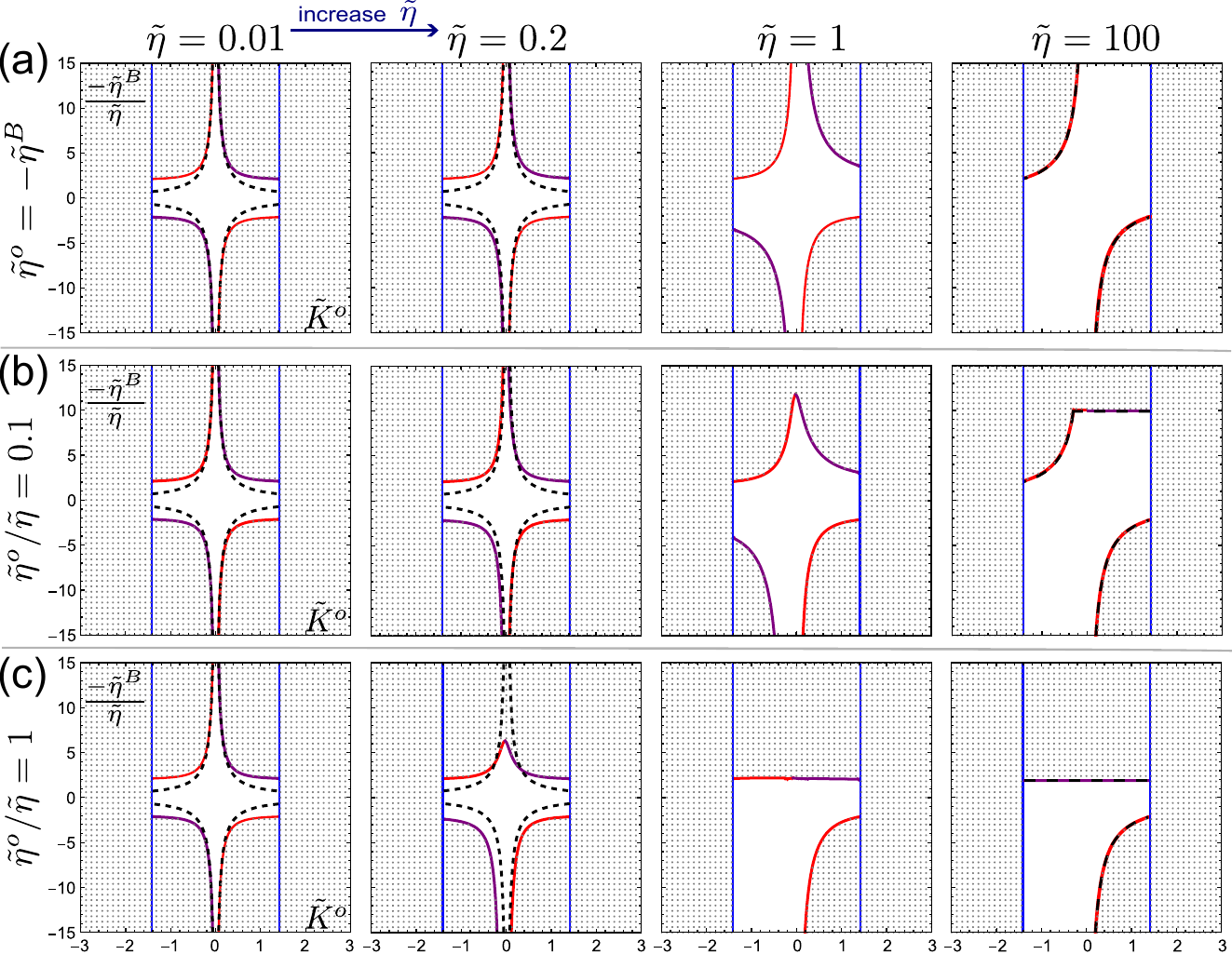}
	\caption{Numerical confirmation of the instability boundaries in the $ \tilde{K}^o$ -- $(\tilde\eta^B/\tilde\eta)$ space with increasing viscosity $\tilde\eta$ in each row for various combinations of $\tilde\eta^B$ and $\tilde\eta^o$, (a) to (c), using the predictions: $1+\tilde{B}-\tilde{K}^{o^2}=0$ (blue), Eq.~\eqref{appeq:BD line express} (purple and red) and their asymptotic expressions (black dashed, Eq.~\eqref{appeq:ko express at small viscosity} for the small-$\tilde\eta$ limit and Eqs.~\eqref{appeq:overdamped bd} and \eqref{appeq:overdamped bd disappear} for the large-$\tilde\eta$ (approximately overdamped) limit. Row (a) is the same case as Fig.~2(b)
    in the main text. Gray dots indicate unstable regions, deduced from numerical solution of Eq.~\eqref{appeq:secular uniform torque dimensionless}. All the plots use the same range in axes. } \label{appfig:numInstability}
\end{figure}

In the region $1 +\tilde{B} -\tilde{K}^{o^2}<0$, Eqs.~\eqref{appeq:a2cd}-\eqref{appeq:Q express} still apply. However, Eq.~\eqref{appeq:a2cd} indicates that $cd<0$, which contradicts the requirement that $c\geq0$ and $d\geq0$, which is necessary in order to have all of the solutions stable at the instability boundary, $\mathrm{Im}[s_i\leq 0$. Therefore, there are no instability boundaries in this region. 
On the other hand, instability may occur when $1 +\tilde{B} -\tilde{K}^{o^2}=0$. This guarantees $\mathrm{Im}[s_i]=0$ (see Eq.~\eqref{appeq:q const}).
%
%
Indeed, Eq.~\eqref{appeq:BD line express}, together with the lines $1 +\tilde{B} -\tilde{K}^{o^2}=0$, encloses the stability region, as also numerically verified in Fig.~\ref{appfig:numInstability}. 

Note that for arbitrary $\tilde\eta^o$, the boundaries given by Eq.~\eqref{appeq:BD line express} can intersect to form a continuous curve. This happens when $Q^2 - 4(1+\tilde{B} -\tilde{K}^{o^2}) = 0$. 
This occurs in an asymmetric way  because a non-zero $\tilde\eta^o(\tilde\eta^o+\tilde\eta^B)$ in $Q$ breaks inversion symmetry  $\tilde\eta^B \leftrightarrow -\tilde\eta^B$, and hence also the symmetry $(\tilde{K}^o,\tilde\eta^B)\leftrightarrow(-\tilde{K}^o,-\tilde\eta^B)$. In the case considered in the main text where $\tilde\eta^o=-\tilde\eta^B$ or when $\tilde\eta^o=0$ such symmetry is restored. 

\subsection{Limiting expressions for instability boundaries} \label{app:limiting expression for instability}

Let us derive the asymptotic expressions for the instability boundaries of Eq.~\eqref{appeq:BD line express} in the region $1+\tilde{B}-\tilde{K}^{o^2} > 0$. 

We start with the small $\tilde\eta$ limit, where it is expected that the instability boundaries to get closer to the origin $(\tilde{K}^o,\tilde\eta^B) = (0,0)$. We further assume large $\tilde{B}$ such that $1+\tilde{B} - \tilde{K}^{o^2} \approx 1+\tilde{B}$ in Eq.~\eqref{appeq:BD line express}, which gives
\begin{align}
\tilde{K}^o \frac{\tilde\eta^B}{\tilde\eta}
&=
\tilde{B}
\, ,
\label{appeq:ko express at small viscosity} 
\\ 
\tilde{K}^o \frac{\tilde\eta^B}{\tilde\eta}
&=
-
\big[
\tilde{B}
+
2\tilde\eta^2
+2\tilde\eta^o(\tilde\eta^o+\tilde\eta^B)
\big]
\label{appeq:ko express at small viscosity2}    
\, .
\end{align}
%
%

In the large $\tilde\eta$-limit we have $Q \gg 1$ (see Eq.~\eqref{appeq:Q express}),
%
%
hence, Eq.~\eqref{appeq:BD line express} becomes 
\begin{align}
\tilde{K}^o\frac{\tilde\eta^B}{\tilde\eta}
&=
(2+\tilde{B})
\, ,
\label{appeq:overdamped bd}
\\
\tilde{K}^o\frac{\tilde\eta^B}{\tilde\eta}
&=
-2
\big[
\tilde\eta^2
+\tilde\eta^o(\tilde\eta^o+\tilde\eta^B)
\big]
\label{appeq:overdamped bd disappear}
\, ,
\end{align}
where we have used $Q+[{Q^2-4(1+\tilde{B}-\tilde{K}^{o^2})}]^{1/2}   
\approx 2 Q$ and $Q-[{Q^2-4(1+\tilde{B}-\tilde{K}^{o^2})}]^{1/2}\approx 2(1+\tilde{B}-\tilde{K}^{o^2})/Q$. 
%
%
Note that, despite taking the large $\tilde\eta$ limit, Eq.~\eqref{appeq:overdamped bd disappear}  still accounts for inertial effects. 

To derive the asymptotic expressions in the overdamped limit, we take $\rho\rightarrow 0$ and thus $\{\tilde\eta,\tilde\eta^o,\tilde\eta^B\}\propto \rho^{-1/2}\rightarrow\infty$. 
Equation~\eqref{appeq:overdamped bd} remains unchanged, but Eq.~\eqref{appeq:overdamped bd disappear} becomes independent of the odd coupling $\tilde{K}^o\tilde\eta^B$:
\begin{align}
1+\frac{\tilde\eta^o}{\tilde\eta^2}
\bigg(
\tilde\eta^o+\tilde\eta^B
\bigg)
=0
\label{appeq:overdamped bd disappear true}
\, .
\end{align}
Equations~\eqref{appeq:overdamped bd} and \eqref{appeq:overdamped bd disappear true}, together with $1+\tilde{B}-\tilde{K}^{o^2}=0$, are the instability boundaries in the overdamped limit. We derive these boundaries directly in Sec.~\ref{app:overdamped} below from the secular equation in the overdamped. 
Note that Eq.~\eqref{appeq:overdamped bd disappear true} is equivalent to $Q=0$, which contradicts the requirement $Q>0$ we discussed for the underdamped case. This is allowed because in the overdamped case the secular equation of Eq.~\eqref{appeq:secular gen} becomes quadratic, hence, the solutions $-ci$ and $-di$ disappear, and so does the requirement $Q>0$ in discussion below Eq.~\eqref{appeq:BD line express}). 
Besides, the reduction of Eq.~\eqref{appeq:secular gen} to a quadratic form makes Eqs.~\eqref{appeq:overdamped bd} and~\eqref{appeq:overdamped bd disappear true} also applicable in the region $1+\tilde{B}-\tilde{K}^{o^2}<0$, see arguments in Sec.~\ref{app:ko negative region} and Sec.~\ref{app:overdamped}.


In the case $\tilde\eta^o=-\tilde\eta^B$ that is considered in the main text, the instability boundaries in the small $\tilde\eta$ limit become $\tilde{K}^o\tilde\eta^o/\tilde\eta=\pm\tilde{B}$ (see Eqs.~\eqref{appeq:ko express at small viscosity}-\eqref{appeq:ko express at small viscosity2}), which only differ by their sign. 
As $\tilde\eta$ increases, Eq.~\eqref{appeq:overdamped bd disappear} requires one of the instability boundaries to scale as $\tilde{K}^o\tilde\eta^o/\tilde\eta \sim -2\tilde\eta^2$, while the other boundary is relatively insensitive to $\tilde\eta$. 
In the overdamped limit, there are only two instability boundaries given by Eq.~\eqref{appeq:overdamped bd} and $1+\tilde{B}-\tilde{K}^{o^2}=0$, see Fig.~\ref{appfig:areaRatio}(a), because Eq.~\eqref{appeq:overdamped bd disappear true} can never be satisfied. In other words, Eq.~\eqref{appeq:overdamped bd disappear} is a purely inertial and is invalid in the overdamped limit.

\section{Dynamics in the overdamped limit}\label{app:overdamped}

In the previous section we analyzed the dynamic instabilities in the underdamped case and took $\rho \rightarrow 0$ to obtain the overdamped limit. 
Here, we take the overdamped limit directly in the dynamical matrix. We first reproduce the instability lines found in Sec.~\ref{app:limiting expression for instability} and then examine the nature of wave propagation. 
To this end, we take $\rho=\Lambda=0$ and $A= 2 K^o$  in Eq.~\eqref{appeq:general dynamic matrix} for our odd solid and obtain the overdamped dynamical matrix:
\begin{equation}\label{appeq:dimensionless overdamped dynamic matrix}
\begin{pmatrix}
0
\\    
0
\end{pmatrix}
 =
 \begin{pmatrix}
 1+\tilde{B}-i\omega' &
 \tilde{K}^o-i ( \tilde{\eta}^o 
 +\tilde\eta^B)   \omega'
 \\
\tilde{K}^o+i\tilde\eta^o\omega' 
 &
 1-i\omega'
 \\
 \end{pmatrix}
\begin{pmatrix}
\bar{u}_\mathrm{L}
\\    
\bar{u}_\mathrm{T}
\end{pmatrix}
\, ,
\end{equation} 
where we define the dimensionless quantities: $\omega'=\omega (\eta+\Gamma/k^2) / \mu$, $\{\tilde{B},\tilde{K}^o\}=\{B,K^o\}/\mu$, and $\{\tilde\eta^o,\tilde\eta^B \}/\tilde\eta=\{ \eta^o, \eta^B
 \}/(\eta+\Gamma/k^2)$.
Equation~\eqref{appeq:dimensionless overdamped dynamic matrix} shows that the amplitudes of the eigenmodes couple both the longitudinal and transverse waves  due to the presence of odd elasticity and odd viscosity~\cite{scheibner2020, fruchart2021,surowka2023,markovich2024} where
\begin{align}\label{appeq:overdamped eigenmodes}
\{ \bar{u}_\mathrm{L},\bar{u}_\mathrm{T} \} = \{\tilde{K}^o-i \omega'( \tilde{\eta}^o 
 +\tilde\eta^B) ,  1+\tilde{B}-i\omega' \}
 \, ,
\end{align} 
and $\omega'$ satisfying the secular equation:
\begin{align}\label{appeq:reduced secular overdamped}
-\big(
1+\tilde{B}-\tilde{K}^{o^2}
\big)
+ i 
\big(
2+\tilde{B}
-\tilde{K}^o\frac{\tilde\eta^B}{\tilde\eta}
\big)\omega' 
+
\bigg[
1+\frac{\tilde{\eta}^{o}}{\tilde{\eta}^2}(\tilde{\eta}^o+\tilde{\eta}^B)
\bigg] {\omega'}^2 
=0  \, 
.
\end{align} 

\subsection{Dynamic instability}

For briefness, we denote the real coefficients in Eq.~\eqref{appeq:reduced secular overdamped} as $a_0\equiv 1 +\tilde{B} -\tilde{K}^{o^2}$, $a_1\equiv 2+\tilde{B}
-\tilde{K}^o (\tilde\eta^B/\tilde\eta)$ and $a_2\equiv 1+\tilde{\eta}^{o}(\tilde{\eta}^o+\tilde{\eta}^B)$, and accordingly $\omega' = [-i a_1 \pm (-a_1^2+4a_0 a_2)^{1/2}\ ]/ 2 a_2$. 
The conditions for unstable solutions (i.e., with a positive imaginary part in $\omega'$) can be  derived based on the sign of $a_0$, $a_1$ and $a_2$.

(i) If $a_0>0$, for $a_2>0$, we have $a_0 a_2 > 0$ and therefore $|a_1|>{|-a_1^2+4a_0 a_2|}^{1/2}$. Then $a_1 <0$ gives two unstable solutions, whereas $a_1 >0$ gives two stable solutions.
For $a_2<0$, $a_0 a_2 < 0$ and hence $|a_1|<{|-a_1^2+4a_0 a_2|}^{1/2}$. When $a_1 <0$ the solutions can be written as $-i(|a_1|\pm{|-a_1^2+4a_0 a_2|}^{1/2}\ )/ 2 |a_2|$, giving one stable and one unstable solutions, and similarly for $a_1 >0$.
%
%
Therefore, for $a_0>0$ only the region in which both $a_1>0$ and $a_2>0$ is stable. 

(ii) On the line $a_0=0$, for $a_1 a_2<0$ there is one unstable solution, whereas for $a_1 a_2>0$ there are two stable solutions. 

(iii) In the region where $a_0<0$, we find with similar analysis to (i) that the system is stable only when both $a_1<0$ and $a_2 <0$.

\begin{figure}[t]
	\centering
	\includegraphics[width=16.5 cm]{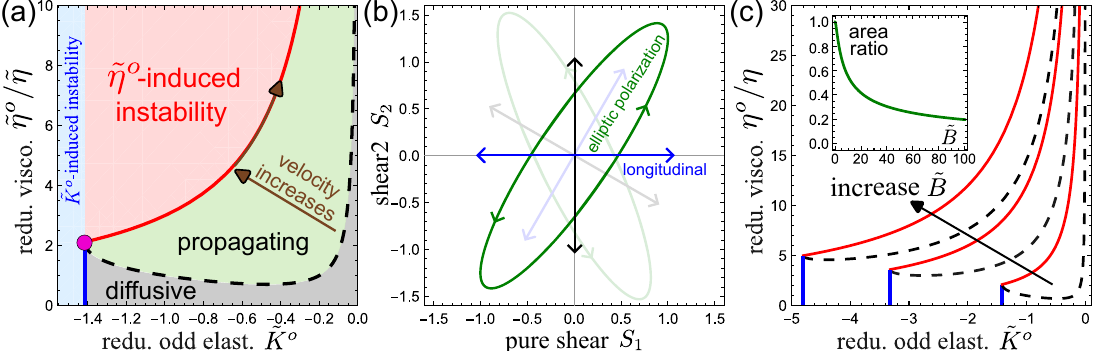}
	\caption{(a) Dynamic phase diagram for the overdamped case in the quadrant $\tilde\eta^o\tilde{K}^o<0$. The instability boundaries (blue and red lines) meet the propagation boundary (dashed) at the tripoint (purple circle). In the propagation region (green), the wave velocity increases along the arrow directions: either by increasing the coupling strength $|\tilde{K}^o\tilde\eta^o|$ or by moving away from the $\tilde{K}^o$-induced instability. The maximal velocity is achieved at the $\tilde\eta^o$-induced instability. (b) Elliptic polarization (tilted green curve) of propagating waves in the shear deformation space. In this space, longitudinal (blue horizontal) and transverse (black vertical) waves form orthogonal basis, which corresponds to pure shear ($\nabla_x u_x-\nabla_y u_y$) and shear2 $(\nabla_x u_y+\nabla_y u_x)$ when waves travel along the x-direction in real space. The shaded curves are an example for a different propagation direction $\{k_x,k_y\}=\{\cos(\pi/12),\sin(\pi/12)\}$, where the basis is rotated by $\pi/6$ radian. Arrows indicate temporal oscillation. Here the temporal damping pre-factor is discarded for clarity. (c) Increasing $\tilde{B}$ expands the stable region, i.e., diffusive (gray) + propagating (green) in (a). Nevertheless, the area ratio of the propagating region to the stable region decreases (inset).} 	\label{appfig:areaRatio}
\end{figure}

The discussion above implies that the instability boundaries of the overdamped case are $a_0=0$,  $a_1=0$ and $a_2=0$, where the last two conditions are the same as Eqs.~\eqref{appeq:overdamped bd} and \eqref{appeq:overdamped bd disappear true} when taking the overdamped limit. 
Note that the region $a_0<0$ can have two instability boundaries $a_1=0$ and $a_2=0$ (see case (iii) above), in contrast with the underdamped case (see Sec.~\ref{app:ko negative region}). This is also consistent with the previous discussion in Sec.~\ref{app:limiting expression for instability}. 
In our simple example of the main text where $\tilde\eta^B=-\tilde\eta^o$, there are only two instability boundaries given by $a_0=0$ (blue line, for $\tilde{K}^o$-induced instability) and $a_1=0$ (red line, for $\tilde\eta^o$-induced instability) as shown in Fig.~\ref{appfig:areaRatio}(a). This is because in this simplified case $a_2=1$ such that the predicted boundary $a_2=0$ can never be reached.   

\subsection{Wave propagation}

A remarkable consequence of odd elasticity is the possibility of wave propagation even for an  overdamped solid~\cite{scheibner2020} in stark contrast to overdamped `passive' solids that do not allow for wave propagation. The onset of wave propagation in these non-reciprocal elastic materials occurs beyond an exceptional line (a line in which two eigenmodes coalesce)~\cite{heiss2012}.

In our case the exceptional line is $-a_1^2+4a_0 a_2=0$. Beyond this line $\omega'$ acquires a real part such that waves propagate. In our viscoelastic solid waves are still attenuated as the imaginary part of $\omega'$ does not vanish, but the existence of the real part also results in phase-shift $\Phi$ between the eigenmode amplitudes (Eq.~\eqref{appeq:overdamped eigenmodes}) of longitudinal and transverse waves. 
%
%
The existence of an exceptional line for our odd viscoelastic solid is a result of the coupling between the odd solid and the odd fluid (via $\tilde\eta^o$ or $\tilde\eta^B$). This is in  contrast to other odd elastic lattice-based solids where $\tilde{K}^o$ alone enables propagation~\cite{scheibner2020}. 

To exemplify the effects of the odd solid-fluid coupling, we focus on the simple example of the main text where $\tilde\eta^B=-\tilde\eta^o$ (hence $a_2=1$) and the solutions of Eq.~\eqref{appeq:reduced secular overdamped} read 
\begin{align}
 \omega' 
 =
 \frac{1}{2}
 \left[
 - i \Bigg( 2+\tilde{B}+\frac{\tilde{K}^o\tilde{\eta}^o}{\tilde{\eta}} \Bigg)
 \pm
 \sqrt{
 4 \Big( 1+\tilde{B}-{\tilde{K}^o}^2  
 \Big)
 -\Big( 2+\tilde{B}+\frac{\tilde{K}^o\tilde{\eta}^o}{\tilde{\eta}} \Big)^2 
 }\
 \right] 
 \, . 
 \end{align}
Therein, the odd coupling $\tilde{K}^o\tilde\eta^o$ is a destabilizing factor that counteracts the stabilizing viscous/frictional damping. When strong enough it induces wave propagation (while the waves are still decaying), and it eventually completely destabilizes the homogeneous system.
The region in which waves are propagating is:
\begin{align}\label{appeq:wave prop condition}
2+\tilde{B}
>
-\tilde{K}^o\frac{\tilde\eta^o}{\tilde\eta}
>
2+\tilde{B}-2\sqrt{1+\tilde{B}-\tilde{K}^{o^2}}
\, .
\end{align}
Here $1+\tilde{B}-\tilde{K}^{o^2}>0$ and the first inequality
$2+\tilde{B}>-\tilde{K}^o\tilde\eta^o/\tilde\eta$ are required for stable waves, while the second inequality is the exceptional line $-\tilde{K}^o\tilde\eta^o/\tilde\eta =2+\tilde{B}-2({1+\tilde{B}-\tilde{K}^{o^2}})^{1/2}
$ (see Fig.~\ref{appfig:areaRatio}(a)).  

Strengthening the odd coupling  $|\tilde{K}^o\tilde\eta^o|$ increases the wave velocity ($\sim \mathrm{Re}[\omega'] = (-a_1^2+4a_0 a_2)^{1/2}/2$) until reaching the $\tilde\eta^o$-induced instability (note that $a_2=1$ in the case we consider). 
The maximal velocity occurs at the instability boundary and increases when moving away from the $\tilde{K}^o$-induced instability, as indicated in Fig.~\ref{appfig:areaRatio}(a). At the tripoint (purple circle), where the instability and propagating boundaries meet, waves no longer propagate or decay because $\omega'=0$.

The coupling $\tilde{K}^o\tilde\eta^o$ also causes the phase-shift in the propagating waves $|\Phi|= \text{Tan}^{-1}[ (-a_1^2+4a_0)^{1/2}/2(1+\tilde{B})]$ in the eigenmode amplitudes $\{ \bar{u}_L,\bar{u}_T \} = \{\tilde{K}^o,  1+\tilde{B}-i\omega' \}$. 
Larger $|\tilde{K}^o\tilde\eta^o|$ affects $|\Phi|$ in a  non-monotonic way, see inset of Fig.~2(c)
in the main text.
Near the propagating boundary (the exceptional line), $|\Phi|$ roughly scales linearly with the wave velocity. 
It increases until reaching a local maximum at $\tilde{K}^o\tilde\eta^o/\tilde\eta = -\tilde{B}-[2\tilde{K}^{o^2}/(1+\tilde{B})]$
before attaining a constant value $\text{Tan}^{-1}[2a_0^{1/2}/(1+\tilde{B})]$ at the instability boundary, which decreases towards the tripoint.

While in 2D there is no typical wave polarization in real space by construction (polarization requires two transverse directions), a similar phenomenon to elliptic polarization arises due to the presence of $\Phi$ in the \textit{shear deformation} space~\cite{scheibner2020}. 
Note that longitudinal and transverse waves create an orthogonal basis in the shear deformation space of pure shear ($\nabla_x u_x-\nabla_y u_y$, $S_1$-axis) and shear2 ($\nabla_x u_y + \nabla_y u_x$, $S_2$-axis), see Fig.~\ref{appfig:areaRatio}(b).
However, the directions of longitudinal and transverse waves in this shear space depend on the wave propagation direction. Only for simple $x$-traveling waves these go along the $S_1$- and $S_2$-axes. 
%
%

When our odd viscoelastic waves start to propagate, the wave eigenmodes couple the longitudinal and transverse modes with the phase-shift $\Phi$, such that in the shear-deformation space they follow an elliptic trajectory (i.e., elliptically polarized) that does not go through the origin, see Fig.~\ref{appfig:areaRatio}(b) (the figure disregards the temporal damping pre-factor). 
%

Since the destabilization effects from the odd coupling $\tilde{K}^o\tilde\eta^o$ drive wave propagation, it is expected that a larger bulk modulus $\tilde{B}$, which stabilizes the system, will decrease the region of wave propagation. 
Indeed, increasing $\tilde{B}$ expands the stable region, while the ratio between the areas of the propagating and stable regions decreases as $2\sqrt{1+\tilde{B}}/(2+\tilde{B})$ (see Fig.~\ref{appfig:areaRatio}(c)). The area ratio is calculated using Eq.~\eqref{appeq:wave prop condition} as the fraction of the two diverging integrals:
\begin{align}
\frac{\int_0^{({1+\tilde{B}})^{1/2}} 
\Big[
{2\sqrt{1+\tilde{B}-\tilde{K}^{o^2}}}/{\tilde{K}^o}
\Big]
d\tilde{K}^o }{\int_0^{({1+\tilde{B}})^{1/2}}  \Big[{(2+\tilde{B})}/{\tilde{K}^o}\Big] d\tilde{K}^o}  
=
\frac{2\sqrt{1+\tilde{B}}}{2+\tilde{B}}
\, .
\end{align} 

\end{document}